%% file: main.tex
\documentclass[english,11pt]{article}
\pdfoutput=1
\usepackage[sort,comma]{natbib}
\usepackage[pagebackref=false,colorlinks,linktocpage=TRUE]{hyperref}
\hypersetup{pdffitwindow=true,linkcolor=blue,citecolor=blue,urlcolor=blue}
\usepackage{amsmath}
\usepackage{bbm}
\usepackage{multirow}

\usepackage{amsfonts}
\usepackage{mathdots}
\usepackage{amsthm}

\usepackage{multirow}
\usepackage{centernot}
\usepackage{todonotes}
\usepackage[ruled,linesnumbered]{algorithm2e}
\usepackage{float}
\usepackage{optidef}
\usepackage{dsfont}

\usepackage{lmodern}
\usepackage{amssymb,amsmath}
\usepackage{ifxetex,ifluatex}
\usepackage{fixltx2e} 
\ifnum 0\ifxetex 1\fi\ifluatex 1\fi=0 
  \usepackage[T1]{fontenc}
  \usepackage[utf8]{inputenc}
\else 
  \ifxetex
    \usepackage{mathspec}
  \else
    \usepackage{fontspec}
  \fi
  \defaultfontfeatures{Ligatures=TeX,Scale=MatchLowercase}
\fi
\IfFileExists{upquote.sty}{\usepackage{upquote}}{}
\IfFileExists{microtype.sty}{%
\usepackage{microtype}
\UseMicrotypeSet[protrusion]{basicmath} 
}{}
\usepackage{hyperref}
\hypersetup{unicode=true,
            pdftitle={pcaNet example output},
            pdfborder={0 0 0},
            breaklinks=true}
\usepackage[margin=1in]{geometry}
\usepackage{color}
\usepackage{fancyvrb}

\DefineVerbatimEnvironment{Highlighting}{Verbatim}{commandchars=\\\{\}}
\usepackage{framed}

\SetKwInOut{Input}{Input}
\SetKwInOut{Output}{Output\,}
\SetKwInOut{Data}{Data}

\DeclareMathOperator*{\argmax}{arg\,max}
\DeclareMathOperator*{\argmin}{arg\,min}

\definecolor{shadecolor}{RGB}{248,248,248}

\usepackage[font={footnotesize,it}, sc, bf]{caption}
\renewenvironment{abstract}{%
  \noindent\textbf\abstractname .\hspace{1pt}
}{
  \endlist \par\bigskip\bigskip
}

\usepackage{lastpage}
\RequirePackage[hyperpageref]{backref}
\renewcommand*{\backref}[1]{} 
\renewcommand*{\backrefalt}[4]{
    \ifcase #1
       No referred.
    \or
       \emph{Referred to on page #2.}
    \else
       \emph{Referred to on pages #2.}
    \fi}

\usepackage[shortlabels]{enumitem}

\usepackage[titletoc,title]{appendix}

\binoppenalty=\maxdimen
\relpenalty=\maxdimen

\usepackage{xcolor}
\definecolor{cambridgebluecore}{RGB}{0, 176, 185}
\definecolor{cambridgebluedark}{RGB}{17, 94, 103}
\definecolor{cambridgebluelight}{RGB}{133, 176, 154}

\definecolor{cambridgeredcore}{RGB}{213, 0, 50}
\definecolor{cambridgereddark}{RGB}{138, 21, 56}
\definecolor{cambridgeredlight}{RGB}{232, 156, 174}

\definecolor{cambridgeblue2core}{RGB}{0, 114, 206}
\definecolor{cambridgeblue2dark}{RGB}{0, 60, 113}
\definecolor{cambridgeblue2light}{RGB}{108, 172, 228}

\definecolor{cambridgeorangecore}{RGB}{232, 119, 34}
\definecolor{cambridgeorangedark}{RGB}{190, 77, 0}
\definecolor{cambridgeorangelight}{RGB}{241, 190, 72}

\definecolor{cambridgegreencore}{RGB}{100, 167, 111}
\definecolor{cambridgegreendark}{RGB}{78, 91, 49}
\definecolor{cambridgegreenlight}{RGB}{183,	191, 16}

\usepackage{amsthm}
\usepackage{thmtools}

\usepackage[]{graphicx}
\graphicspath{{../figures/}}
\usepackage{subcaption}
\usepackage{epstopdf}
\usepackage{tikz}
\tikzset{font={\fontsize{9pt}{8}\selectfont}}
\usetikzlibrary{positioning}

\title{Two-step penalised logistic regression for multi-omic data with an application to cardiometabolic syndrome.}
\author{Alessandra Cabassi, Denis Seyres, Mattia Frontini, Paul D. W. Kirk}

\begin{document}

\begin{center}
{\LARGE\bf Two-step penalised logistic regression for multi-omic data with an application to cardiometabolic syndrome}
\end{center}
\medskip
\begin{center}
{\large Alessandra Cabassi$^{1}$, Denis Seyres$^{2,3,4}$, Mattia Frontini$^{3,4,5,6}$, and Paul D. W. Kirk$^{1,7}$ \\[15pt]

\emph{$^{1}$MRC Biostatistics Unit, University of Cambridge, UK}\\
\emph{$^{2}$National Institute for Health Research BioResource, Cambridge University Hospitals, UK}\\
\emph{$^{3}$Department of Haematology, University of Cambridge, UK}\\
\emph{$^{4}$NHS Blood and Transplant, Cambridge Biomedical Campus, Cambridge, UK}\\
\emph{$^{5}$Institute of Biomedical \& Clinical Science, College of Medicine and Health,\\ University of Exeter Medical School, UK}\\
\emph{$^{6}$British Heart Foundation Centre of Excellence, Cambridge Biomedical Campus, UK}\\
\emph{$^{7}$Cambridge Institute of Therapeutic Immunology \& Infectious Disease,\\ University of Cambridge, UK}\\
}

\end{center}

\bigskip

\begin{center}
Preprint, \today
\end{center}
\bigskip\bigskip

\begin{abstract}\newline
\textbf{Summary:} Building classification models that predict a binary class label on the basis of high dimensional multi-omics datasets poses several challenges, due to the typically widely differing characteristics of the data  layers in terms of number of predictors, type of data, and levels of noise. Previous research has shown that applying classical logistic regression with elastic-net penalty to these datasets can lead to poor results \citep{liu2018data}. We implement a two-step approach to multi-omic logistic regression in which variable selection is performed on each layer separately and a predictive model is then built using the variables selected in the first step. Here, our approach is compared to other methods that have been developed for the same purpose, and we adapt existing software for multi-omic linear regression \citep{zhao2020structured} to the logistic regression setting. Extensive simulation studies show that our approach should be preferred if the goal is to select as many relevant predictors as possible, as well as achieving prediction performances comparable to those of the best competitors. Our motivating example is a cardiometabolic syndrome dataset comprising eight 'omic data types for 2 extreme phenotype groups (10 obese and 10 lipodystrophy individuals) and 185 blood donors. Our proposed approach allows us to identify features that characterise cardiometabolic syndrome at the molecular level.\\
\textbf{Availability:} \textsf{R} code is available at \url{https://github.com/acabassi/logistic-regression-for-multi-omic-data}. \\
\textbf{Contact:} \href{alessandra.cabassi@mrc-bsu.cam.ac.uk}{alessandra.cabassi@mrc-bsu.cam.ac.uk}, \href{paul.kirk@mrc-bsu.cam.ac.uk}{paul.kirk@mrc-bsu.cam.ac.uk}.\\
\end{abstract}

\section{Introduction}
In this work, we focus on the problem of making predictions for a binary variable using multiple high-dimensional 'omic layers. In the context of precision medicine, this can be used for example, to sort patients into low- and high-risk groups for a certain disease, as we show in our application. It is well established that building predictive models for datasets with a large number of variables $p$ compared to the number of statistical units $n$ requires extra care, since classical statistical methods designed for large $n$ small $p$ situations may fail or overfit the data. Combining multiple large $p$, small $n$ datasets of different types raises new questions. The large $p$ small $n$ problem is exacerbated by the presence of multiple datasets. Moreover, different datasets can have different scales, number of covariates, and fractions of covariates that are associated with the outcome of interest. Multi-omic datasets can also have varying numbers of correlated variables both within and across 'omic layers. Furthermore, it may be important to balance the predictive performance of the model with the stability of the set of selected variables \citep{kirk2010discussion, meinshausen2010stability, kirk2011retro,kirk2013balancing}.

Many variable selection and binary predictive models exist (see Section~\ref{sec:literature-review} for a brief review). Here, we focus on penalised likelihood approaches, which are scalable, widely used and have the advantage of having efficient implementations available. In particular, we make use of the elastic-net, which allows us to be flexible in terms of how we treat correlations among the variables. In this context, multiple studies have shown that it can be beneficial to apply different penalties to each 'omic layer \citep{boulesteix2017ipf, liu2018data}. The problem with these approaches is then how to choose the values of the penalty parameters. Indeed, while classical penalised likelihood approaches employ cross-validation to select the parameters of the model, the computational cost can become prohibitive when the number of parameters increases. For this reason, \citet{zhao2020structured} resort to a Gaussian process approximation to the prediction error surface in order to more efficiently find the optimal parameters of the model. Another important question is whether or not one wants, or needs, to ensure that all layers contribute to the predictive model, irrespective of their relative dimensionality.

In our motivating example, we are interested in finding a molecular signature of cardiometabolic disease in each of the eight available layers. For this reason, we propose a two-step approach in which the variable selection step is performed on each 'omic layer separately, so that the signal from smaller datasets is not obscured by those of the larger ones. In the second step, all the selected variables are stacked together to fit a ridge-penalised logistic regression model. We compare this approach to simply applying elastic-net on the full dataset, to a newly developed integrative method that fits a regression model on all data types together, but assigning different penalty factors to each of them \citep{zhao2020structured}, and to a simple, univariate approach.

We consider a wide range of simulation studies. In each simulation setting there are two data types with varying characteristics, as well as a smaller data type that only contains features that are known to be associated with the outcome of interest, and therefore are not penalised. In real data applications, this corresponds to having two 'omic datasets and a small set of clinical parameters. The simulation studies show that, depending on the goal of the analysis, different integrative methods should be preferred. If the objective is to make accurate predictions and, at the same time, identify the highest possible number of variables that are relevant for the problem at hand, then our approach is the most suitable.

Our motivating example is a multi-omic dataset collected and analysed in \citet{seyres2020transcriptional} to understand and identify the molecular characterisation of cardiometabolic syndrome. We analyse eight different types of 'omic data from 185 blood donors, as well as 10 obese and 10 lipodystrophy individuals. As in \citet{seyres2020transcriptional}, we use these data to identify putative signatures of cardiometabolic syndrome and build a predictive model to determine the probability of belonging to the obese group. Here, we go beyond the analysis previously performed by investigating the impact of the choice of the elastic-net parameter on the estimated probabilities and the variables selected, by comparing to a rank aggregation approach, and by additionally constructing a predictive model to determine the probability of belonging to the lipodystrophy group.

\section{Methods}
First, we briefly recall the basics of penalised logistic regression in Section \ref{sec:penalised-logistic-regression}. Then, a review of the predictive models of this type that have been used to integrate multiple 'omic datasets is given in Section \ref{sec:literature-review}. In Section \ref{sec:epsgo} we give the details of the EPSGO (\emph{efficient parameter selection via global optimisation}) algorithm, which is extensively used in the rest of this work. In Section \ref{sec:separate-en} we suggest two novel approaches to integrate multiple 'omic datasets in the framework of logistic regression.

\subsection{Penalised logistic regression}
\label{sec:penalised-logistic-regression}
In traditional logistic regression settings, one has a dataset $X \in \mathbb{R}^{N \times P}$ made by $N$ observations $\mathbf{x}_n$, $n = 1, \dots, N$, for which a set of $p$ variables has been measured, and by a set of binary responses $\mathbf{y} = [y_1, \dots, y_N] \in \{0, 1\}^N$, one for each observation in $X$. The goal is then to build a model that predicts the probability that the response $y$ corresponding to a new observation $\mathbf{x}$ is equal to one \citep{hastie2009elements,cramer2002origins}. This is done via the \emph{logistic function}
\begin{equation}
	\text{Pr}(Y = 1 | \mathbf{x}) =  \frac{e^{\beta_0 + \boldsymbol{\beta}\mathbf{x}}}{1 + e^{\beta_0 + \boldsymbol{\beta}\mathbf{x}}},
	\label{eq:logistic-regr-first-def}
\end{equation}
where $\beta_0$ and $\boldsymbol{\beta} = [\beta_1, \dots, \beta_p]$ are the so-called regression coefficients. In particular, $\beta_0$ is the \emph{intercept} of the model and each $\beta_p$ is the coefficient corresponding to the $p$th variable (i.e. column) in $X$.

In the presence of large numbers of predictors, the estimates of the regression coefficients $\beta_0$ and $\boldsymbol{\beta}$ given by solving the optimisation problem above are highly variable. To avoid this problem shrinkage methods are often used.
In what follows, we focus on the \emph{elastic-net} (EN):
\begin{equation} \label{eq:elasticnet}
	\underset{\boldsymbol{\beta_0} \in \mathbb{R},\ \boldsymbol{\beta} \in \mathbb{R}^P}{\min} \ - l(X, \mathbf{y} ; \beta_0, \boldsymbol{\beta}) + \lambda \left[(1 - \alpha)  \|\boldsymbol{\beta}\|_2 + \alpha \| \boldsymbol{\beta} \|_1 \right]
\end{equation}
Here, $\|\cdot\|_1$ and $\|\cdot\|_2$ represent the $l_1$ and $l_2$ norms respectively, $\alpha$ is the weight assigned to the $l_1$ penalty, and $1-\alpha$ is the weight assigned to the $l_2$ penalty, and $\lambda$ is a parameter used to determined the strength of the penalty \citep{zou2005regularization}. The parameter $\lambda$ is usually selected via cross-validation (CV; \citealp{kohavi1995study}), while $\alpha$ can either be tuned using CV or chosen so as to give the desired number of selected variables.

\subsection{Literature review}
\label{sec:literature-review}
As mentioned in the introduction, the integration of multiple 'omic datasets in the context of prediction cannot be done via the classical methods for penalised logistic regression such as those presented in the previous section, but requires the development of novel statistical methods. The main ideas behind the methods available in the literature are illustrated below. For simplicity, we refer to the different 'omic datasets as data layers.

One of the first examples of predictive models for multi-omic datasets is that of \citet{zhao2015combining}. First, they apply LASSO regression to a multi-omic dataset in order to do variable selection in each layer separately, then they use the selected variables in a $l_2$-penalised Cox regression model.

\citet{boulesteix2017ipf} instead developed a bespoke penalised regression method for multi-omic data. It is similar to a LASSO regression, but it assigns a different penalty to each layer. This approach is called IPF-LASSO: \emph{integrative LASSO with penalty factors}. Denoting by $M$ the number of data layers and by $X_m$ each layer's data matrix, where $m=1, \dots, M$, Boulesteix \emph{et al.}, like us, are interested in those situations where each layer has observations for the same $N$ individuals and a different set of $P_m$ features, i.e. $X_m\in\mathbb{R}^{N \times {P_m}}$ and the rows in each matrix $X_m$ correspond to the same statistical units. Let $\beta_j^{(m)}$ be the regression coefficient for the $j$th feature of the $m$th layer. IPF-LASSO tries to find the optimal set of coefficients $\boldsymbol{\beta} = [\beta_1^{(1)}, \dots, \beta_{P_1}^{(1)}, \dots, \beta_1^{(M)}, \dots, \beta_{P_M}^{(M)}]$ such that
\begin{equation}
		\underset{\beta_0 \in \mathbb{R},\ \boldsymbol{\beta} \in \mathbb{R}^{P_1 + \dots + P_M}}{\min}  \ - l(X, \mathbf{y} ; \beta_0, \boldsymbol{\beta}) + \sum_{m=1}^M \lambda_m \|\boldsymbol{\beta}^{(m)} \|_1.
			\label{eq:ipf-lasso}
\end{equation}
Boulesteix \emph{et al.} suggest to choose the penalty parameters by a double CV approach. First, for each of candidate set of penalties $\lambda_2, \dots, \lambda_M$, all the predictors are rescaled as follows:
\begin{equation}
x_{ij}^* = \frac{x_{ij}^{(m)}}{\lambda_m/\lambda_1},
\end{equation}
where $i = 1, \dots, n$, $j = 1, \dots, P_{m}$, and $m=1,\dots,M$ (note that the features in the first layer remain unchanged). Then, thanks to this, the same penalty $\lambda_1$ can be applied to all the scaled variables, and the parameter $\lambda_1$ is estimated via CV in the standard way. The candidate set of penalties $\lambda_2, \dots, \lambda_M$ that gives the best prediction performance is then selected together with the corresponding value of $\lambda_1$ found via CV. How to choose the candidate penalty factors $\lambda_2, \dots, \lambda_M$ remains an open question; the authors  pick a grid of predefined values $2^k, k = -a, -(a-1), \dots, 0, \dots, a-1, a$ where $a$ is an integer that varies between 3 and 6 depending on the application. The limitation of this approach is that the computational burden increases very quickly with the number of layers $M$, as the number of candidate sets grows exponentially with $M$, making it impossible to explore a large set of possibilities for the penalty terms. Boulesteix \emph{et al.} apply this method to a wide range of simulation settings, as well as real datasets on acute myeloid leukaemia \citep{tcga2013genomic} and breast cancer \citep{hatzis2011genomic} where the outcomes are overall survival time and relapse-free survival time respectively.

Similarly to what Boulesteix \emph{et al.} did for LASSO, \citet{liu2018data} show that, if the number of informative features is not the same in each dataset, fitting EN regression models with different penalties for each dataset yields better predictions than having a single global penalty. They do so by defining a \emph{multi-tuning parameter elastic-net} regression (MTP-EN). For simplicity, let
\begin{equation}
	N(\beta) = (1-\alpha) \| \boldsymbol{\beta} \|_2 + \alpha \| \boldsymbol{\beta} \|_1.
\end{equation}
Then, the regression parameters of MTP-EN are found by solving the penalised regression problem
\begin{equation}
\underset{\beta_0 \in \mathbb{R},\ \boldsymbol{\beta} \in \mathbb{R}^{P_1 + \dots + P_M}}{\min} \ - l(X, \mathbf{y} ; \beta_0, \boldsymbol{\beta}) + \lambda_1 N (\boldsymbol{\beta}_1) + \dots + \lambda_M N (\boldsymbol{\beta}_M)
\end{equation} 
This corresponds to fitting a weighted EN model
\begin{equation}
	\underset{\beta_0 \in \mathbb{R},\ \boldsymbol{\beta} \in \mathbb{R}^{P_1 + \dots + P_M}}{\min} \ - l(X, \mathbf{y} ; \beta_0, \boldsymbol{\beta}) + \lambda_1 N_w (\boldsymbol{\beta})
	\label{eq:mtp-en}
\end{equation}
where 
\begin{equation}
N_w(\boldsymbol{\beta}) = \alpha \sum_{p=1}^{P_1+\dots+P_M} w_p |\beta_p| + (1 - \alpha) \sum_{p=1}^{P_1+\dots+P_M} w_p \beta_p^2,
\end{equation}
and the weights $w_p$ are $\mathbf{w} = [1, \dots, 1, \lambda_2/\lambda_1, \dots, \lambda_2/\lambda_1, \dots, \lambda_M/\lambda_1, \dots, \lambda_M/\lambda_1]$.

Both IPF-LASSO and MTP-EN can be easily fitted using the ``glmnet'' \textsf{R} package \citep{friedman2010regularization}, specifying the weights in the \verb|penalty.factor| argument of the \verb|cv.glmnet| function. However, there remains the problem of choosing the penalties of all 'omic layers except the first one.

IPF-LASSO has recently been extended by \citet{zhao2020structured} to combine it with the tree-guided group LASSO of \citet{kim2012tree} for which the grid search-type approach proposed by \citet{boulesteix2017ipf} is not a viable option, given its high computational cost. Therefore, they use the \emph{efficient parameter selection via global optimisation} (EPSGO) algorithm of \citet{frohlich2005efficient} instead (details of the algorithm are given in Section \ref{sec:epsgo}). In the same manuscript, \citet{zhao2020structured} also give a more flexible formulation of the structured penalised regression model of \citet{liu2018data} that allows different values of the parameter $\alpha$ for each layer:
\begin{equation}
\underset{\boldsymbol{\beta_0} \in \mathbb{R},\ \boldsymbol{\beta} \in \mathbb{R}^{P_1 + \dots + P_M}}{\min} \ - l(X, \mathbf{y} ; \beta_0, \boldsymbol{\beta}) + \lambda_1 N_1 (\boldsymbol{\beta}_1) +  \dots  + \lambda_M N_M (\boldsymbol{\beta}_M)
\end{equation} 
where $N_m(\boldsymbol{\beta}_M) = (1 - \alpha_m)\| \boldsymbol{\beta}_m \|_2  + \alpha_m \| \boldsymbol{\beta}_m \|_1$.

\citet{zhao2020structured} call the model of \citet{liu2018data} \emph{sIPF-EN}  (where the ``s'' stands for ``simple''), and this new, more general one, \emph{IPF-EN}. This is the naming convention that is used in the remainder of this manuscript.

\subsection{Efficient parameter selection via global optimisation}
\label{sec:epsgo}
We provide a short introduction to online Gaussian processes, which are central to the EPSGO algorithm, and then describe the EPSGO algorithm in more detail.

\subsubsection{Online Gaussian processes}

A \emph{Gaussian process} (GP) is a collection of random variables, any finite number of which have a joint Gaussian distribution \citep{rasmussen2006gaussian}.

A GP is fully specified by its mean function $\mu : \mathcal{X} \to \mathbb{R}$ and covariance function $\kappa: \mathcal{X} \times \mathcal{X} \to \mathbb{R}$. Here we assume that $\mathcal{X} \subseteq \mathbb{R}^D$. Given a finite sample of points $\mathbf{x} = \left[x_1, \dots, x_N\right]$,  the GP provides a probabilistic model for the function $q(\mathbf{x}) $. For this reason, Gaussian processes are often used as priors on functions. We indicate this by
\begin{equation}
	q \sim \mathcal{GP} (\mu, \kappa).
\end{equation}
For simplicity, the mean function is often considered to take value zero on the entire domain, without loss of generality. 

Typically one does not observe exactly $q(\mathbf{x})$, but has instead available noisy observations of $q(\mathbf{x})$, which we denote as
\begin{equation}
	y = q(\mathbf{x}) + \epsilon,
\end{equation}
where $\epsilon$ is independent and identically distributed (i.i.d.) noise with variance $\sigma^2_{\epsilon}$. Equivalently, we can write
\begin{equation}
\mathbf{y} | \mathbf{x} \sim \mathcal{N} \left(\mathbf{0},k(\mathbf{x}, \mathbf{x}) + \sigma^2_{\epsilon} I \right)
\end{equation}
where $I$ is the identity matrix.

Using standard properties of Gaussian distributions, one can show that,  given a set of $N$ data points $\mathcal{D}_N = \left\{ d_1, \dots, d_N \right\}$, where $d_n = (x_n, y_n)$, the predictive distribution conditioned of $\mathcal{D}_N$ is also Gaussian:
\begin{equation}
q(\mathbf{x}_{\text{new}}) | \mathbf{x}_{\text{new}}, \mathcal{D}_N \sim \mathcal{N} \left(\mu_N (\mathbf{x}_{\text{new}}), K_N(\mathbf{x}_{\text{new}}, \mathbf{x}_{\text{new}})\right).
\end{equation}
for any finite set of points $\mathbf{x}_{\text{new}} = [x^{\text{new}}_1, \dots, x^{\text{new}}_N ]$ where the conditional mean vector $\mu_N$ and covariance matrix $K_N$, are available in closed form \citep[Chapter~2]{rasmussen2006gaussian}. From the definition of GP, it follows that the posterior distribution $q$ is a Gaussian process:
\begin{equation}
q | \mathcal{D}_{\tau} \sim \mathcal{GP}(\mu_{\tau}, k_{\tau}).
\end{equation}

The predictive distribution can also be updated in an iterative fashion using Bayes' rule. Having observed $N$ data points, once a new data point $d_{t+1}$ is available, the updated predictive distribution is:
\begin{equation}
	p_{N+1} (y^{\text{new}} | d_{N+1}) = \frac{p(d_{N+1} | y^{\text{new}})\hat{p}(y^{\text{new}})}{\int p(d_{N+1} | y)\hat{p}(y)dy}.
\end{equation}
Here the integral in the denominator is intractable, so the predictive distribution after observing $N$ points is denoted by $\hat{p}_N$ to indicate that only an approximation is available.

Both the expected model value $\hat{\mu}_N(x)$ and the estimated variance of the model $\hat{\sigma}_N^2(x)$ can be evaluated by recursive formulae that can be easily updated as soon as a new data point is available \citep{csato2002sparse}.

\subsubsection{The EPSGO algorithm}
The EPSGO algorithm was initially developed by \citet{frohlich2005efficient} to efficiently tune the parameters of support vector machines (SVMs) and subsequently used by \citet{sill2014c060} to select the parameters $\alpha$ and $\lambda$ of EN (Equation \ref{eq:elasticnet}). The implementation of Sill \emph{et al.} is used in the \textsf{R} package ``IPFStructPenalty'' of \citet{zhao2020structured} to tune the parameters of sIPF-EN and IPF-EN.

The idea is to reframe the task of tuning the model parameters as an optimisation problem. Denoting by $\mathcal{X}$ the parameter space of the model of interest, and by $q: \mathcal{X} \subseteq \mathbb{R}^D \to \mathbb{R}$ a measure of the quality of the model, in the case of logistic regression models, this can be the out-of-sample misclassification rate (MR). The goal is to find the parameters $\mathbf{x}^*$ such that
\begin{equation}
	\mathbf{x}^* = \underset{\mathbf{x \in \mathcal{X}}}{\argmin}\ q(\mathbf{x}).
\end{equation}
To do so, the EPSGO algorithm models the prior on the error surface $q$ on the parameter space $\mathcal{X}$ as a Gaussian process.

In the first step of the algorithm, some points are sampled from the parameter space that are used to fit an online GP. In order to obtain a good coverage of the parameter space $\mathcal{X}$, the Latin hypercube sampling strategy of \citet{mckay1979comparison} is used. The recommended number of points to be sampled at this stage is $N = 10D$.

Once the online GP has been fit on a set of $N$ points,  the improvement function is defined as
\begin{equation}
	I(\mathbf{x}) = \max \left\lbrace q_{\text{min}} - Y, 0 \right\rbrace.
\end{equation}
where $q_{\text{min}}$ indicates the smallest value of the function $q$ observed up to the current iteration and $Y \sim \mathcal{N}(\hat{\mu}_N, \hat{\sigma}_N)$. A new point $\mathbf{x}^{\text{new}}$ in the parameter space $\mathcal{X}$ is chosen so as to maximise the expected improvement criterion of \citet{jones1998efficient}:
\begin{equation}
	\mathbf{x}^{\text{new}} = \argmax_{\mathbf{x} \in \mathcal{X}} \mathbb{E}[I(\mathbf{x})].
\end{equation}
and the online GP is updated after evaluating the error surface at $\mathbf{x}^{\text{new}}$. This procedure is repeated until convergence is reached.

\subsection{Penalised logistic regression for multi-omic data}
\label{sec:separate-en}
We propose two ways of doing penalised logistic regression on multi-omic data: (1) \emph{separate EN on each layer with fixed $\alpha$}; and (2) \emph{separate EN on each layer where $\alpha$ is selected via EPSGO}.

\paragraph{1. Separate EN on each layer with fixed $\boldsymbol{\alpha}$} First, a variable selection step is performed on each 'omic layer separately using EN with a fixed value of $\alpha$. The selected variables are then used to build a predictive model using ridge-penalised logistic regression. All regression models are fitted using the ``glmnet'' \textsf{R} package and the values of $\lambda$ in each model are selected via CV. The value of $\alpha$ should be chosen depending on the particular application; here we explore how the performances of the method change for different values of $\alpha$. For our simulation studies and real data analysis we find $\alpha=0.1$ to be a reasonable value.

\paragraph{2. Separate EN on each layer, $\boldsymbol{\alpha}$ selected via EPSGO} The difference between this method and the previous one is that the EPSGO algorithm is used in the first step to pick an optimal value for $\alpha$ in each 'omic layer. This can be convenient when the user does not have a particular preference for the value of $\alpha$. However, we show in Section \ref{sec:cms-multivariate-signature-identification} that this approach is not always preferable to the previous one.

\section{Simulation study}
\label{sec:ipf-simulations}
We perform a simulation study in order to compare the two approaches  presented in the previous section to their main competitors: \emph{na\"ive EN} and \emph{sIPF-EN}, detailed below. To this end, we modify the implementations of na\"ive EN and sIPF-EN of the \textsf{R} package ``IPFStructPenalty'', which currently only handles linear regression, in order to do logistic regression. The two other methods are implemented from scratch, heavily relying on the ``glmnet'' and ``IPFStructPenalty'' \textsf{R} packages. We also consider a univariate approach. The code used to produce all the results presented below is available at \url{https://github.com/acabassi/logistic-regression-for-multi-omic-data}.

\paragraph{Na\"{i}ve EN} This is the original EN algorithm (Equation \ref{eq:elasticnet}) applied to all the 'omic layers stacked together. We make use of the EPSGO algorithm to automatically select the best value of $\alpha$, while $\lambda$ is chosen via CV.

\paragraph{sIPF-EN} As mentioned in Section \ref{sec:literature-review}, this is a variation of EN that assigns different penalty factors $\lambda$ to each layer, but selects the same value of $\alpha$ for each of them \citep{zhao2020structured}.

\paragraph{Univariate approach} For each 'omic variable, a logistic regression model is built where the only predictors are the variable of interest, and, where appropriate, any covariates that are known to be related to the outcome and therefore are always included in the model. If the null hypothesis that the regression coefficient of the variable of interest should be zero is rejected, then that variable is selected. A ridge-penalised regression model is then built using all the selected variables as well as the covariates that are always included in the model.

\subsection{Simulation settings}
Our simulation settings are similar to those of \citet{boulesteix2017ipf}. We generate three layers of data for each experiment, with $N=100$ observations each. The first layer represents a set of clinical covariates that are known to be related to the outcome of interest, and for this reason are not penalised. The other two layers represent two 'omic datasets with varying numbers of covariates and proportions of covariates that are correlated with the outcome. We denote the number of non-penalised covariates by $P_N$, the number of variables in the first and second penalised layers by $P_1$ and $P_2$ respectively. Each has a small number of relevant variables, denoted by $P_1^r$ and $P_2^r$ respectively.

In each dataset, the responses are drawn independently from a Bernoulli distribution with parameter $\tau = 0.5$. The variables are then drawn from the following multivariate Gaussian distributions:
\begin{align}
	\left[X_1, \dots, X_{P_N + P_1 + P_2}\right]^T | Y = 0 & \sim \mathcal{MN} (\boldsymbol{0}_{P_N + P_1 + P_2}, \Sigma), \\
	\left[X_1, \dots, X_{P_N + P_1 + P_2}\right]^T | Y = 1 & \sim \mathcal{MN} (\boldsymbol{\mu}_{P_N + P_1 + P_2}, \Sigma),
\end{align}
where 
\begin{equation}
\boldsymbol{\mu} = \left[\beta_1, \dots, \beta_1, 0, \dots, 0, \beta_2, \dots, \beta_2, 0, \dots, 0 \right]
\end{equation}
with $P_1^r$ elements of $\boldsymbol{\mu}$ equal to $\beta_1$ and $P_2^r$ elements equal to $\beta_2$.
The covariance matrix $\Sigma$ is either the identity matrix
\begin{equation} \label{eq:diagonal-matrix}
\Sigma_0 = \mathbb{I}_{P_N + P_1+P_2}
\end{equation} 
or a block diagonal matrix similar to the one considered in the simulation studies of \citet{boulesteix2017ipf} and \citet{zhao2020structured} that we indicate with $\Sigma_1$. The penalised layers have blocks of correlated variables both within and across layers. All the non-penalised covariates are correlated, but uncorrelated to the penalised ones. That is
\begin{equation} \label{eq:block-matrix-with-nonpen}
\Sigma_1 = 
\left[
\begin{array}{c | cccc | cccc}
N & & & & & & & & \\
\hline
& A_1 &  & & & B_{12} & & & \\
& & A_1 & & & &B_{12} & & \\
& & & \dots & & & & \dots & \\
& & & & A_1 & & & & B_{12}\\
\hline
& B_{21} & & & &  A_2 & & & \\
& &B_{21} & & & & A_2 & &\\
& & & \dots & & & & \dots & \\
& & & &B_{21} & & & & A_2 \\
\end{array}
\right]
\end{equation}
where $b=10$, $N$, $A_1$ and $A_2$ are matrices of size $P_N \times P_N$, $P_1/b \times P_1/b$, and $P_2/b \times P_2/b$ respectively with ones on the diagonal and all other elements equal to $\rho$ and $B_{12}$ and $B_{21}$ are matrices of size $P_1/b \times P_2/b$ and $P_2/b \times P_1/b$ respectively with all elements equal to $\rho$.

We consider the same sets of values for $P_1, P_2, P_1^r, P_2^r, \beta_1, \beta_2$ as Boulesteix \emph{et al.}, reported in Table \ref{tab:ipf-simulation-settings}. Moreover, we set $P_N = 2$ and $\beta_N = \beta_1$. The value of $\rho$ is equal to 0.4 in all simulation settings, as in Boulesteix \emph{et al.}

\begin{table}[h!]
\centering
\begin{tabular}{l | c c c c c c c c}
& $P_N$ &  $P_1$  & $P_2$ & $P_1^r$ & $P_2^r$ & $\beta_N$ & $\beta_1$ & $\beta_2$ \\
\hline
Setting A & 2 & 1000 & 1000 & 10 & 10 & 0.5 & 0.5 & 0.5 \\
Setting B & 2 & 100 & 1000 & 3 & 30 & 0.5 & 0.5 & 0.5\\
Setting C & 2 & 100 & 1000 & 10 & 10 & 0.5 & 0.5 & 0.5\\
Setting D & 2 & 100 & 1000 & 20 & 0 & 0.3 & 0.3 & -\\
Setting E & 2 & 20 & 1000 & 3 & 10 & 1 & 1 & 0.3\\
Setting F & 2 & 20 & 1000 & 15 & 3 & 0.5 & 0.5 & 0.5\\ 
\end{tabular}
\caption[Simulation settings.]{Values of $P_N, P_1, P_2, P_1^r, P_2^r,  \beta_N, \beta_1, \beta_2$ used for the simulation study.}
\label{tab:ipf-simulation-settings}
\end{table}

In the Supplementary Material we consider three additional sets of simulation settings. In the first one, only the two 'omic layers are included in the regression. In the other two, we consider again the same simulation scenarios presented here, but with $P_N = 10$ and $P_N = 100$. We also compare these methods to a different univariate selection method followed by a ridge regression on the selected variables.

\subsection{Simulation results}

\begin{figure}
\centering
\includegraphics[width=.95\textwidth]{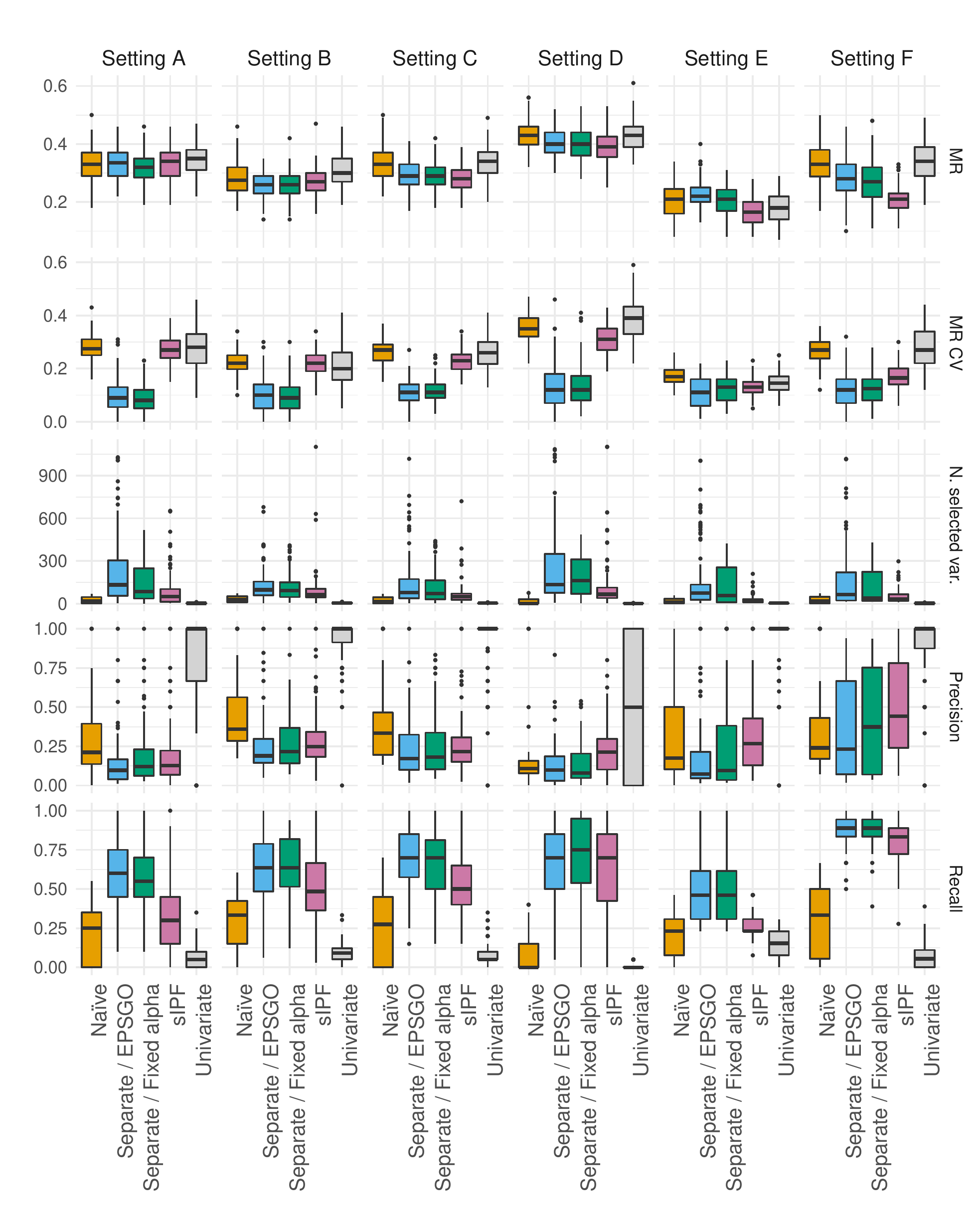}
\vspace{.2cm}
\caption[Simulation: EN for multi-omic data, diagonal covariance.]{Simulation study comparing  different variants of EN for multi-omic data. The covariance matrix used here is the diagonal matrix $\Sigma_0$. ``MR'' is the out-of-sample misclassification rate, ``MR CV'' the within-sample misclassification rate. The non-penalised covariates are not included when computing precision and recall.}
\label{fig:ipf-simulation-diagonalcov-fewnonpen}
\end{figure}

\begin{figure}
\centering
\includegraphics[width=.95\textwidth]{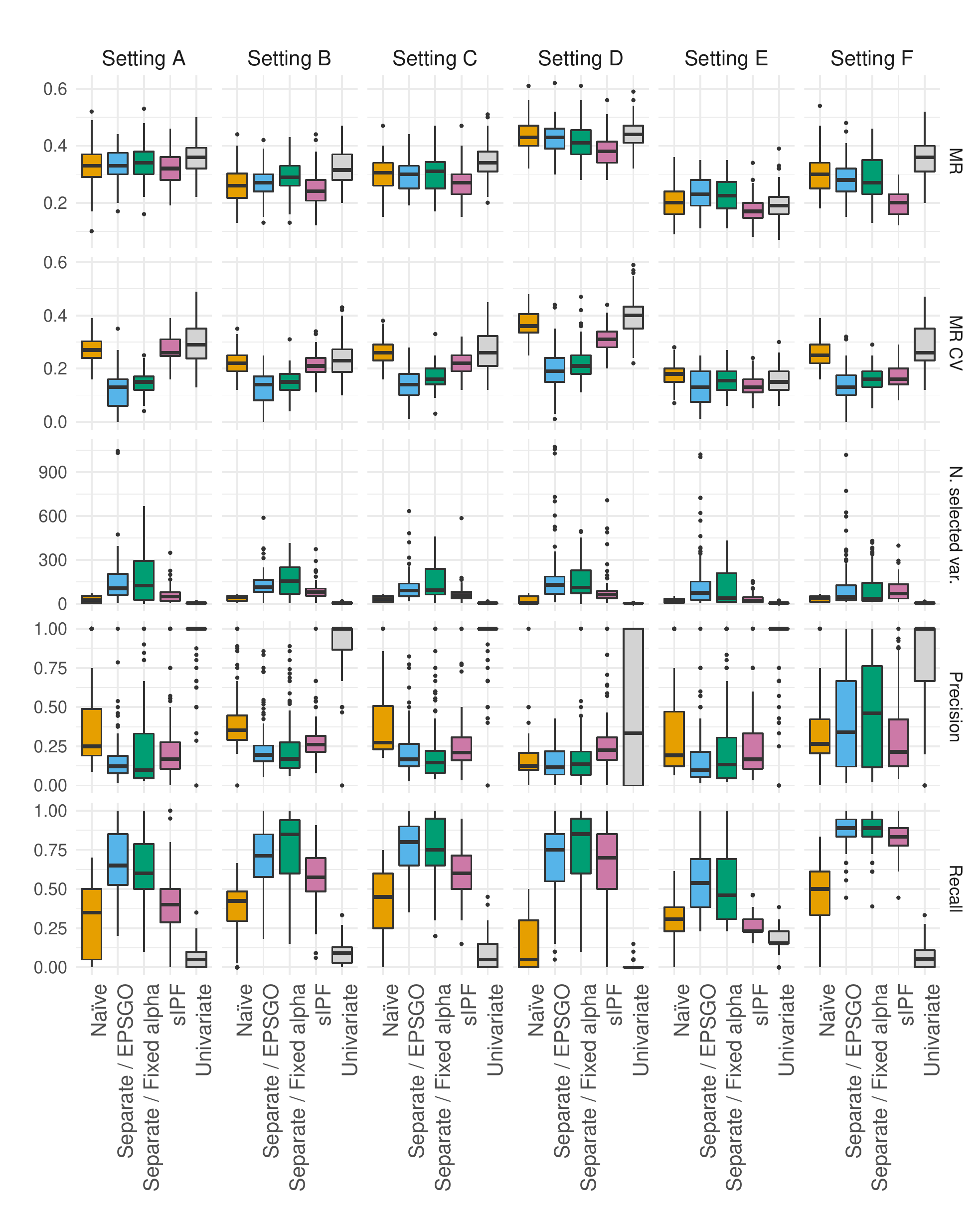}
\vspace{.2cm}
\caption[Simulation: EN for multi-omic data, block diagonal covariance.]{Simulation study comparing  different variants of EN for multi-omic data. The covariance matrix used here is the block matrix $\Sigma_1$. ``MR'' is the out-of-sample misclassification rate, ``MR CV'' the within-sample misclassification rate. The non-penalised covariates are not included when computing precision and recall.}
\label{fig:ifp-simulation-nondiagcov-fewnonpen}
\end{figure}

Figures \ref{fig:ipf-simulation-diagonalcov-fewnonpen} and \ref{fig:ifp-simulation-nondiagcov-fewnonpen} show the outcome of the simulation studies. For each setting and each algorithm, we report the following quantities: the MR on the test set, the MR on the training set,  the number of selected variables minus the number of non-penalised covariates, the proportion of selected variables that are among the relevant ones, excluding the non-penalised covariates (precision), the proportion of relevant variables that are selected by the algorithm, excluding the non-penalised covariates (recall).

Figure \ref{fig:ipf-simulation-diagonalcov-fewnonpen} shows that, when the covariates are uncorrelated, all methods have comparable out-of-sample MRs, except in settings E and F where the MR is slightly higher for the na\"ive approach and slightly lower for sIPF-EN. The within-sample MR is lower for the separate with EPSGO and separate with fixed $\alpha$ methods, suggesting that those two might be prone to overfitting. Concerning the precision, there is no clear pattern throughout settings. On the contrary, the two instances of separate regression on each layer consistently show higher values of the recall. Unsurprisingly, the same two algorithms also select the highest number of variables in all settings. This behaviour is opposite to that of the univariate method, which selects a very low number of variables and therefore has values of the recall always close to zero.

In Figure \ref{fig:ifp-simulation-nondiagcov-fewnonpen}, we see that if the covariates are correlated, sIPF-EN has the lowest MR, thanks to the fact that, contrarily to the other methods, it takes into account the correlation between data layers. The only other method that does this is na\"{i}ve EN, however, assigning the same penalty to all layers puts this method at a disadvantage in the settings where the two layers are highly unbalanced (i.e. settings D, E and F). In those settings, the MR of the two-step approaches is comparable if not better than that of na\"{i}ve-EN. Again, the within-sample MR suggests that the two algorithms that perform variable selection on each layer separately may be overfitting. As above, the same two methods select the highest number of variables. This is reflected in lower precision and higher recall, on average. As expected, the univariate approach has the worst performance overall.

\section{Application to cardiometabolic syndrome data}
\label{sec:application}

Cardiometabolic syndrome (CMS) is a combination of metabolic dysfunctions such as abdominal obesity, high levels of fasting glucose, elevated blood pressure, low-level inflammation, high level of ``bad’’ cholesterol, and low level of ``good'' cholesterol. While the exact causes of CMS are not known, it has been shown to be associated with higher risk of type 2 diabetes and cardiovascular disease \citep{grundy2005diagnosis}.

Here we seek to identify a set of molecular features that characterise CMS by integrating multiple ’omic layers. This can give insights into the molecular mechanisms driving the development of this disease and identify relevant biological markers. Moreover, these can be used to stratify the undiagnosed population for their probability of being affected by CMS.

We consider the data of \citet{seyres2020transcriptional}, which contain multiple 'omic datasets as well as a set of anthropometric and biochemical parameters for:
\begin{itemize}
\item 185 non-obese and non-diabetic blood donors, recruited amongst the UK’s National Health Service Blood and Transplant donors;
\item 10 patients affected by familial partial lipodystrophy syndrome, which is characterised by loss of subcutaneous fat, cared for by the National Severe Insulin Resistance Service at Addenbrooke's Hospital in Cambridge;
\item 10 morbidly obese individuals (BMI greater than 40), who were referred for bariatric surgery by the Obesity Clinic of Addenbrooke's Hospital in Cambridge.
\end{itemize}

For each individual, ChIP-seq, RNA-seq data, and DNA methylation data were collected from two types of white blood cells: monocytes and neutrophils. In addition to that, metabolites and lipids were measured from plasma. The ChIP-seq dataset contains 25,600 and 26,300 peaks for monocytes and neutrophils respectively. The RNA-seq dataset contains 10,433 genes for monocytes and 20,597 for neutrophils. The DNA methylation dataset has values for 26,214 CpG sites in the monocytes and 21,442 in neutrophils. The metabolite and lipid datasets contain 988 and 123 features respectively.

Based on their clinical parameters, we select 16 donors as controls. These are considered to be healthy people and, together are referred to as ``controls'' in the remainder of this manuscript. More details, including data processing and consideration of batch effects, can be found in \citet{seyres2020transcriptional}.

\subsection{Multivariate signature identification}
\label{sec:cms-multivariate}
This section is dedicated to the multivariate analysis of the CMS data. In Section \ref{sec:cms-multivariate-signature-identification} we select the variables that help distinguish obese individuals from healthy people, using the penalised logistic regression methods presented in Section \ref{sec:literature-review}. In Section \ref{sec:cms-multivariate-risk-prediction} we use the selected variables to estimate the probability of each blood donor of belonging to the class of obese individuals. In Section \ref{sec:lipo-vs-controls} we repeat the analysis, this time comparing lipodystrophy patients to healthy donors.

\subsubsection{Signature identification}
\label{sec:cms-multivariate-signature-identification}
Because we are interested in identifying the variables that help discriminate between healthy people and CMS patients, we choose the two algorithms that have the highest recall in our simulation settings, which are those that perform variable selection by training a separate EN model on each layer. Moreover, contrarily to the other methods, these have the advantage of selecting a set of features in each layer that together are predictive of patient status. This allows us to identify a molecular signature of CMS in each layer.

In this section, first we explain how the available samples are divided into training and test sets. Then, we present the results obtained with a fixed value of $\alpha$, explaining the rationale behind the choice of the value 0.1. Finally, we comment on the results obtained choosing the value of $\alpha$ as suggested by \citet{zhao2020structured} and explain why fixing the value of $\alpha$ turns out to be more convenient for our application.

\paragraph{Training and test sets}
We consider the following comparisons: obese individuals versus controls and lipodystrophy patients versus controls. Each of these defines a different split of our training set into healthy people (\emph{controls}, labelled by ``0'' in our logistic regression) and individuals affected by lipodystrophy or obesity (\emph{cases}, label ``1'') and helps extracting the most relevant features for each comparison. In this section we only present the results obtained for the first comparison; results for the second comparison are reported in Section \ref{sec:lipo-vs-controls}.

\paragraph{Separate EN on each layer with fixed $\alpha$}
We use separate EN on each layer with fixed $\alpha$ to identify putative multivariate signatures that discriminate between the considered groups. Before doing so, we centre and scale each dataset so that all variables have mean 0 and variance 1 across the individuals in which they were measured.

\begin{figure}
\centering
\begin{subfigure}{\textwidth}
\centering
\includegraphics[width=.49\textwidth]{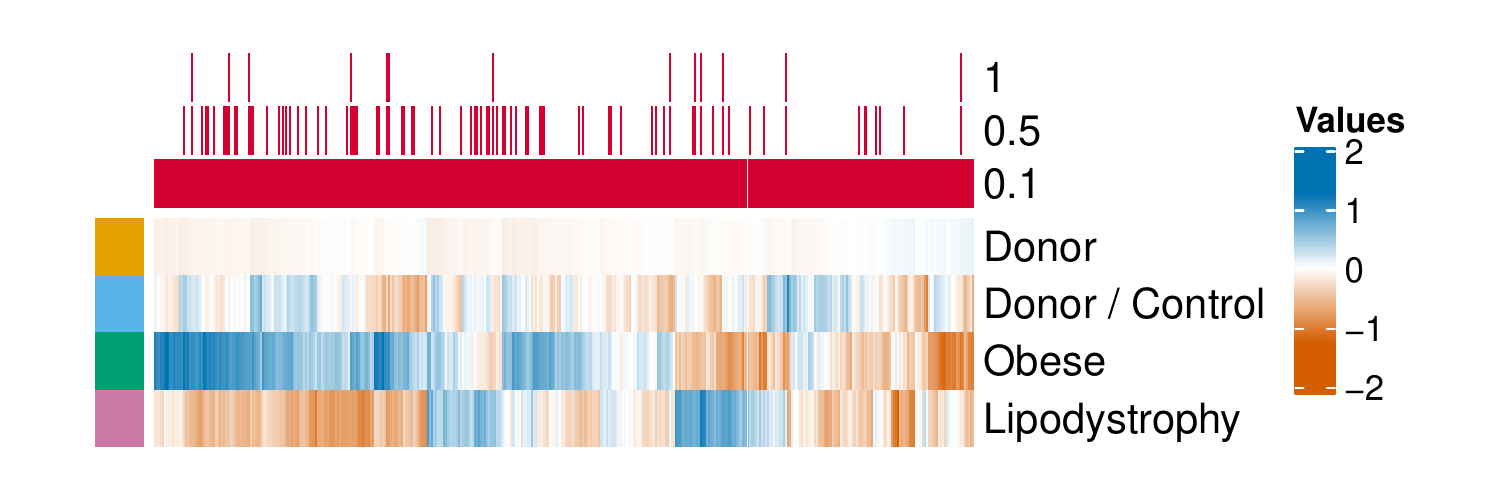}
\includegraphics[width=.49\textwidth]{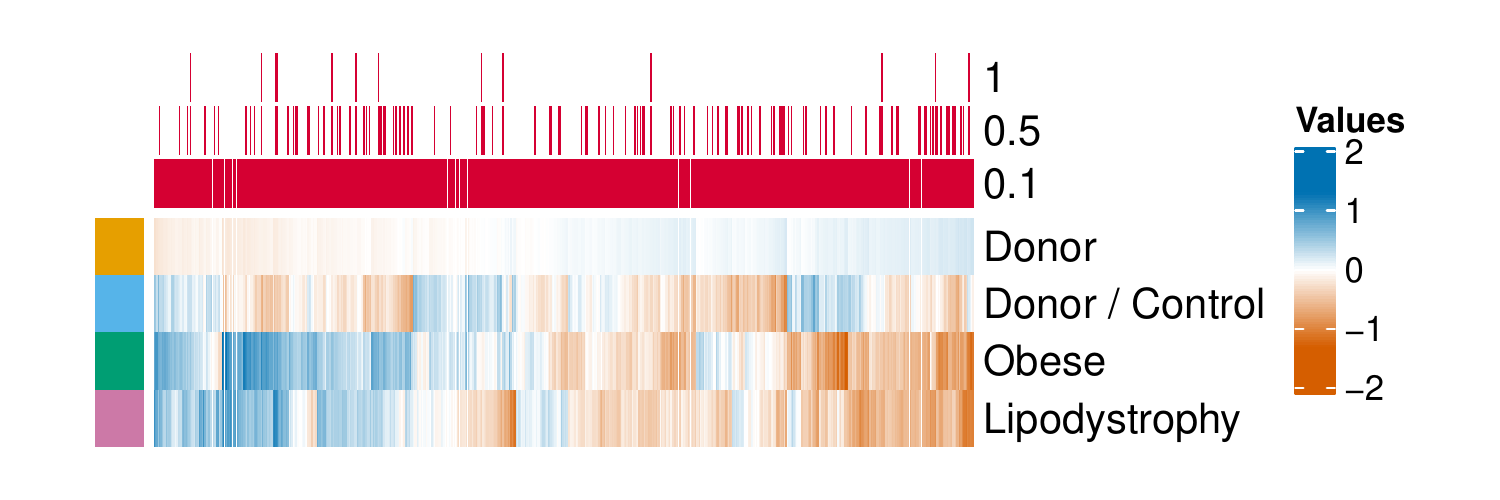}
\caption{ChIP-seq, monocytes and neutrophils.}
\end{subfigure}
\begin{subfigure}{\textwidth}
\centering
\includegraphics[width=.49\textwidth]{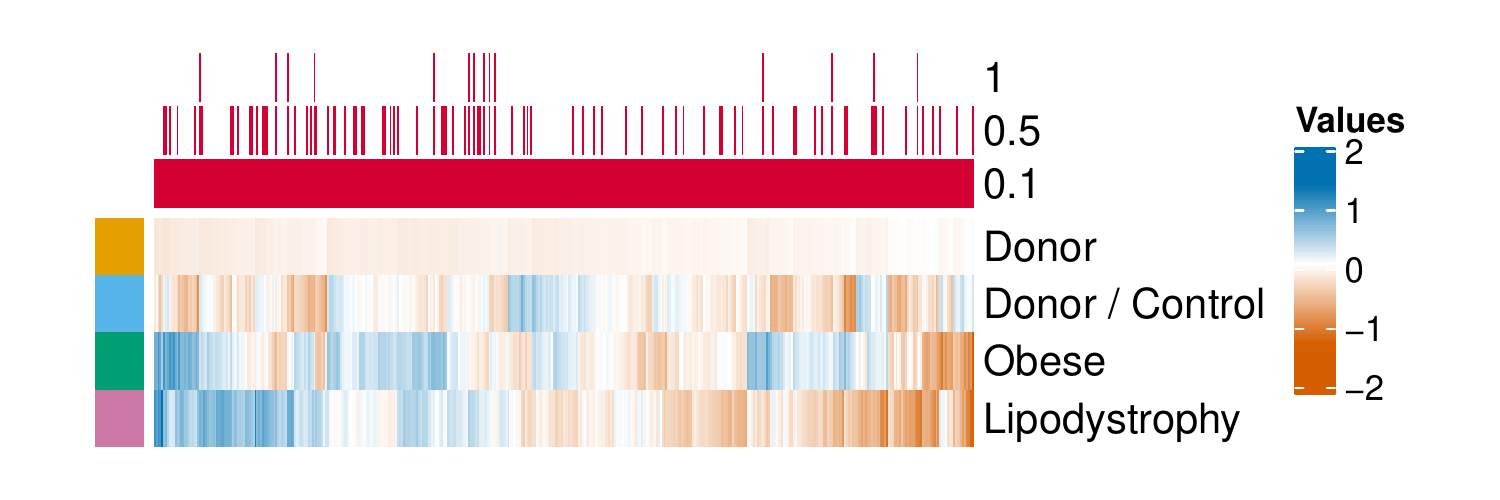}
\includegraphics[width=.49\textwidth]{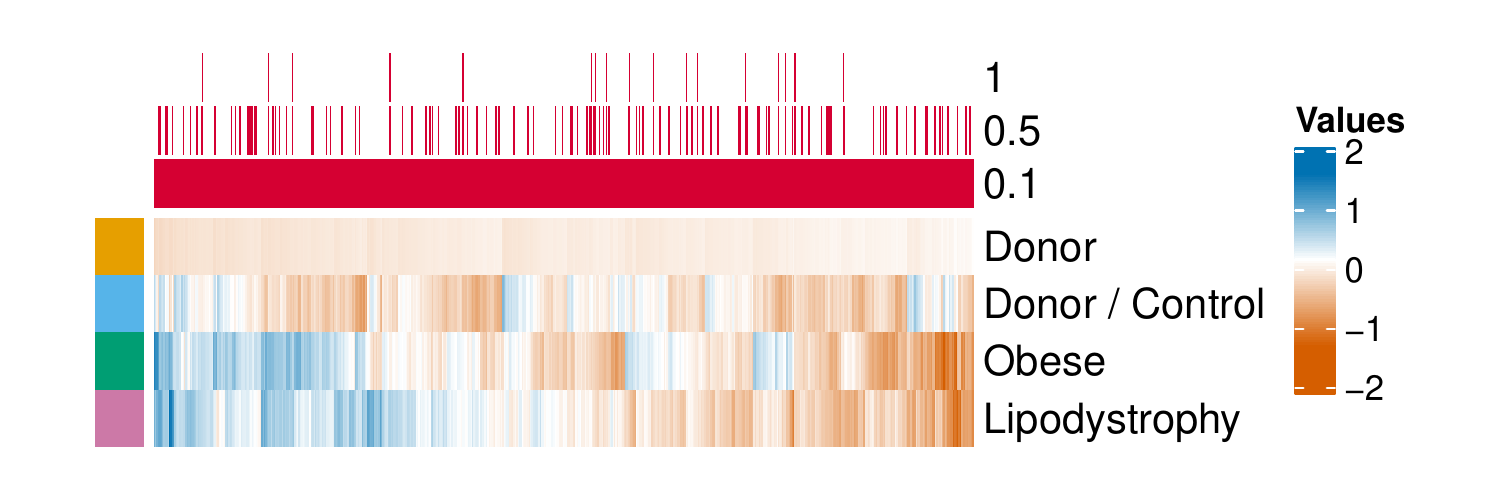}
\caption{RNA-seq, monocytes and neutrophils.}
\end{subfigure}
\begin{subfigure}{\textwidth}
\centering
\includegraphics[width=.49\textwidth]{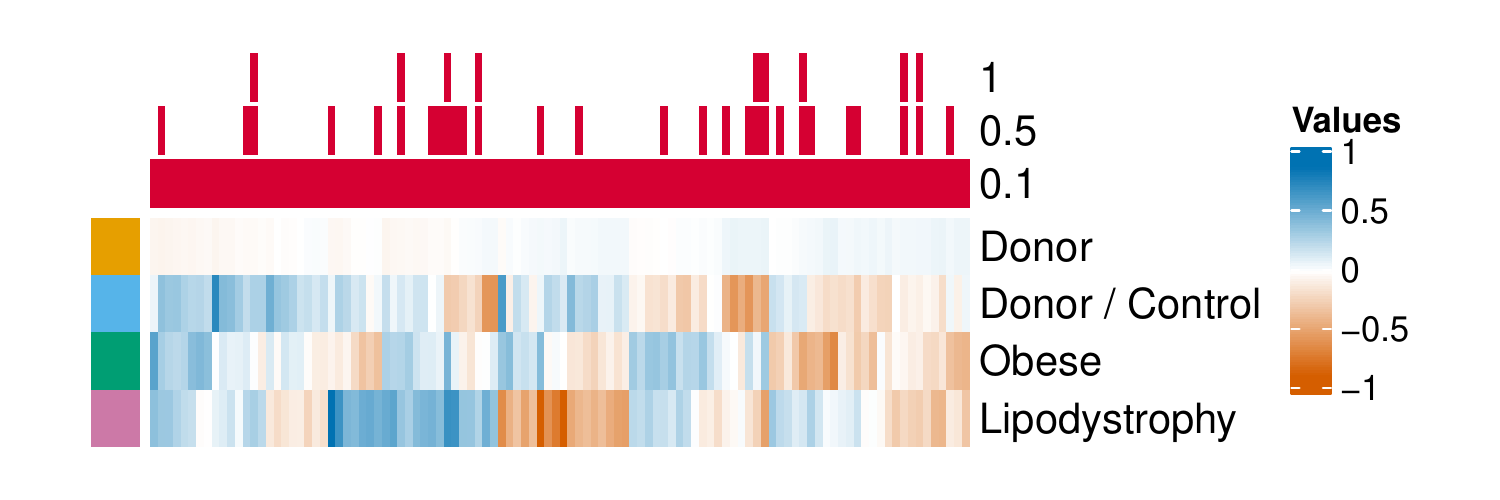}
\includegraphics[width=.49\textwidth]{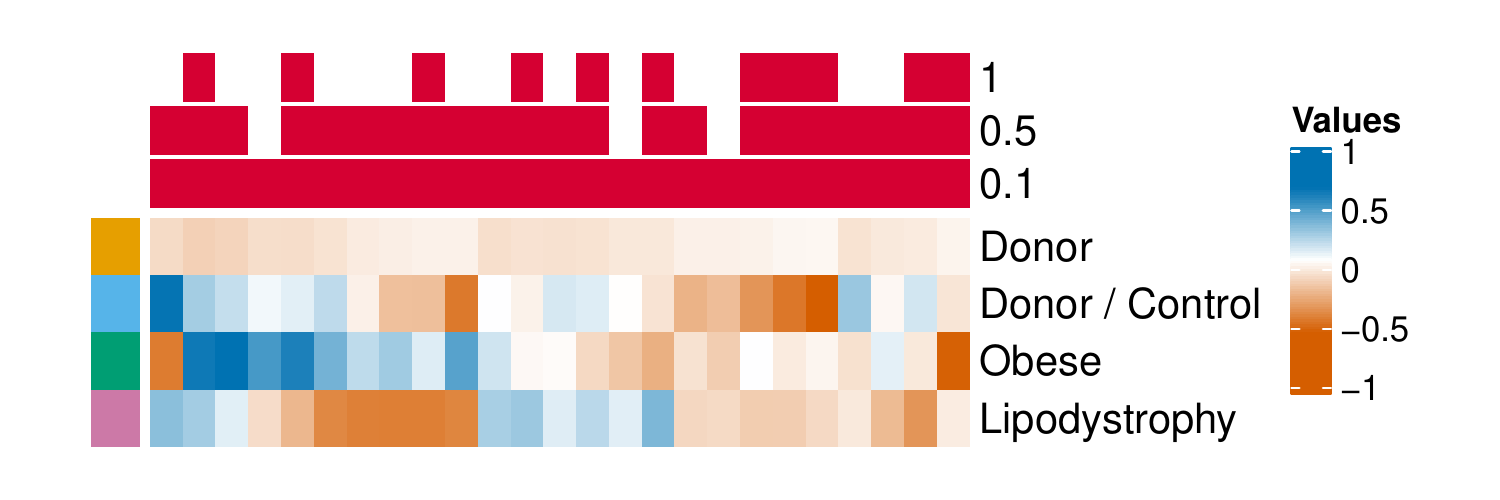}
\caption{Methylation, monocytes and neutrophils.}
\end{subfigure}
\begin{subfigure}{.49\textwidth}
\centering
\includegraphics[width=\textwidth]{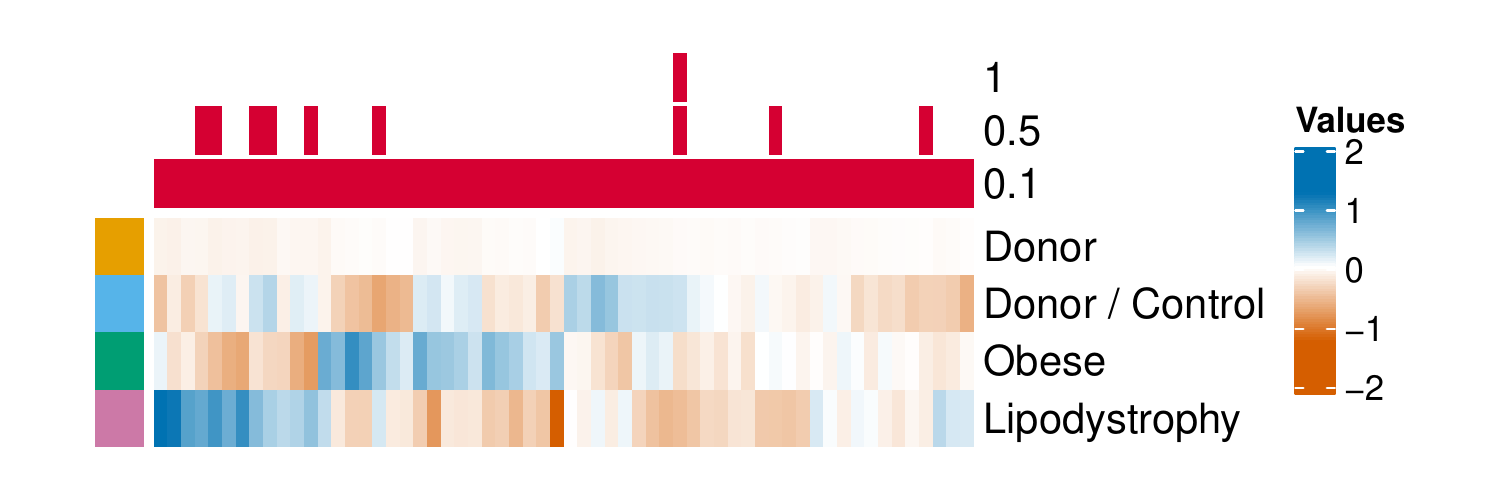}
\caption{Metabolites.}
\end{subfigure}
\begin{subfigure}{.49\textwidth}
\centering
\includegraphics[width=\textwidth]{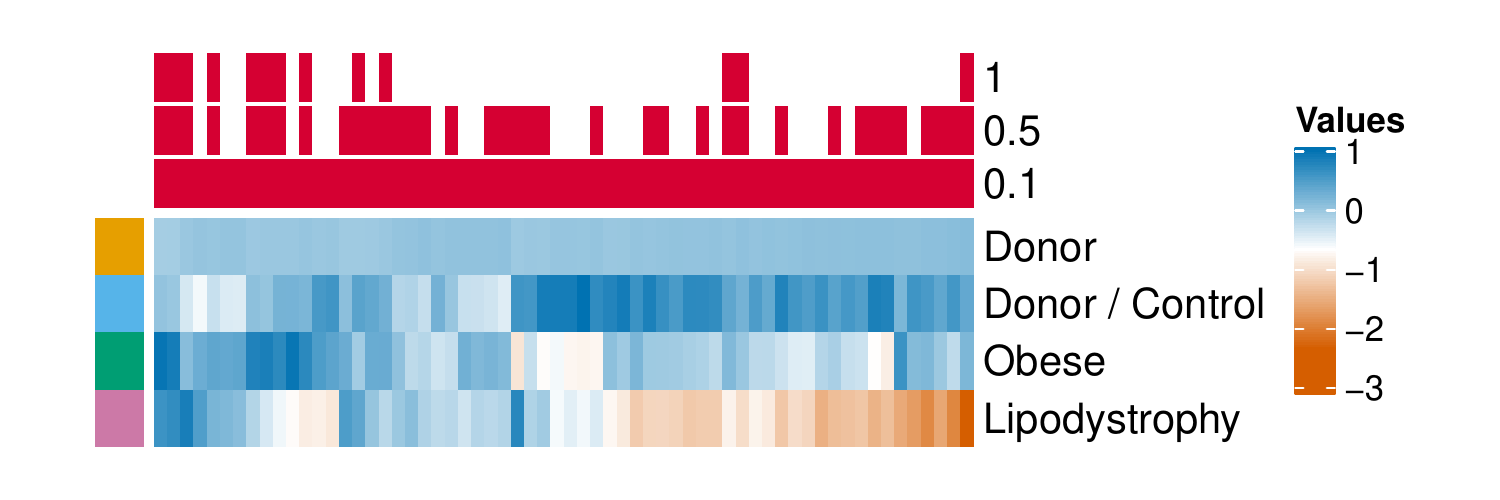}
\vspace{-.4cm}
\caption{Lipids.}
\end{subfigure}
\caption{Variables selected with separate EN and fixed $\alpha$, for different values of the parameter $\alpha$. The red bars indicate which variables are selected with $\alpha = 1, 0.5$ and $0.1$ respectively. Below are shown the average values of those variables for the donors and controls and for the lipodystrophy and obese individuals. Only the variables selected for at least one of those values are reported here.}
\label{fig:bari-vs-healthy-choice-of-alpha}
\end{figure}

The training set is formed by the donors who have been selected as controls and the obese individuals. We use 10-fold CV as suggested by \citet{zou2005regularization}. To do so, we use the \verb|cv.glmnet| function of the \textsf{R} package ``\verb|glmnet|''.
Since different CV splits result in different subsets of selected variables, we repeat the CV procedure 1000 times for each layer, and then consider the largest set of selected variables and the set of variables that is selected most often. We notice that the latter only contains a few variables, if any, and that many of the selected variables are in common between the two (see Supplementary Material), so we decide to keep the former. Repeating the analysis for $\alpha = 1$, $0.5$ and $0.1$, which roughly corresponds to having high, medium, and low amount of selection, the two highest values of a lead to selecting very few, if any, variables in most layers (Figure \ref{fig:bari-vs-healthy-choice-of-alpha}). For this reason, we decide to pick $\alpha=0.1$. Note that, in this setting, despite giving relative weight of only 10\% to the LASSO penalty, a tiny percentage of the available variables is selected. 

Figure \ref{fig:bari-vs-healthy-choice-of-alpha} shows, for each 'omic layer, the average value of each selected variable for each category of people: donors who have been selected as controls, the remaining donors, obese individuals and lipodystrophy patients. As we might expect, the average values taken by each variable have opposite signs in the two sets of people used in the training set. Perhaps more interestingly, we note that, while the other donors have average values that are close to zero, lipodystrophy patients take extreme values on most of the selected variables. On top of those values, red bars indicate which variables are selected for each value of $\alpha$. The variables that are not selected for any value of $\alpha$ are not shown.

\paragraph{Comparison with separate EN with $\alpha$ selected by EPSGO}
We now apply the same strategy as above, except that we let the EPSGO algorithm choose $\alpha$ so as to minimise the MR. The first step of the EPSGO algorithm fails on the lipid data. For this reason the results presented here only comprise the remaining seven 'omic layers.

Table \ref{tab:separate-en-comparison} shows the number of selected variables in each layer by each method, as well as the selected value of $\alpha$ and the number of selected variables that are in common between the two algorithms. The optimal values of $\alpha$ selected via error surface optimisation are all higher than $0.1$, except for the metabolite data. Consequently, fewer variables are selected compared to fixing $\alpha=0.1$. While in some cases this may be a desirable feature, here it makes it difficult to identify molecular signatures, especially for the ChIP-seq data of monocytes, where only one variable is selected.

Therefore, the automatic selection of $\alpha$ via the EPSGO algorithm presents two main disadvantages in applications like ours. The first and most important one is that the algorithm may not always work, as it is the case for the lipid data. Secondly, selecting the EN parameter so as to minimise the average out-of-sample error can be quite convenient in some cases, it may not be the best choice in applications where the goal is to select a reasonable number of predictive features in each dataset.

\begin{table}
\centering
\begin{tabular}{l c c c c}

&    \begin{tabular}{@{}c@{}}\textbf{{\#} variables} \\ $\alpha=0.1$\end{tabular} &\begin{tabular}{@{}c@{}}\textbf{{\#} variables} \\ \textbf{EPSGO}\end{tabular} & \begin{tabular}{@{}c@{}} $\alpha$ \\ \textbf{EPSGO}\end{tabular}& \textbf{$\cap$}\\
\hline
ChIP-seq monocytes & 428 & 1 & 0.59 & 1\\
ChIP-seq neutrophils &611 & 40 & 0.75 & 29\\
RNA-seq monocytes & 425 & 111 & 0.96 & 21\\
RNA-seq neutrophils & 592 & 219 & 0.13 & 82\\
Methylation monocytes & 106 & 31 & 0.75 & 0\\
Methylation neutrophils & 25 & 54& 0.59 & 5\\
Metabolites & 60 & 195 & 0.02 & 55\\
Lipids & 62 & - & - & -\\
\hline
\end{tabular}
\caption[Comparison of two-step EN methods.]{Comparison of separate EN methods. From left to right, are reported: the number of selected variables when $\alpha=1$, the number of selected variables when $\alpha$ is selected using EPSGO, the value of $\alpha$ picked by EPSGO, and the number of variables that are selected both with fixed $\alpha$ and with $\alpha$ selected via EPSGO.}
\label{tab:separate-en-comparison}
\end{table}

\subsubsection{Probability of being affected by cardiometabolic syndrome}
\label{sec:cms-multivariate-risk-prediction}

After performing variable selection on each layer separately, we train a ridge-penalised logistic regression model on the matrix formed by all the variables selected in each layer to compute the probability of belonging to the \emph{case} group for each individual. The training and test sets are the same as in Section \ref{sec:cms-multivariate-signature-identification}. Again, we present here the results obtained for the comparison between obese individuals and controls; the results for the lipodystrophy patients versus controls case can be found in Section \ref{sec:lipo-vs-controls}.

\begin{figure}
	\centering
	\begin{subfigure}{.49\textwidth}
		\centering
		\includegraphics[width=\textwidth]{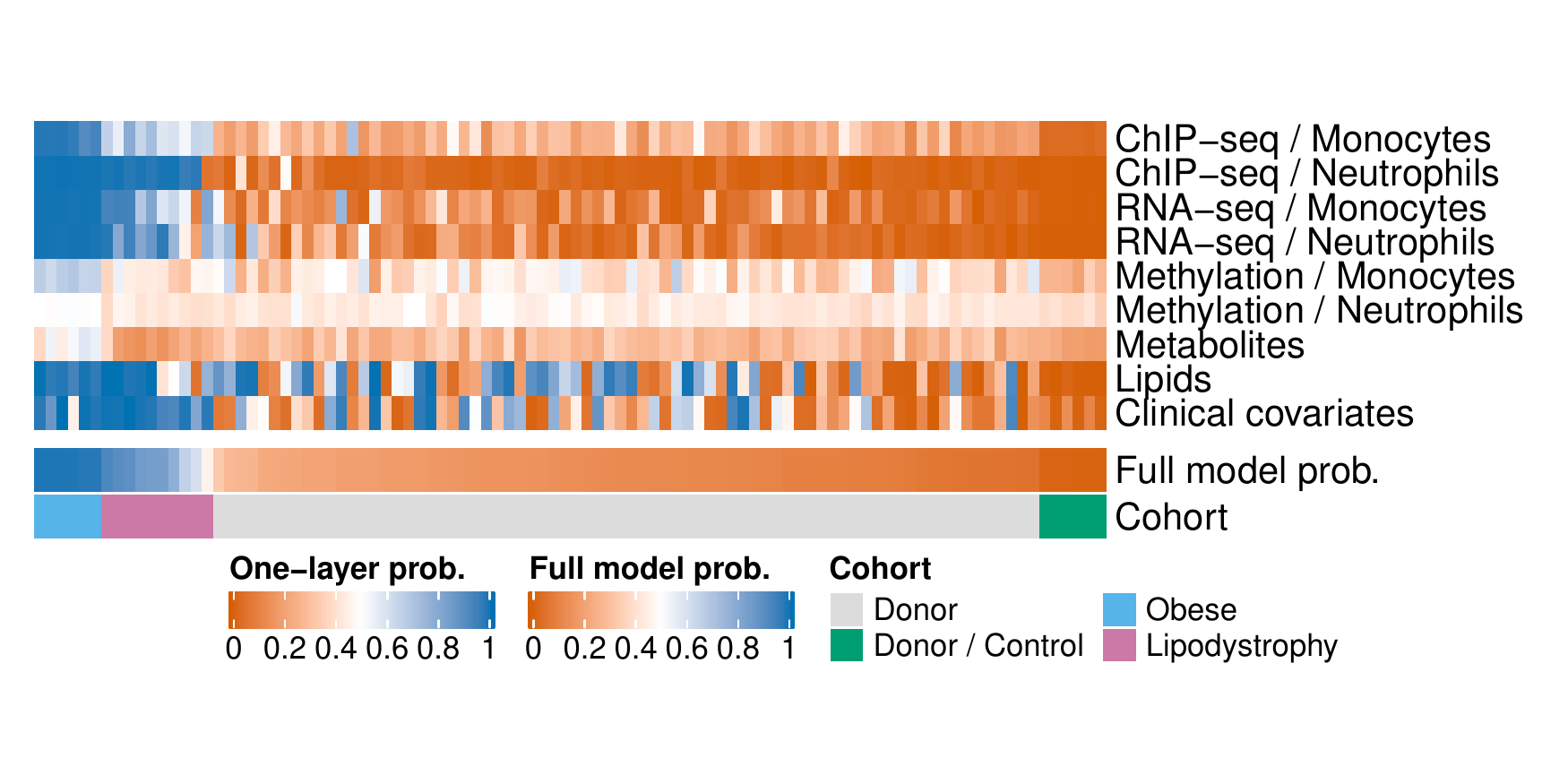}
		\vspace{-.5cm}
		\caption{Probability of being a \emph{case}, $\alpha=0.1$.}
		\label{fig:multivariate-predictive-modelling-obese-alpha01}
	\end{subfigure}
	\begin{subfigure}{.49\textwidth}
		\centering
		\includegraphics[width=\textwidth]{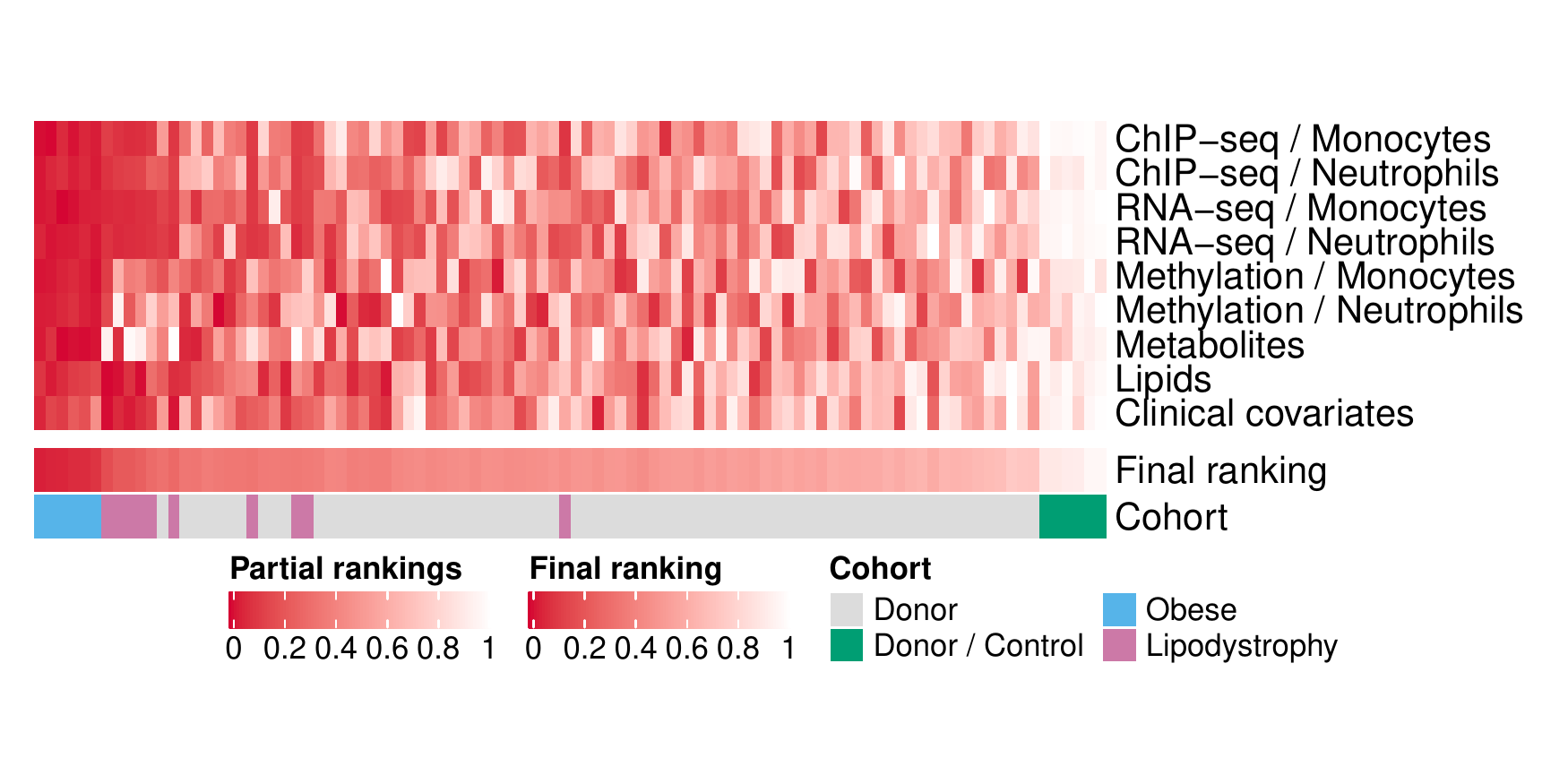}
		\vspace{-.5cm}
		\caption{Combined rankings, $\alpha=0.1$.}
		\label{fig:combined-ranks-obese-alpha01}
	\end{subfigure}
	\begin{subfigure}{.49\textwidth}
		\centering
		\includegraphics[width=\textwidth]{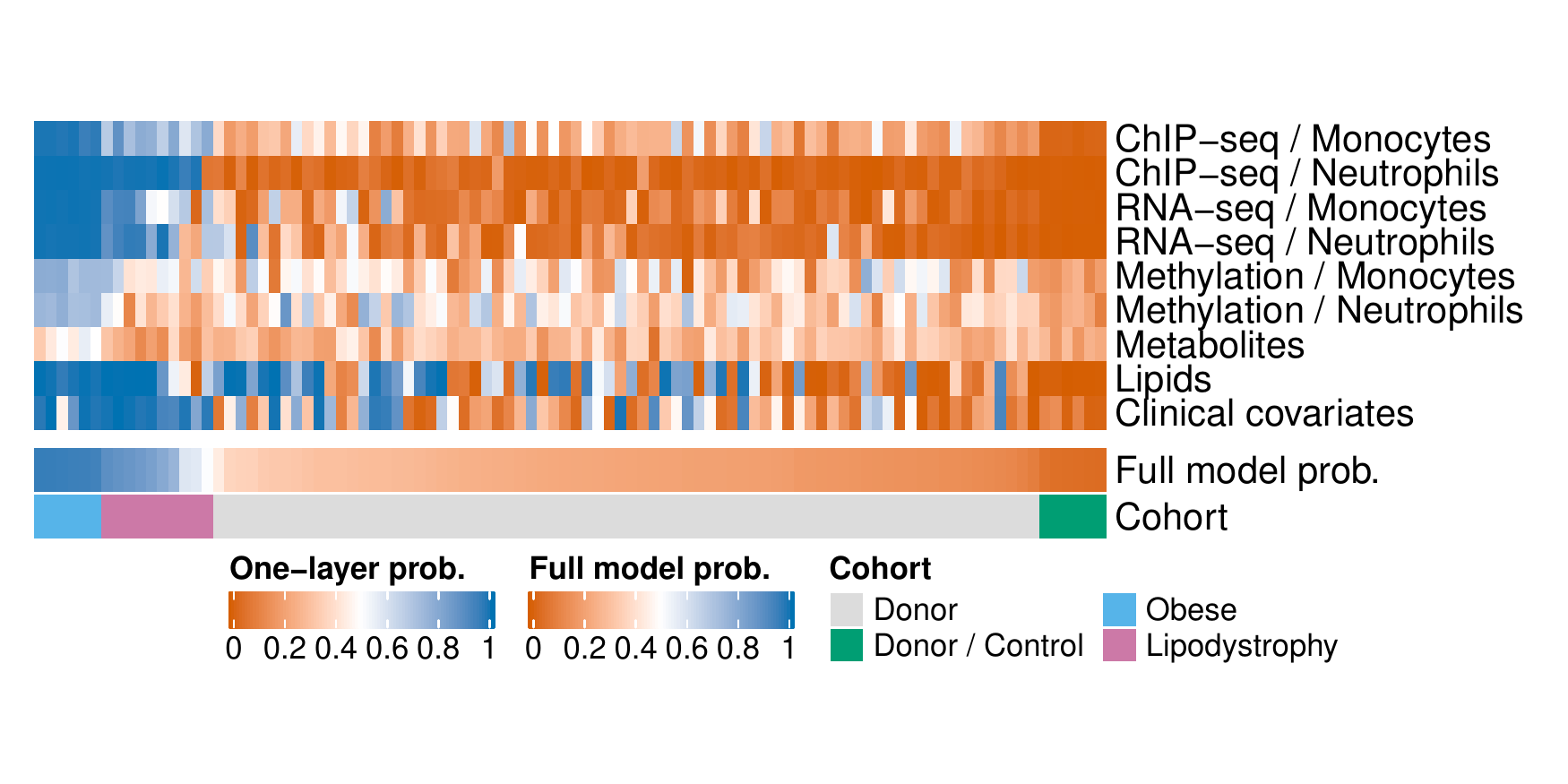}
		\vspace{-.5cm}
		\caption{Probability of being a \emph{case}, $\alpha=0.5$.}
		\label{fig:multivariate-predictive-modelling-obese-alpha05}
	\end{subfigure}
	\begin{subfigure}{.49\textwidth}
		\centering
		\includegraphics[width=\textwidth]{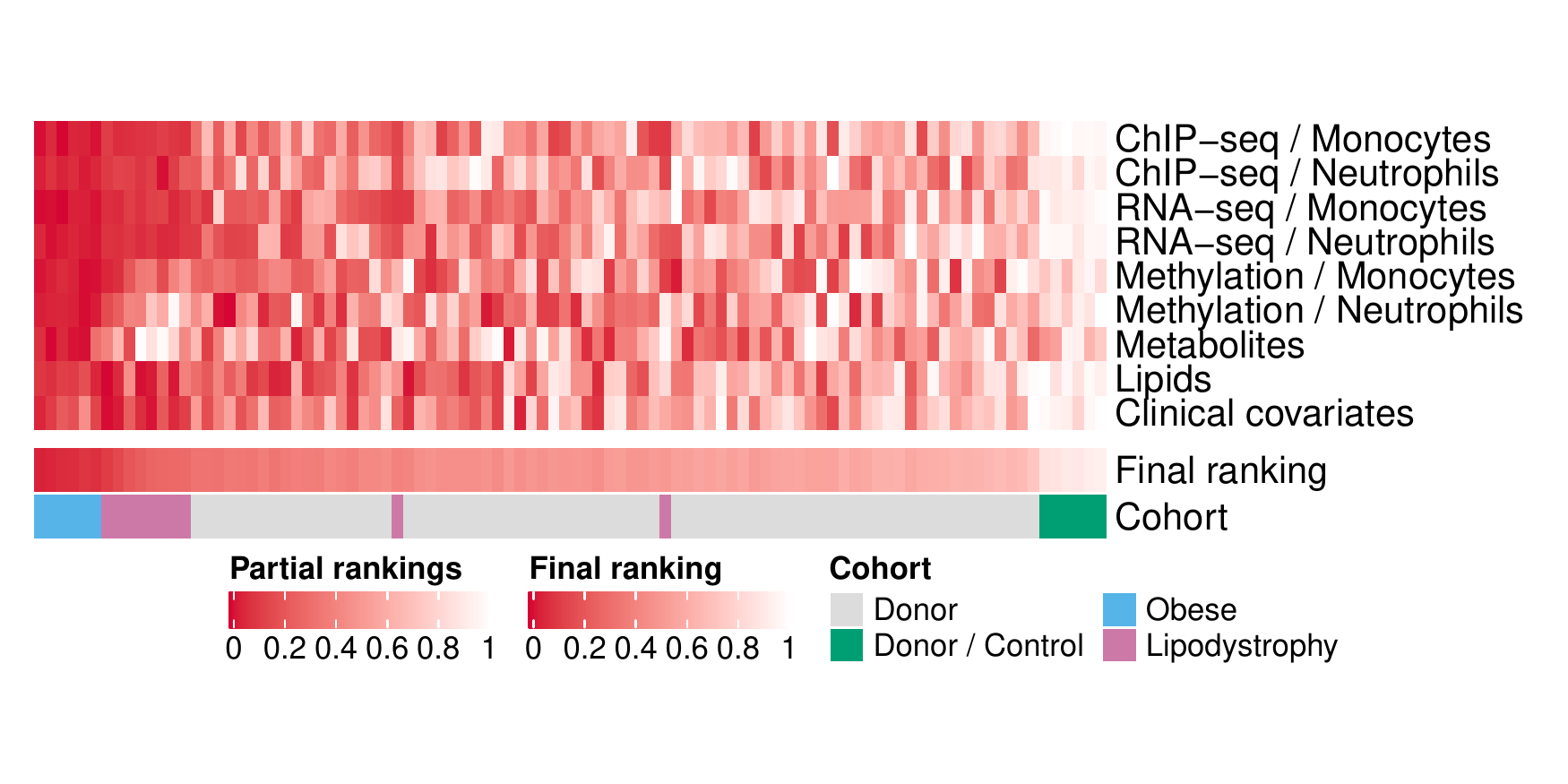}
		\vspace{-.5cm}
		\caption{Combined rankings, $\alpha=0.5$.}
		\label{fig:combined-ranks-obese-alpha05}
	\end{subfigure}
	\begin{subfigure}{.49\textwidth}
		\centering
		\includegraphics[width=\textwidth]{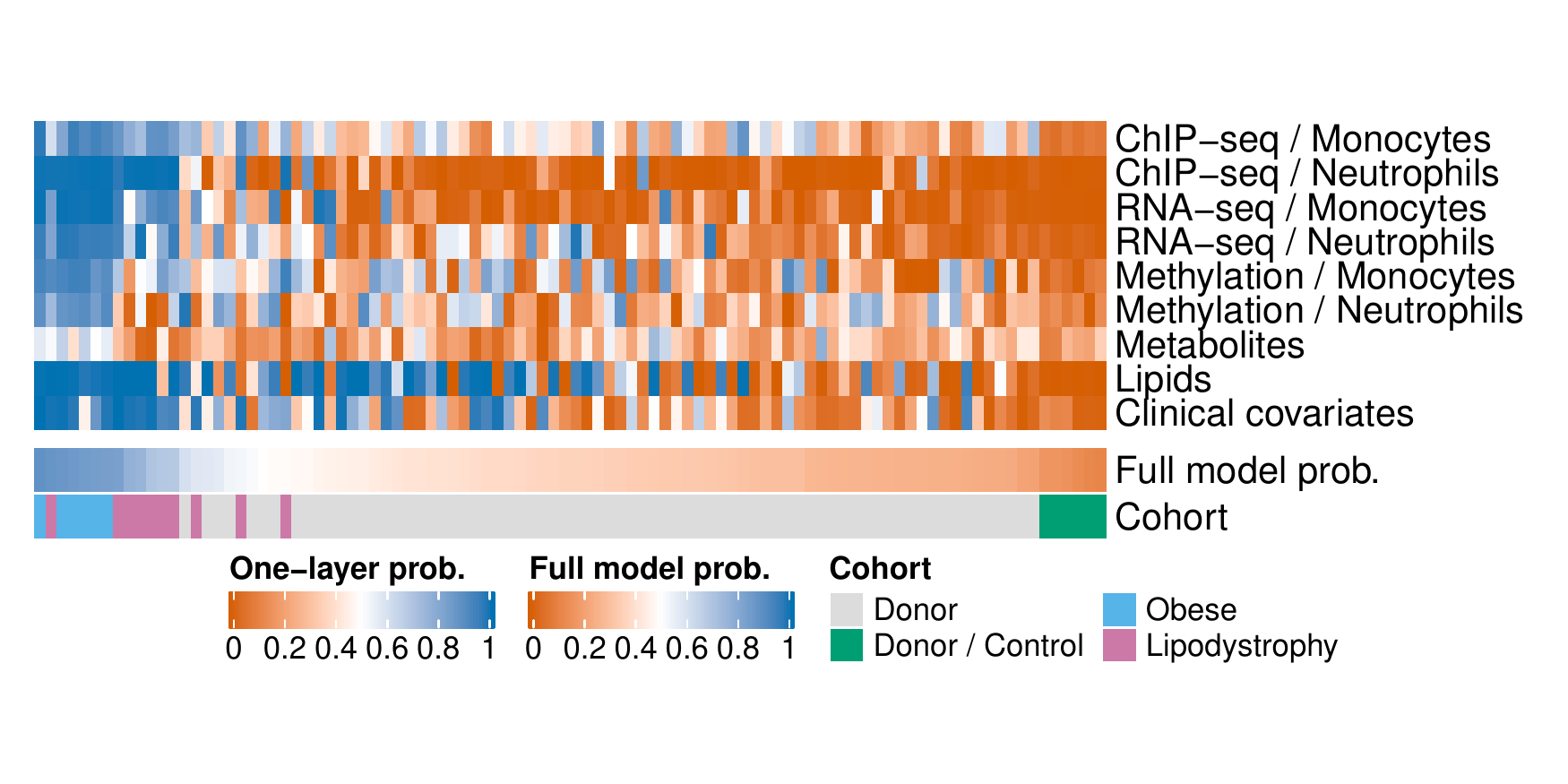}
		\vspace{-.5cm}
		\caption{Probability of being a \emph{case}, $\alpha=1$.}
		\label{fig:multivariate-predictive-modelling-obese-alpha1}
	\end{subfigure}
	\begin{subfigure}{.49\textwidth}
		\centering
		\includegraphics[width=\textwidth]{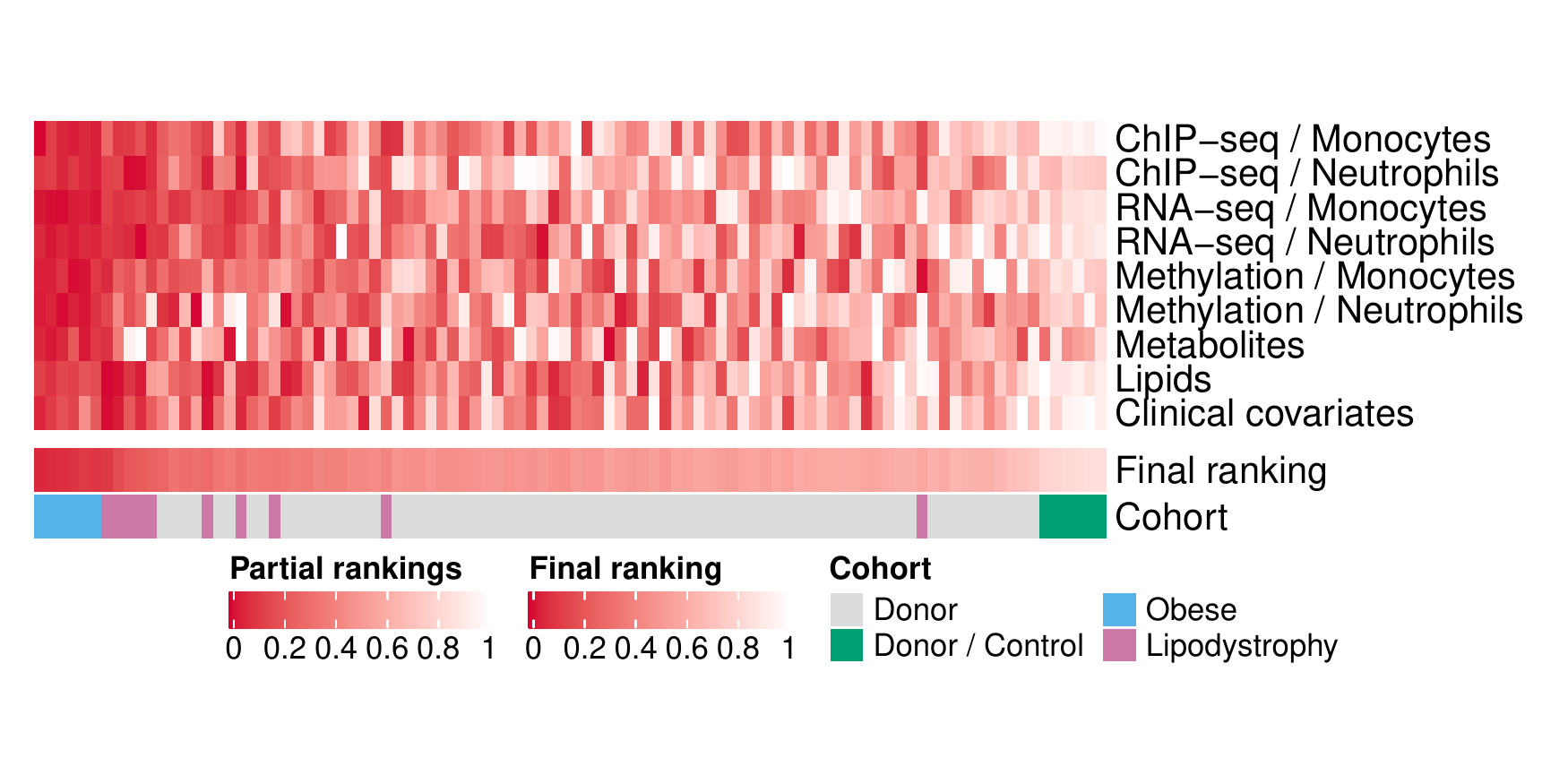}
		\vspace{-.5cm}
		\caption{Combined rankings, $\alpha=1$.}
		\label{fig:combined-ranks-obese-alpha01}
	\end{subfigure}
\caption[CMS probabilities and rankings: obese individuals versus controls.]{Probabilities of belonging to the class of obese individuals and ranking of each person according to those probabilities. Both quantities are shown on each dataset separately and considering all the data types jointly. The model is trained on the obese individuals and control donors. Each column corresponds to one of the individuals who have no missing data, each row corresponds to one of the layers. The columns are sorted by probability of being a case in (a), (c), and (e) and final ranking in (b), (d), and (f). All rankings are divided by the total number of observations.}\label{fig:multivariate-predictive-modelling-obese}
\end{figure}

Figure \ref{fig:multivariate-predictive-modelling-obese} shows, for each person, the probability of belonging to the extreme phenotype group (which in this case comprises the obese individuals). The probabilities estimated on each layer separately are also reported. These are derived by fitting a logistic regression with $l_2$ penalty on the selected variables only. The ChIP-seq and RNA-seq data give similar predictions, while the lipidomics dataset produces slightly different ones. For the methylation and metabolomics datasets, the probabilities of being a case do not differ greatly among individuals.

It is interesting to note that the lipodystrophy patients have higher probabilities of belonging to the same class as the obese individuals than the blood donors. This suggests that, on the molecular level, lipodystrophy patients are more similar to obese people than the average person. This is not surprising, as those two conditions are characterised by similar biochemical and clinical profiles. On the other hand, some blood donors have very similar predicted values to the obese and lipodystrophy individuals. This may indicate that blood donors can show similar characteristics to those in the extreme phenotype groups, which could provide insights into the pathogenesis of CMS.

These peculiarities of the results can be better observed by assigning a ranking to each person from 1 to 96 based on their probability of belonging to the class with label ``1'', where the person with rank 1 has the highest probability. We do this based on the probabilities estimated on each layer separately, and then take the average as the aggregated rank for each person. Note that this combined ranking does not correspond to the ranking implied by the probabilities of class membership obtained using the full ridge-penalised model (Figures \ref{fig:multivariate-predictive-modelling-obese-alpha01}, \ref{fig:multivariate-predictive-modelling-obese-alpha05}, and \ref{fig:multivariate-predictive-modelling-obese-alpha1}).

Many other ways of combining the rankings could have been considered. The literature on rank aggregation is vast; the first efforts on this topic date back to the XVIII century \citep{borda1781memoire}. This is still a thriving field in modern times, with a wide range of rank aggregation methods being developed for different types of applications, including genomic and multi-omic studies \citep{blangiardo2007statistical, lin2009integration}. An overview is provided by \citet{lin2010rank}. Due to the fact that, as we have seen, data available for these studies often have missing values, the focus has recently shifted to methods that can handle partial rankings \citep{aerts2006gene, kolde2012robust}. A comparative study of such methods has been performed by \citet{li2019comparative}. However, we find that, in this simple case, taking the average ranks is a sensible choice.

Figure \ref{fig:multivariate-predictive-modelling-obese} also shows the rankings obtained on each dataset separately and those computed by combining all the partial rankings together. Again, some of the lipodystrophy patients score similarly to the obese individuals. Moreover, we find some donors among the lipodystrophy patients.

\subsection{Lipodystrophy patients versus control donors}
\label{sec:lipo-vs-controls}
\begin{figure}
	\centering
	\begin{subfigure}{\textwidth}
		\centering
		\includegraphics[width=.49\textwidth]{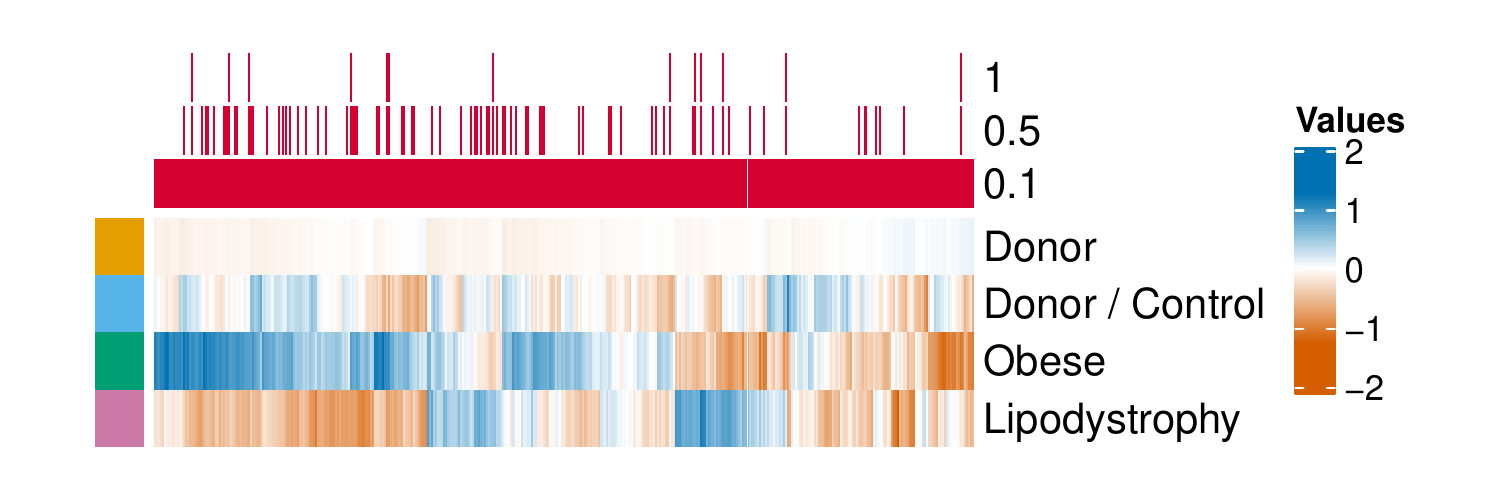}
		\includegraphics[width=.49\textwidth]{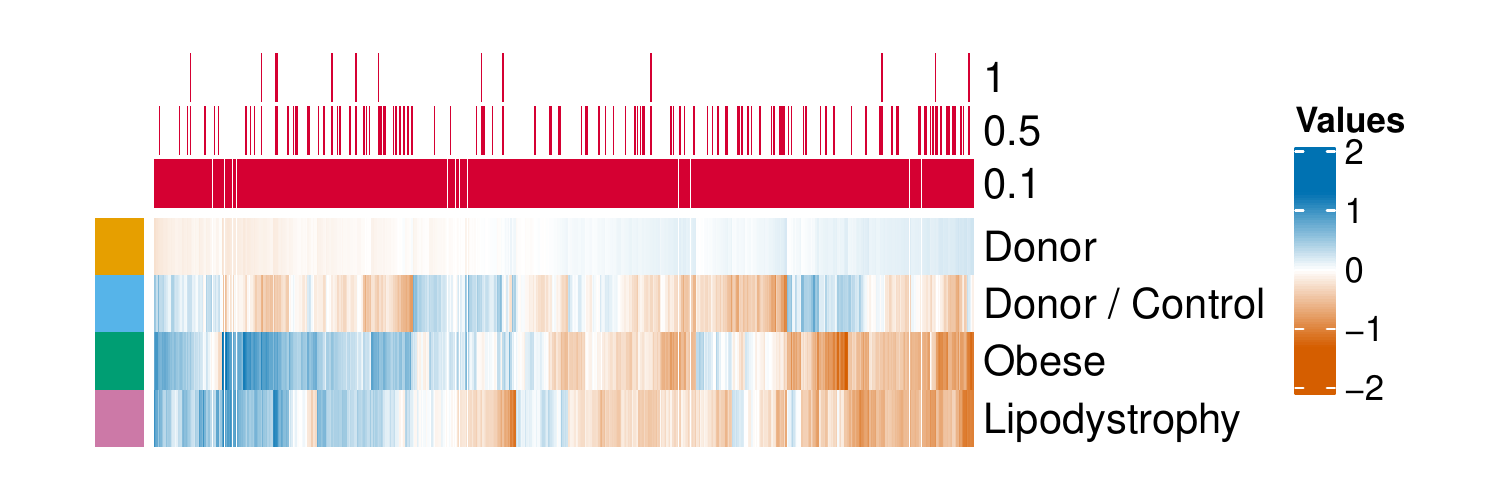}
		\caption{ChIP-seq, monocytes and neutrophils.}
	\end{subfigure}
	\begin{subfigure}{\textwidth}
		\centering
		\includegraphics[width=.49\textwidth]{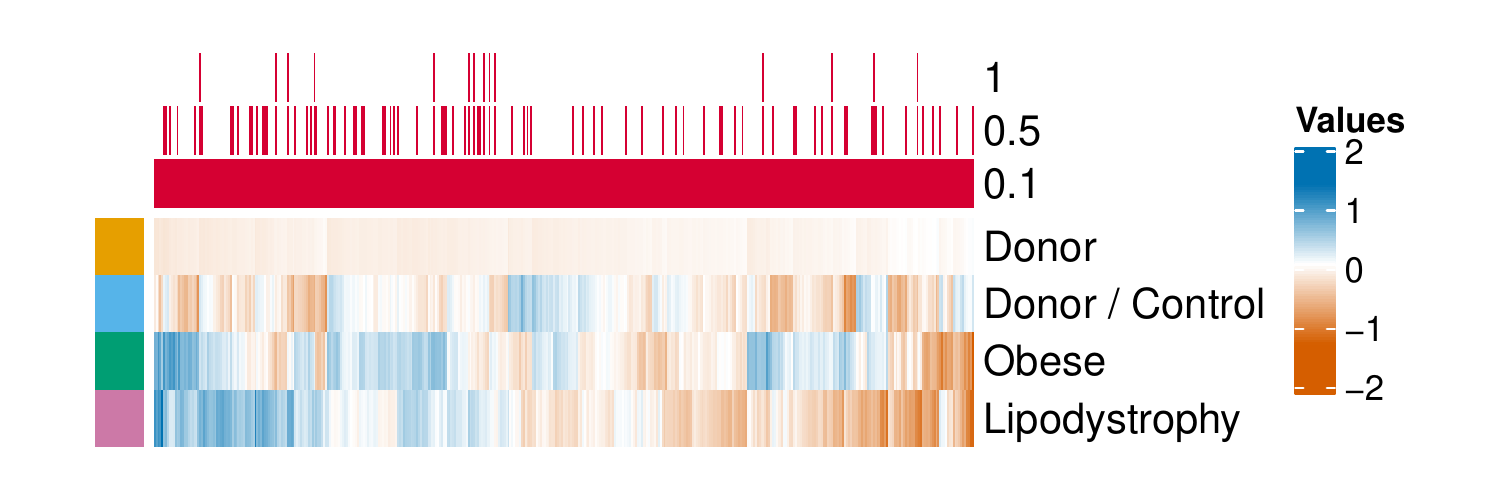}
		\includegraphics[width=.49\textwidth]{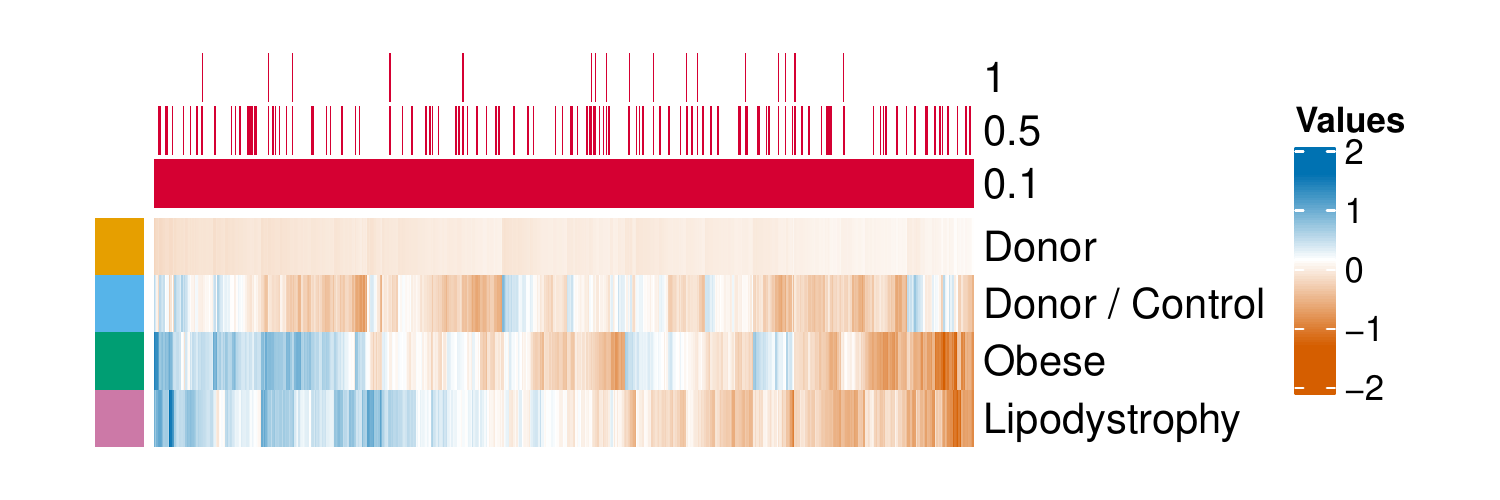}
		\caption{RNA-seq, monocytes and neutrophils.}
	\end{subfigure}
	\begin{subfigure}{\textwidth}
		\centering
		\includegraphics[width=.49\textwidth]{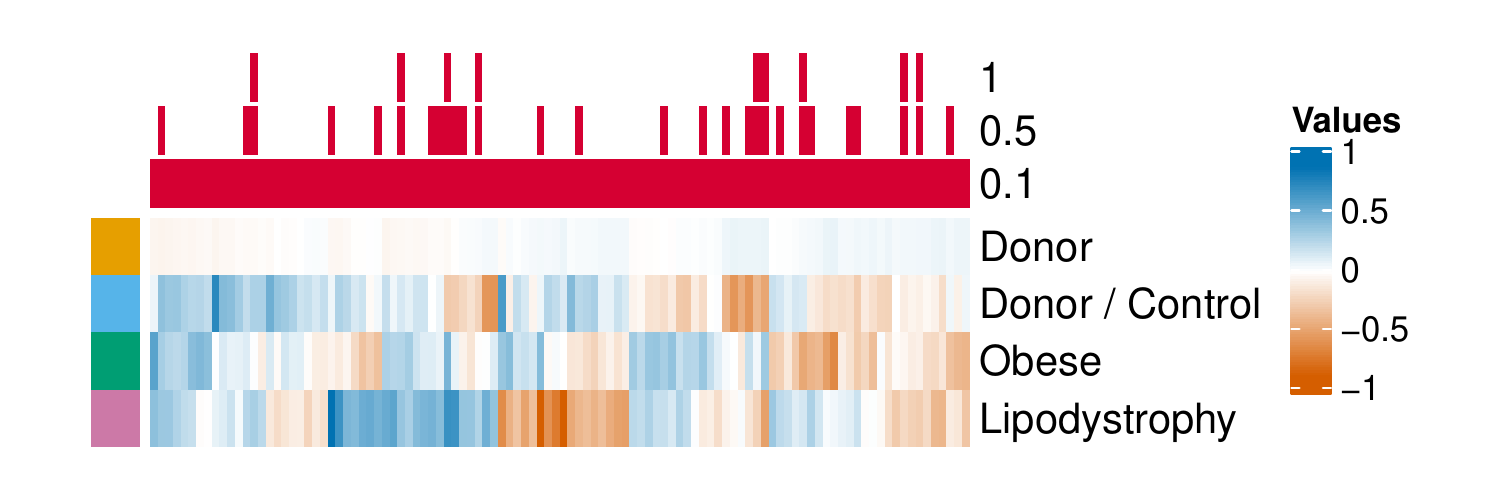}
		\includegraphics[width=.49\textwidth]{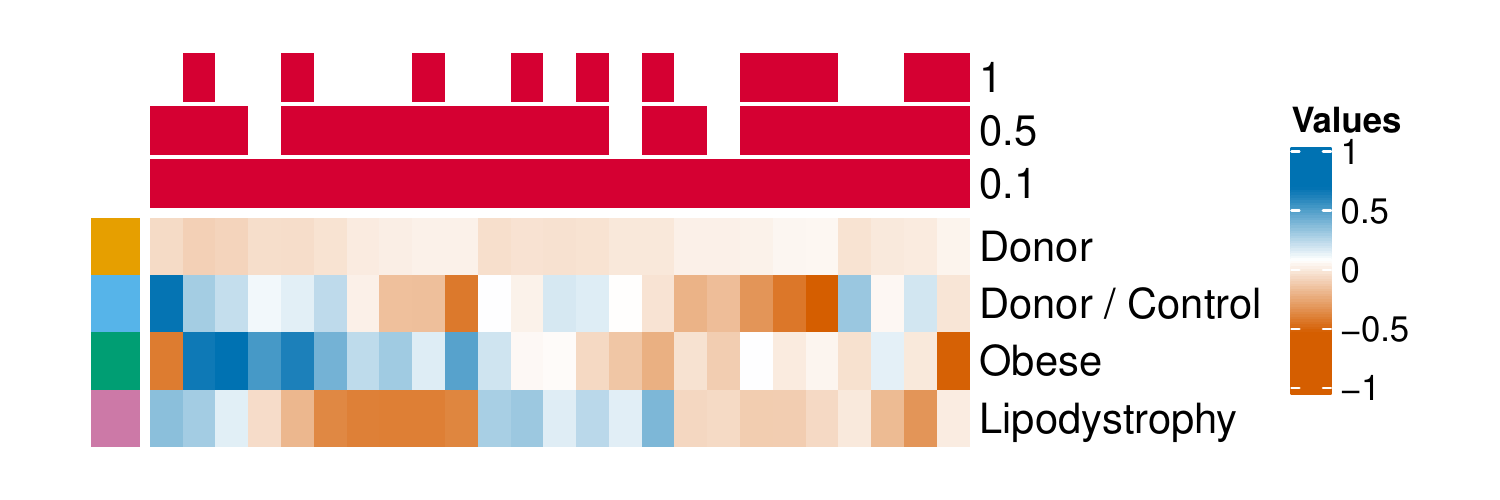}
		\caption{Methylation, monocytes and neutrophils.}	
	\end{subfigure}
	\begin{subfigure}{.49\textwidth}
		\centering
		\includegraphics[width=\textwidth]{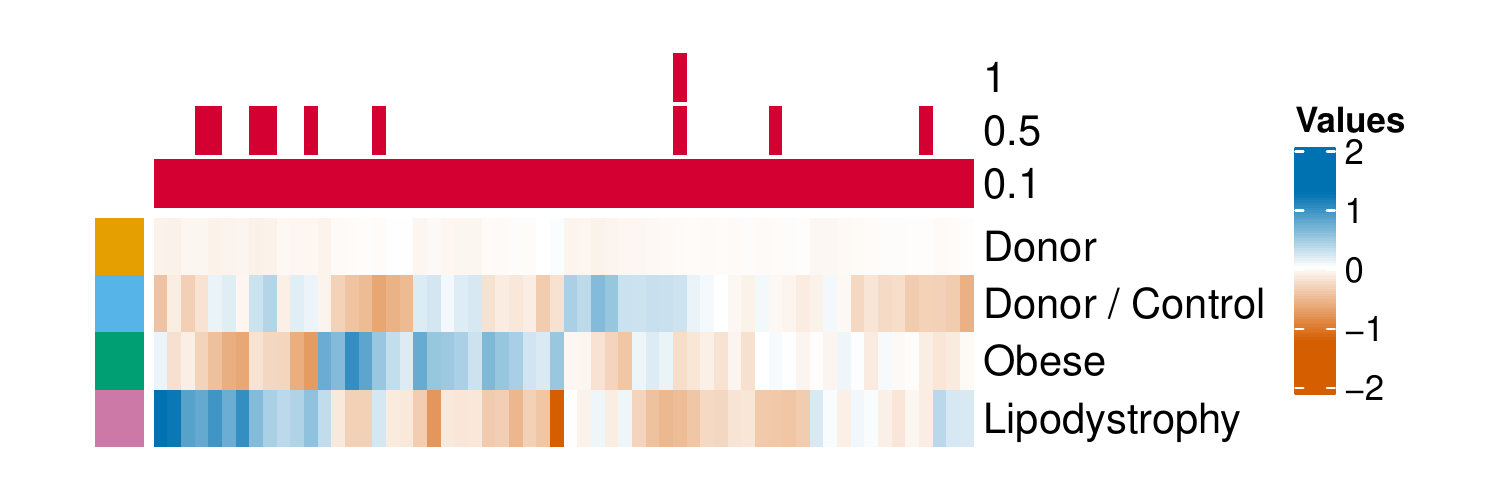}
		\caption{Metabolites.}
	\end{subfigure}
	\begin{subfigure}{.49\textwidth}
		\centering	
		\includegraphics[width=\textwidth]{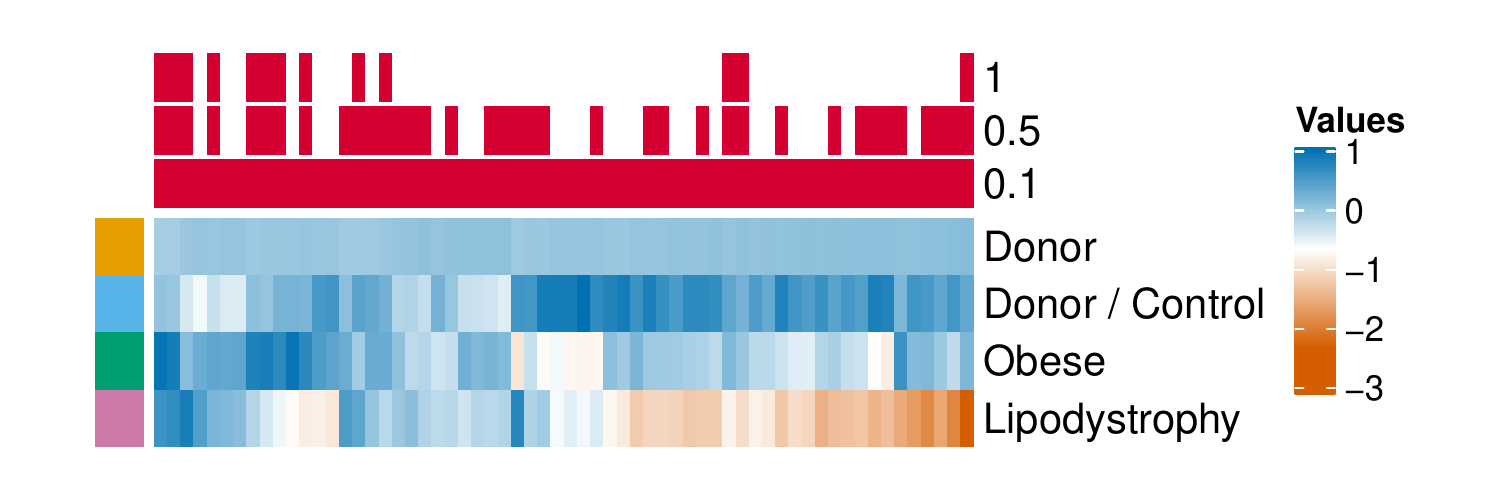}
		\caption{Lipids.}
	\end{subfigure}
	\caption[Multivariate signature: lipodystrophy patients versus controls.]{Signature identification. Penalised logistic regression model trained on the lipodystrophy patients and control donors. Average values of the selected variables for each group of people. Each column corresponds to one of the selected variables.}
\label{fig:signature-identification-lipodystrophy}
\end{figure}

Figure \ref{fig:signature-identification-lipodystrophy} shows the average values of the selected variables for the second comparison (lipodystrophy patients versus control donors). Interestingly, lipodystrophy and obese individuals have similar average values in neutrophils for ChIP-seq and RNA-seq data, but the same is not true for the other datasets.

\begin{figure}
	\centering
	\begin{subfigure}{.49\textwidth}
		\centering
		\includegraphics[width=\textwidth]{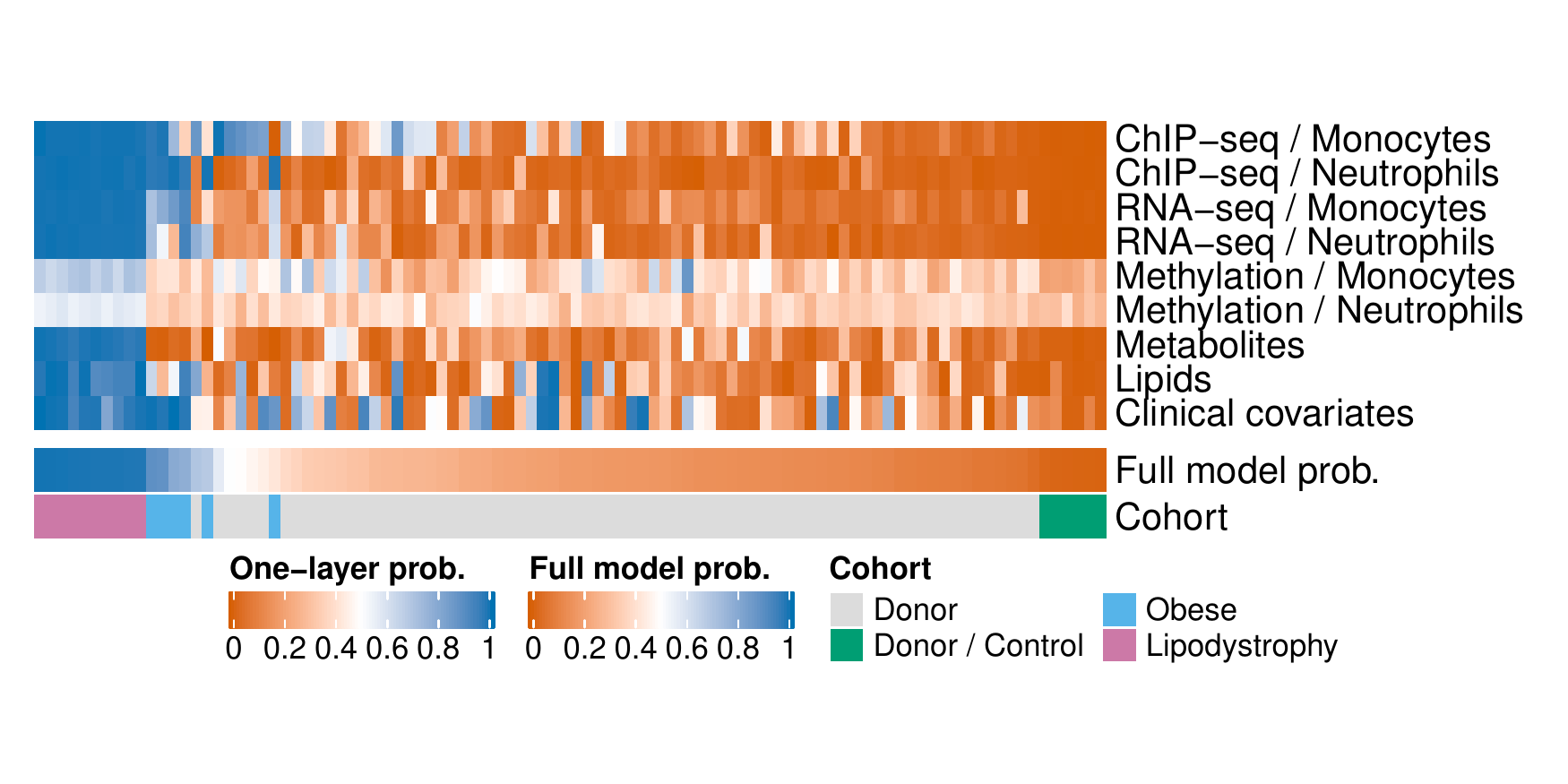}
		\vspace{-.5cm}		
		\caption{Probability of being a \emph{case}, $\alpha=0.1$.}
	\end{subfigure}
	\begin{subfigure}{.49\textwidth}
		\centering
		\includegraphics[width=\textwidth]{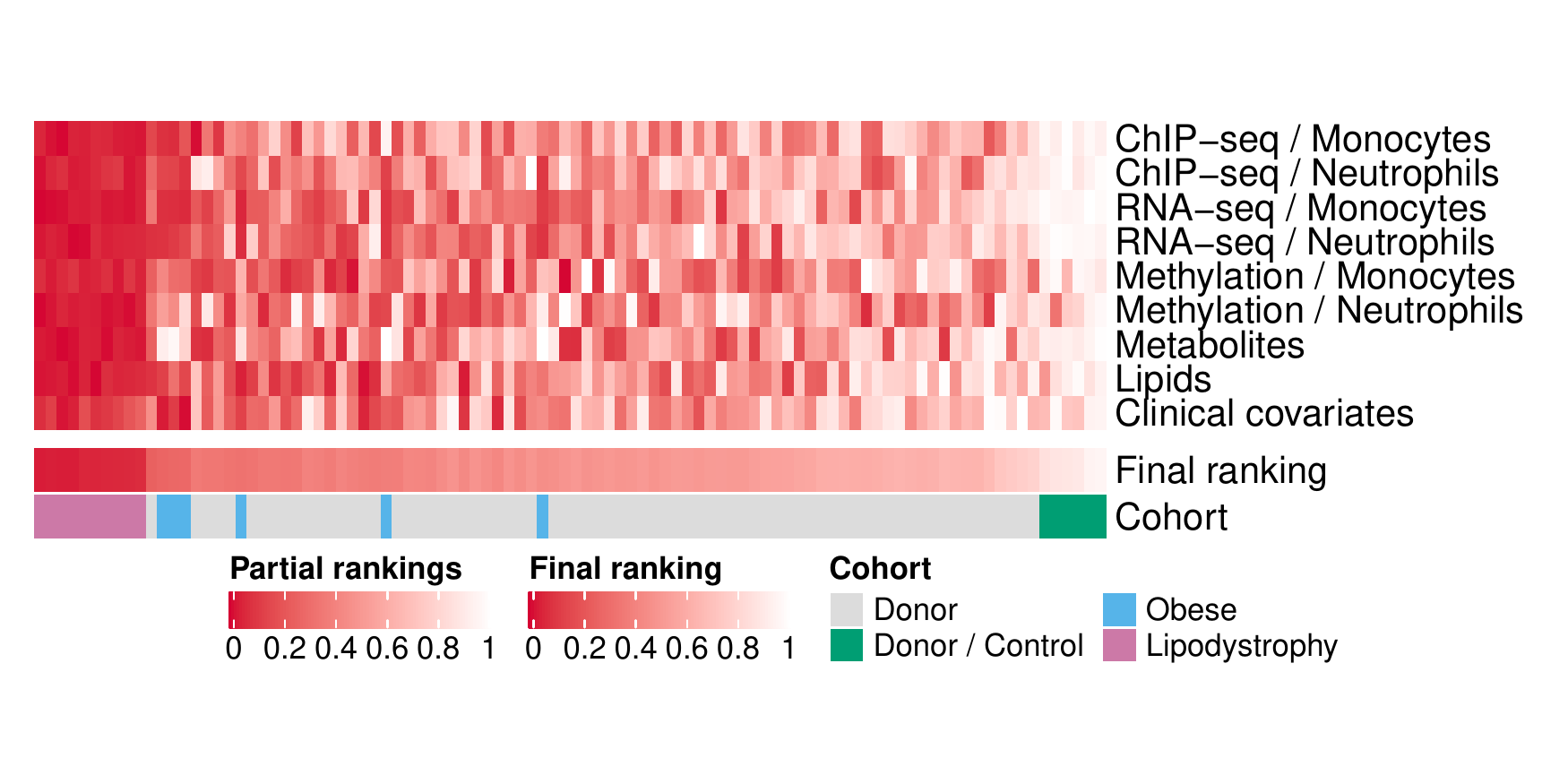}
		\vspace{-.5cm}
		\caption{Combined rankings, $\alpha=0.1$.}
	\end{subfigure}
		\begin{subfigure}{.49\textwidth}
		\centering
		\includegraphics[width=\textwidth]{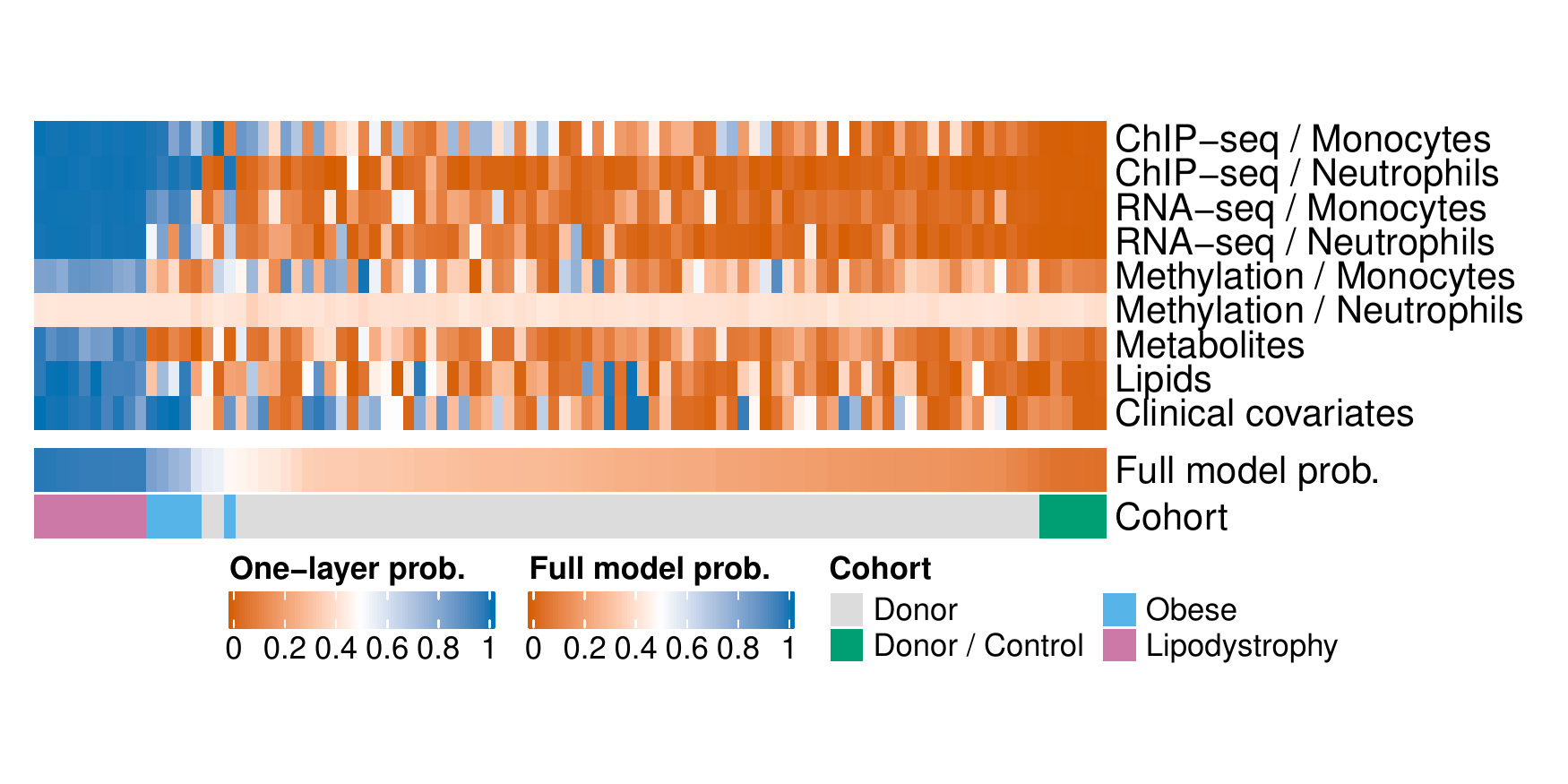}
		\vspace{-.5cm}		
		\caption{Probability of being a \emph{case}, $\alpha=0.5$.}
	\end{subfigure}
	\begin{subfigure}{.49\textwidth}
		\centering
		\includegraphics[width=\textwidth]{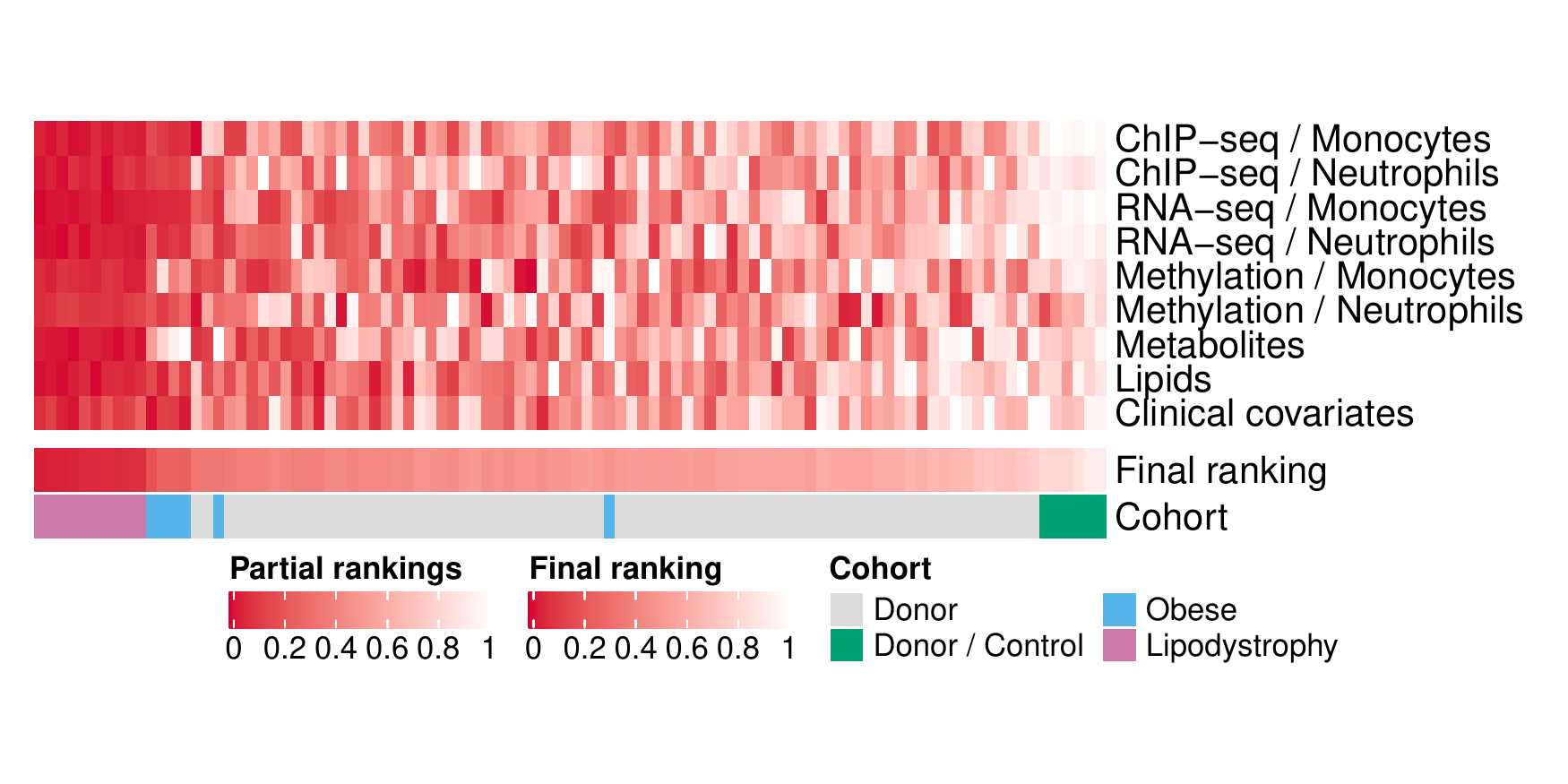}
		\vspace{-.5cm}
		\caption{Combined rankings, $\alpha=0.5$.}
	\end{subfigure}
		\begin{subfigure}{.49\textwidth}
		\centering
		\includegraphics[width=\textwidth]{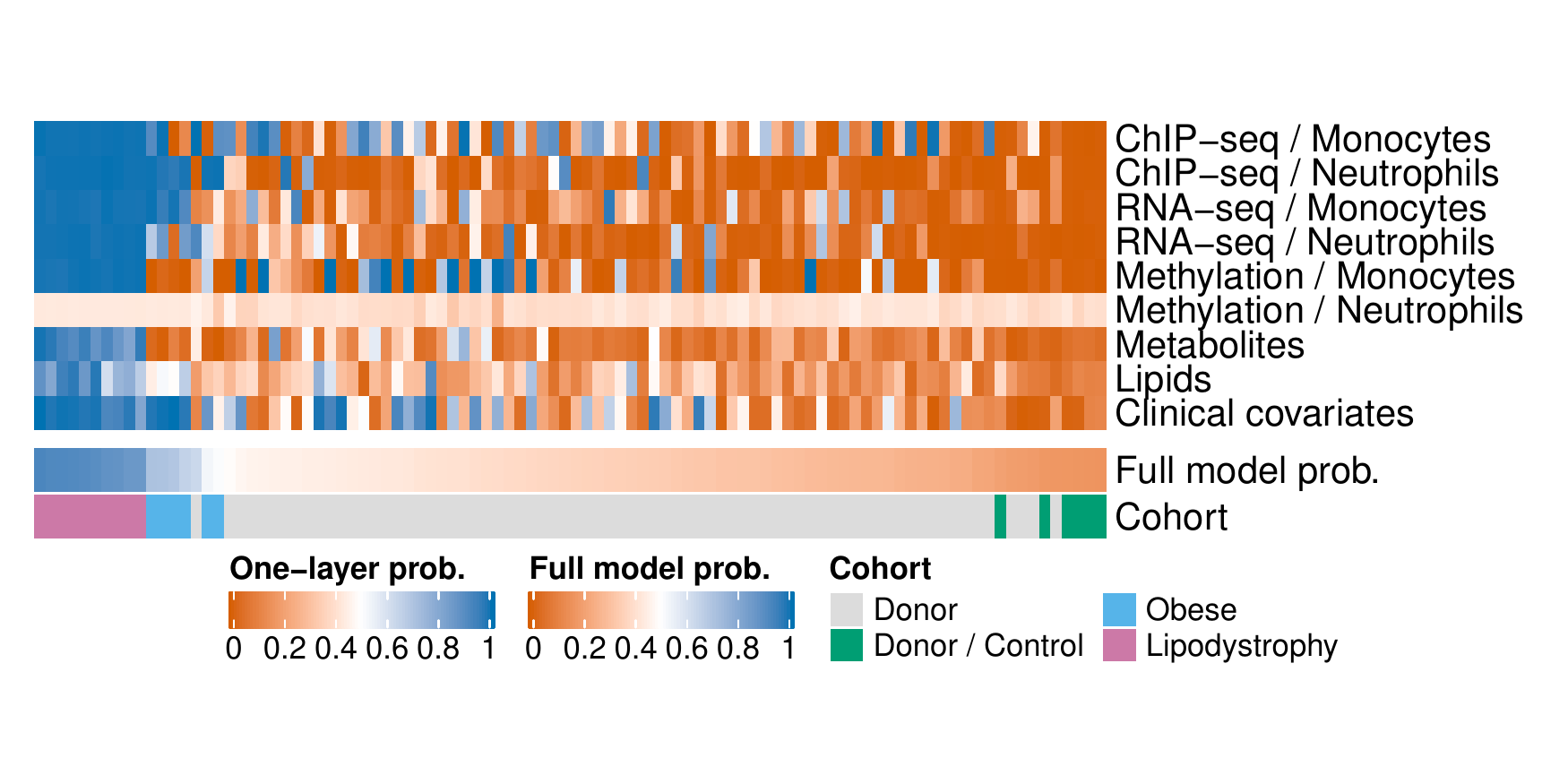}
		\vspace{-.5cm}		
		\caption{Probability of being a \emph{case}, $\alpha=1$.}
	\end{subfigure}
	\begin{subfigure}{.49\textwidth}
		\centering
		\includegraphics[width=\textwidth]{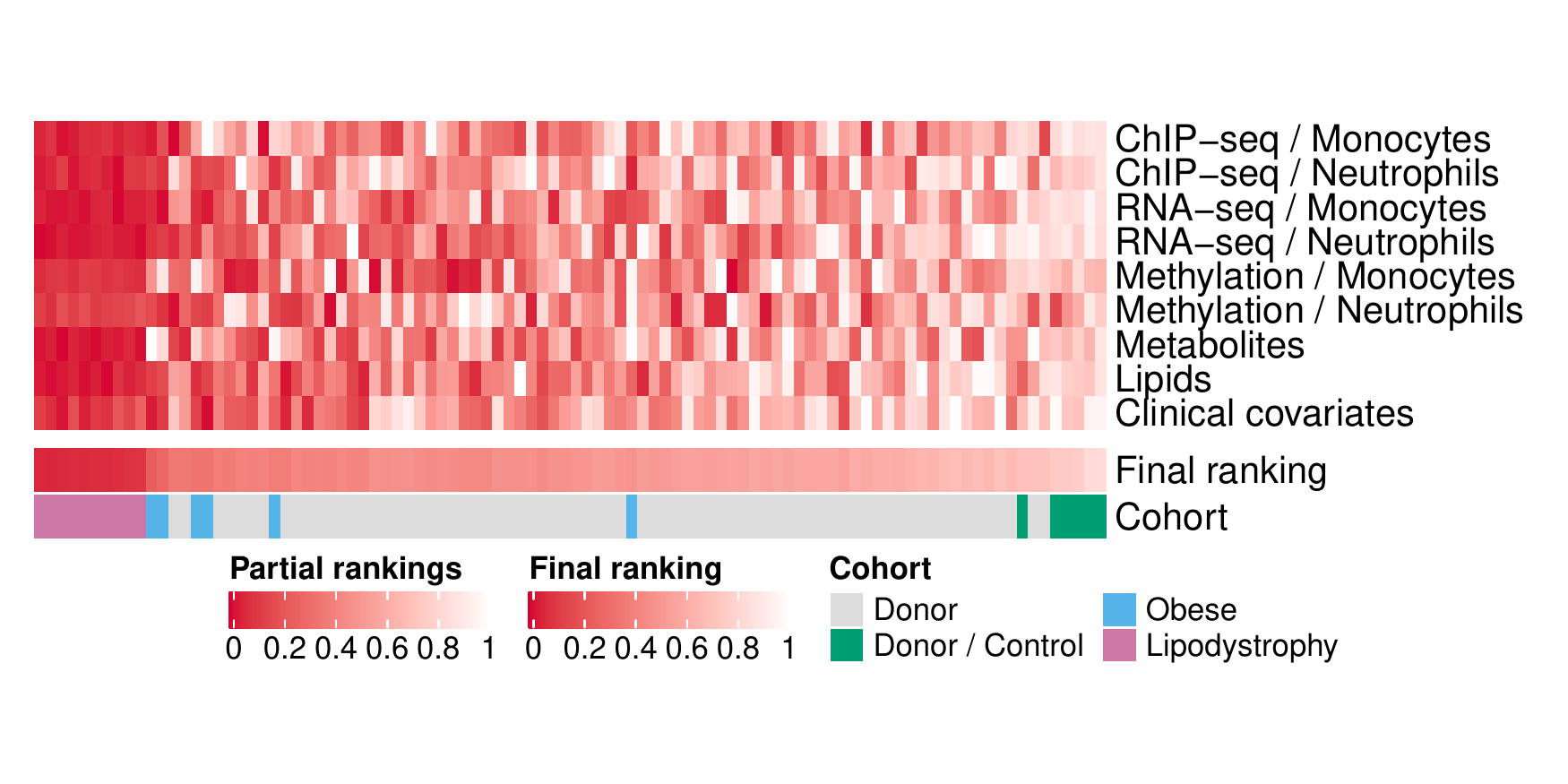}
		\vspace{-.5cm}
		\caption{Combined rankings, $\alpha=1$.}
	\end{subfigure}
	\caption[CMS probabilities and rankings: lipodystrophy patients versus controls.]{Probabilities of belonging to the class of lipodystrophy patients and ranking of each person according to those probabilities. Both quantities are shown on each dataset separately and considering all the data types jointly. The model is trained on the lipodystrophy patients and control donors. Each column corresponds to one of the individuals who have no missing data, each row corresponds to one of the layers. The columns are sorted by probability of being a case in (a), (c), and (e) and final ranking in (b), (d), and (f). All rankings are divided by the total number of observations.}
\label{fig:probabilities-ranks-lipodystrophy}
\end{figure}

In Figure \ref{fig:probabilities-ranks-lipodystrophy} are reported the probabilities of being a \emph{case}, i.e. lipodystrophy patient, for each individual and each layer, as well as those given by a ridge regression on the clinical covariates. The probability given by the full model including all the selected variables is also reported. Moreover, the rankings of each person by probability of being a \emph{case} for each layer and the set of clinical covariates are reported, together with the final, average ranking. The results are comparable to those obtained when using the obese individuals for the training set, where all patients have high probabilities and rankings.

\subsection{Interpretation of results}

While it might be interesting to use models like the one presented here to identify donors affected by CMS, the results presented in this work cannot be used for diagnostic purposes and should be considered as explorative. This is because the sample size is quite small, especially compared to the large number of covariates considered. Widely accepted practices for studies where prediction models are developed, validated or updated for prognostic or diagnostic purposes are those outlined by \citet{moons2015transparent}, which are summarised into the TRIPOD checklist. Moreover, \citet{riley2020calculating} give precise indications for sample size calculation in clinical prediction models. To build a predictive model that can be used in the clinic, one would need to collect data in compliance with those guidelines.

\section{Conclusion}
\label{sec:conclusion}

We proposed two new ways of building binary predictive models for multi-omic datasets with the aim of retrieving as many relevant predictive variables as possible. The variables are selected via EN on each 'omic layer separately, allowing the user to either have the freedom to choose the amount of selection, manually selecting the value of the parameter $\alpha$, or automatically selecting the value of $\alpha$ that minimises the MR. We compared these two methods to the two main competitor methods for multi-omic logistic regression that use a EN-type penalty, and to a univariate approach. From our simulation studies, we concluded that there is no one-size-fits-all approach that is able to achieve low MR in all settings as well as high precision and recall. However, our two suggested methods give higher values of the recall in all simulation settings, so they should be preferred in contexts where the interpretability of the model is key. We presented a real data analysis of cardiometabolic syndrome data, showing that our approach is able to discriminate between healthy donors and CMS patients and that it selects variables that are predictive of this syndrome, at least in two of the eight available layers. Molecular signatures of CMS are identified in each omic layer, which can be used to guide the identification of new diagnosis and treatment strategies.

\section*{Funding}
A. Cabassi and P. D. W. Kirk are supported by the MRC [MC\_UU\_00002/13]. M. Frontini is supported by a BHF Sr Basic Science Fellowship [FS/18/53/33863]. This work was supported by the National Institute for Health Research [Cambridge Biomedical Research Centre at the Cambridge University Hospitals NHS Foundation Trust] [*]. *The views expressed are those of the authors and not necessarily those of the NHS, the NIHR or the Department of Health and Social Care.

\bibliographystyle{apalike}
\addcontentsline{toc}{section}{Bibliography}
\bibliography{main}

\end{document}


\begin{center}
{\LARGE\bf Two-step penalised logistic regression for multi-omic data with an application to cardiometabolic syndrome}
\end{center}
\medskip
\begin{center}
{\large Alessandra Cabassi$^{1}$, Denis Seyres$^{2,3,4}$, Mattia Frontini$^{3,4,5,6}$, and Paul D. W. Kirk$^{1,7}$ \\[15pt]

\emph{$^{1}$MRC Biostatistics Unit, University of Cambridge, UK}\\
\emph{$^{2}$National Institute for Health Research BioResource, Cambridge University Hospitals, UK}\\
\emph{$^{3}$Department of Haematology, University of Cambridge, UK}\\
\emph{$^{4}$NHS Blood and Transplant, Cambridge Biomedical Campus, Cambridge, UK}\\
\emph{$^{5}$Institute of Biomedical \& Clinical Science, College of Medicine and Health,\\ University of Exeter Medical School, UK}\\
\emph{$^{6}$British Heart Foundation Centre of Excellence, Cambridge Biomedical Campus, UK}\\
\emph{$^{7}$Cambridge Institute of Therapeutic Immunology \& Infectious Disease,\\ University of Cambridge, UK}\\
}

\end{center}

\bigskip

\begin{center}
Supplementary Material
\end{center}
\bigskip\bigskip

\appendix

In this Supplementary Material, we include details that were omitted from the main paper for the sake of brevity.
In Section \ref{sec:additional-simulations} we report the results of additional simulation studies and give more details about the choice of the parameter $\alpha$ for the first of our proposed approaches, where $\alpha$ is kept fixed.
In Section \ref{sec:additional-data-analysis} we present the outcome of the real data analysis presenting some additional output of the comparison between obese patients and healthy donors.

\section{Additional simulation studies}
\label{sec:additional-simulations}

We present here the results of two additional simulation settings. In the first one, only two penalised layers are combined. The second one is similar to the one presented in Section 3.2 of the main paper, with the number of penalised covariates increased from two to ten and 100.

\subsection{Penalised covariates only}
The parameters of the additional simulation study presented here are the same as those presented in Table 1 of the main paper, except that here $P_N = 0$ in all settings. 
In this case the two covariance matrices simplify to:
\begin{equation} \label{eq:diagonal-matrix-with-nonpen}
\Sigma_0 = \mathbb{I}_{P_1+P_2}
\end{equation}
and
\begin{equation} \label{eq:block-matrix}
\Sigma_1 = 
\left[
\begin{array}{cccc | cccc}
A_1 &  & & & B_{12} & & & \\
& A_1 & & & &B_{12} & & \\
& & \dots & & & & \dots & \\
& & & A_1 & & & & B_{12}\\
\hline
B_{21} & & & &  A_2 & & & \\
&B_{21} & & & & A_2 & &\\
& & \dots & & & & \dots & \\
& & &B_{21} & & & & A_2 \\
\end{array}
\right]
\end{equation}

Since there are no covariates that are not penalised here, we also add the results obtained with a different univariate method. The idea is that for each  variable a Mann-Whitney test is performed: if significant differences are observed between the two classes with confidence level 0.05, after adjusting for multiplicity using the Benjamini-Hochberg procedure, then that variable is selected. The variables selected in this way are then used to fit a ridge-penalised regression model. 

In addition to that, for each method in each setting, we report the number of datasets for which each method is run successfully. When sIPF it fails, this is due to the fact that all evaluations of the error surface have the same value. Results for the datasets affected by this problem are missing; these are only a small fraction of the total number of the generated datasets.
Moreover, the 2-step methods that fit a ridge regression model on the selected variables fail when only one variable is selected, since the ``glmnet'' implementation of logistic regression does not work with only one regressor. 

\begin{figure}
\centering
\includegraphics[width=.95\textwidth]{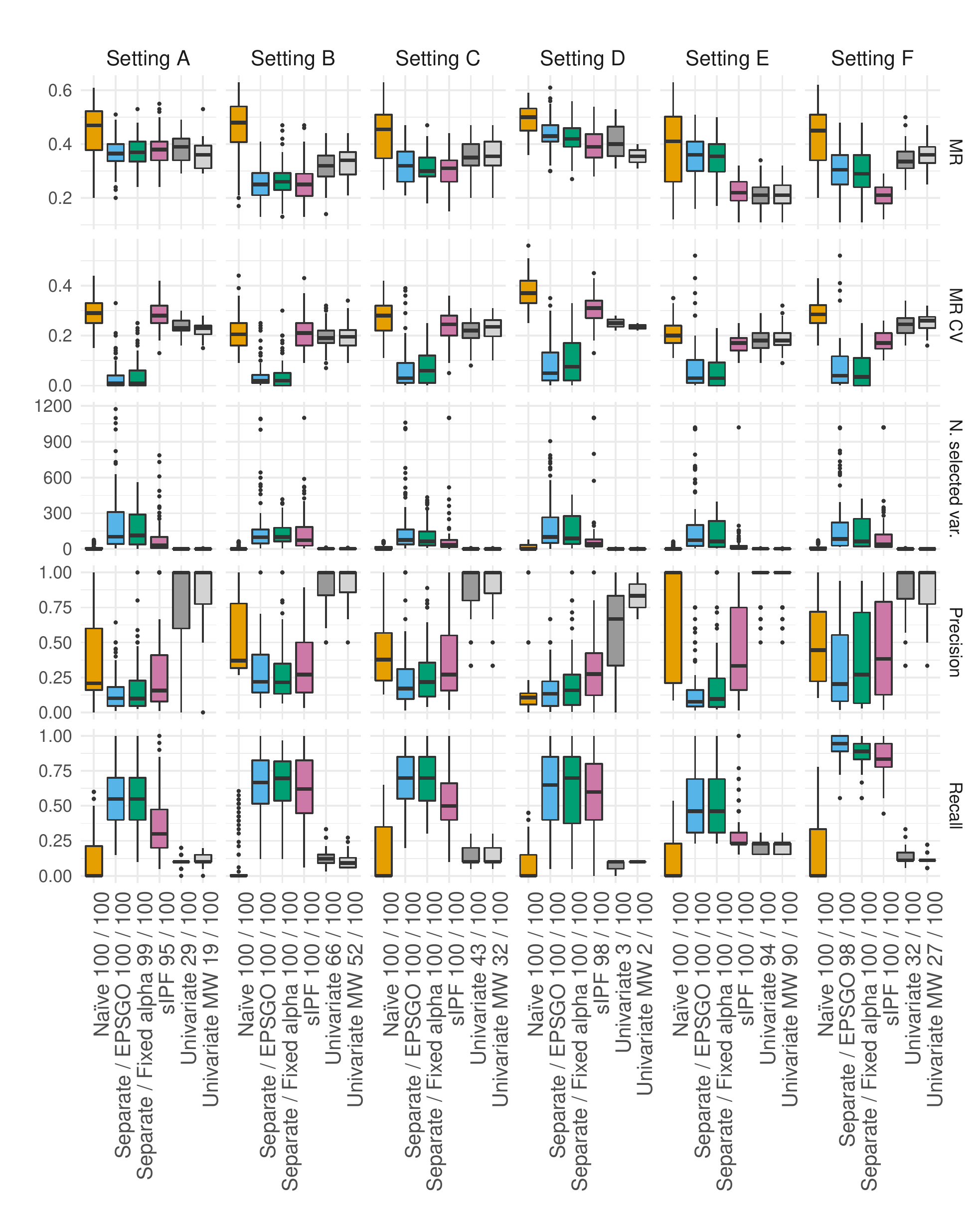}
\caption[Simulation: diagonal covariance, penalised layers only.]{Simulation study comparing  different variants of elastic-net for multi-omic data and two univariate approaches.
The covariance matrix used here is the diagonal matrix $\Sigma_0$ and $P_N=0$.
``MR'' is the out-of-sample misclassification rate, ``MR CV'' the within-sample misclassification rate. 
``Univariate'' refers to the univariate method presented in the main paper, ``Univariate MW'' indicates the univariate Mann-Whitney test.}
\label{fig:diagonalcov-allpen}
\end{figure}

\begin{figure}
\centering
\includegraphics[width=.95\textwidth]{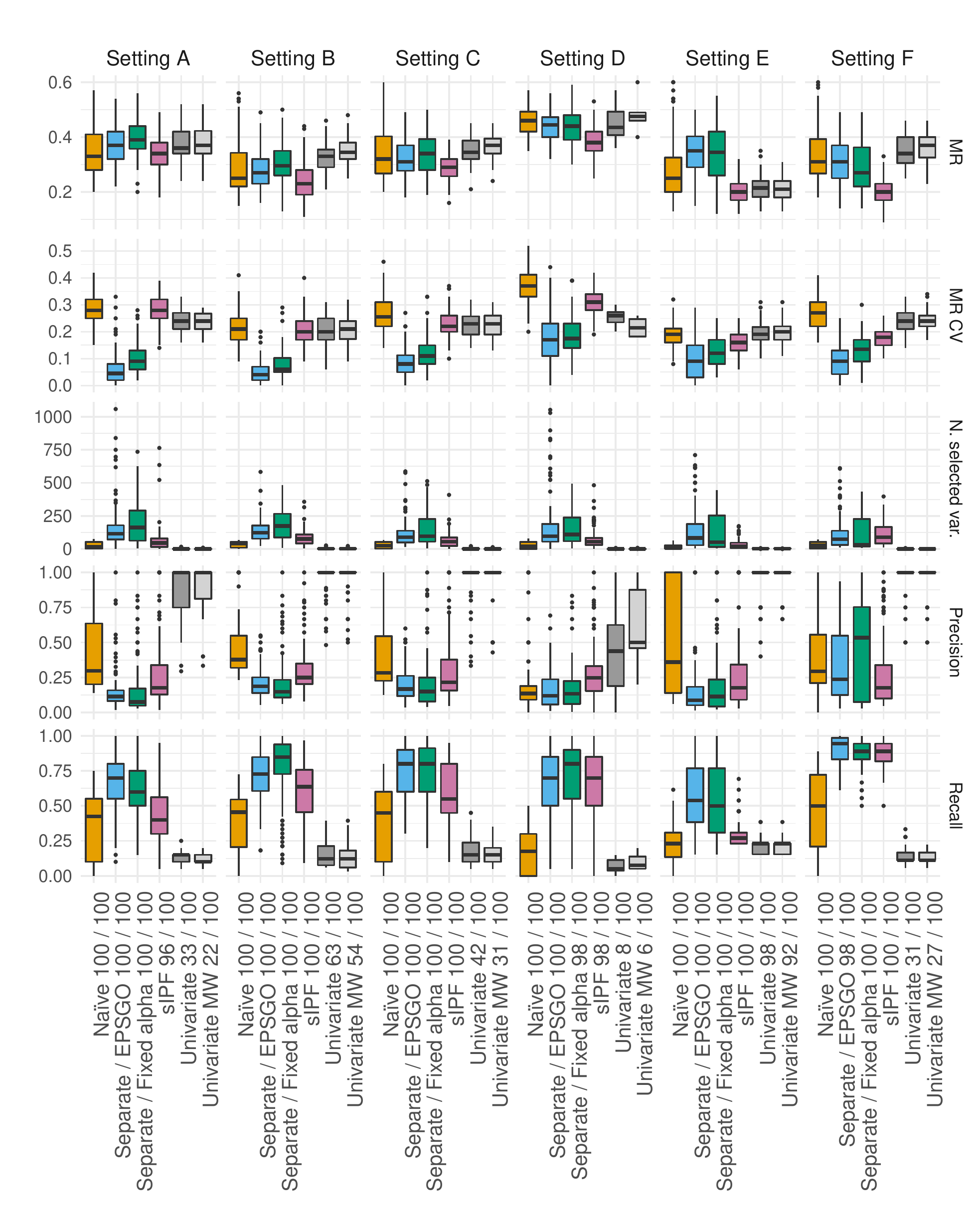}
\caption[Simulation: non-diagonal covariance, penalised layers only.]{Simulation study comparing  different variants of elastic-net for multi-omic data and two univariate approaches.
The covariance matrix used here is the block matrix $\Sigma_1$ and $P_N=0$.
``MR'' is the out-of-sample misclassification rate, ``MR CV'' the within-sample misclassification rate. ``Univariate'' refers to the univariate method presented in the main paper, ``Univariate MW'' indicates the univariate Mann-Whitney test.}
\label{fig:nondiagcov-allpen}
\end{figure}

The results are shown in Figures \ref{fig:diagonalcov-allpen} and \ref{fig:nondiagcov-allpen}.
In the case of diagonal covariance, na\"{i}ve EN has higher MR, both within and out-of-sample. The number of variables selected by this method is extremely low, so it is not surprising to observe higher precision and lower recall compared to the other methods. The low number of selections is due to the fact that na\"{i}ve-EN, like sIPF-EN, it is able to identify sets of features from different layers that together are predictive of outcome status.
The separate EN-type methods have very low values of within-sample MR, but similar values of out-of-sample MR to sIPF-EN in the first four settings. In the last two settings, which are highly unbalanced, sIPF-EN greatly outperforms the other two algorithms both in terms of out-of-sample MR and precision.
In the non-diagonal covariance case, similar outcomes are observed. The main difference is that na\"{i}ve-EN has lower MN and higher recall in this setting.
In both cases, the two univariate methods behave similarly, selecting fewer variable than the other methods (as expected) and achieving a low MR only in Setting E.

\subsection{Higher number of  non-penalised covariates}

\begin{figure}
\centering
\includegraphics[width=.95\textwidth]{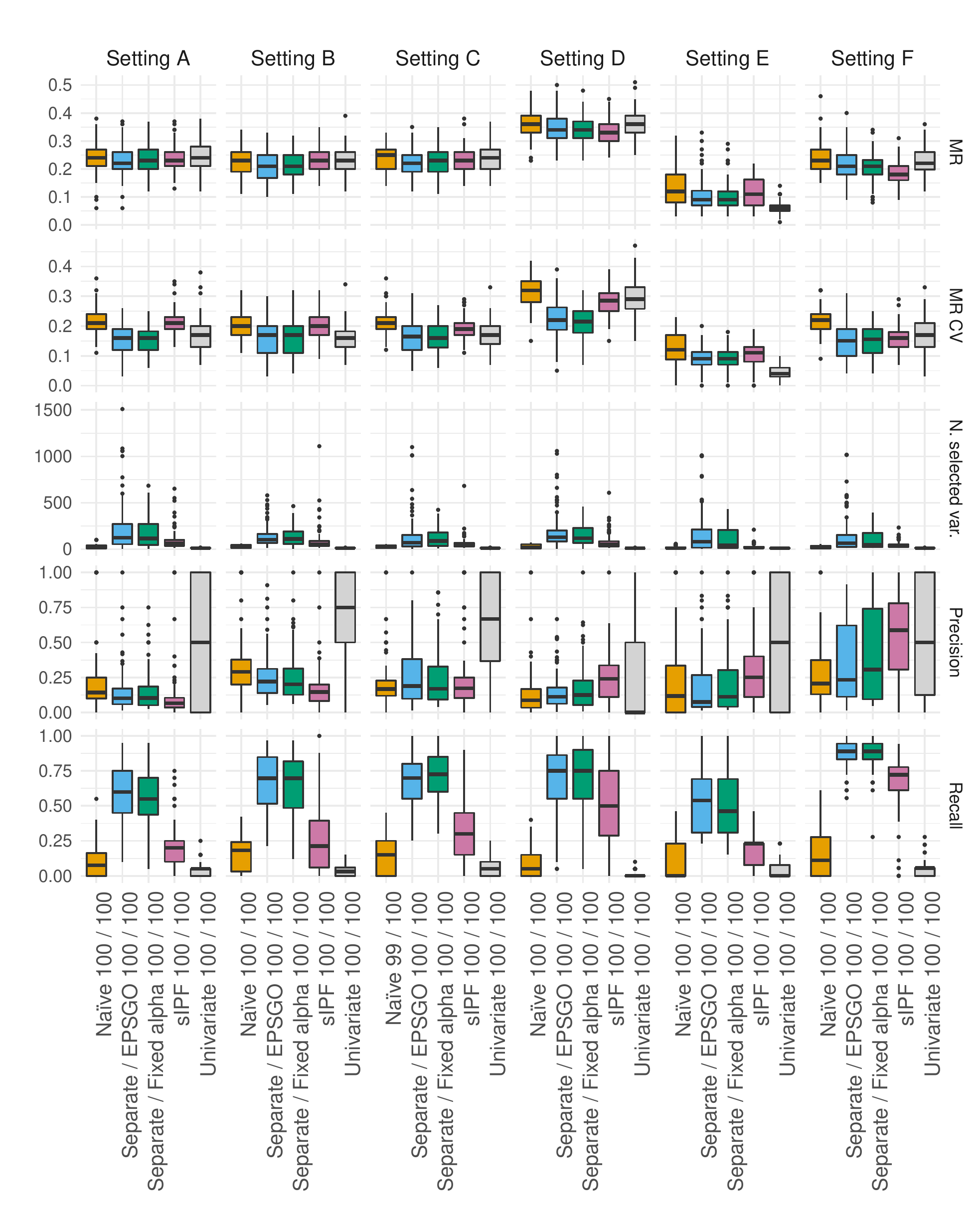}
\caption[Simulation: diagonal covariance, 10 non-penalised covariates.]{Simulation study comparing  different variants of elastic-net for multi-omic data.
The covariance matrix used here is the diagonal matrix $\Sigma_0$ and $P_N=10$.
``MR'' is the out-of-sample misclassification rate, ``MR CV'' the within-sample misclassification rate. The non-penalised covariates are not included when computing precision and recall.}
\label{fig:diagonalcov-few-nonpen}
\end{figure}

\begin{figure}
\centering
\includegraphics[width=.95\textwidth]{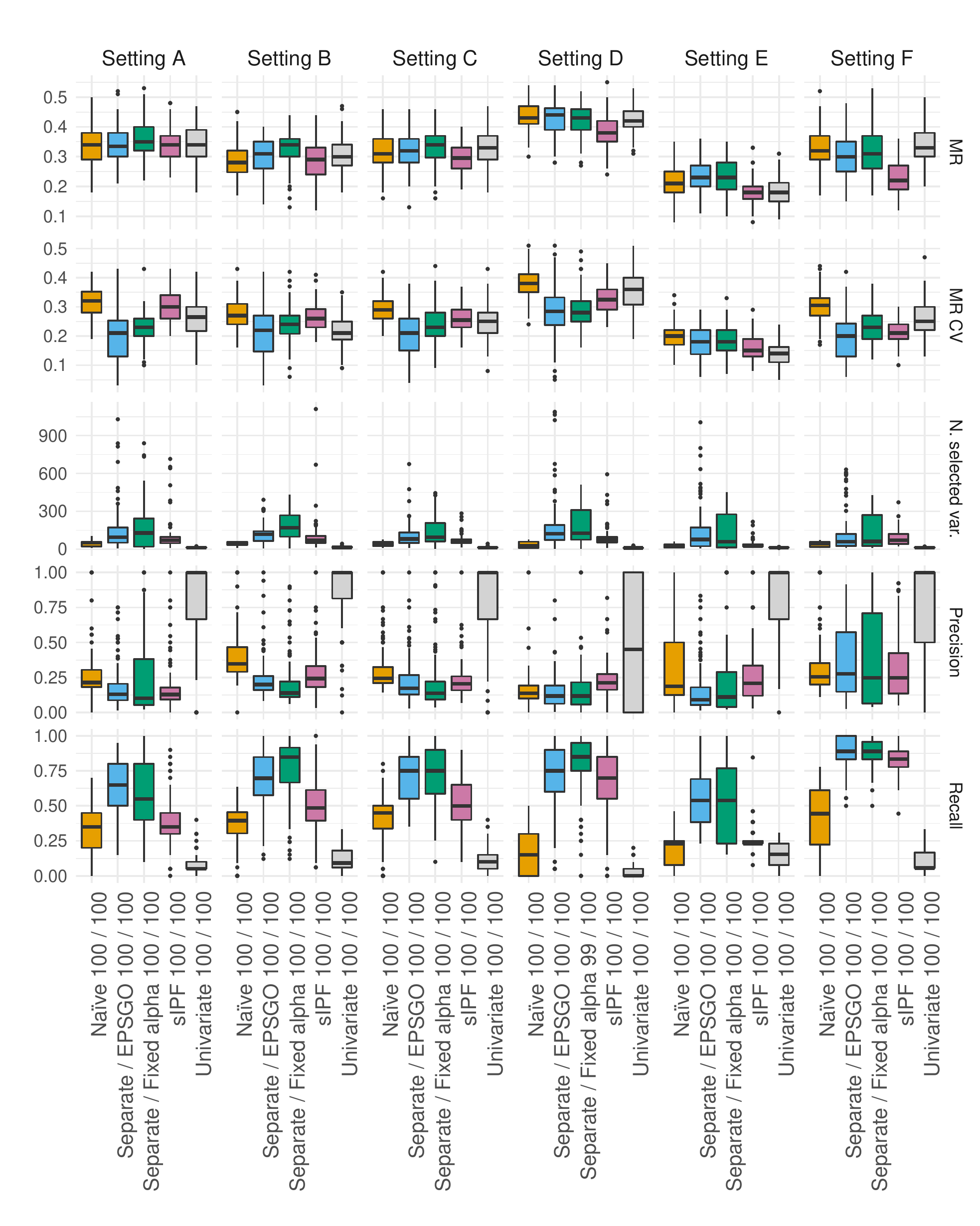}
\caption[Simulation: block diagonal covariance, 10 non-penalised covariates.]{Simulation study comparing  different variants of elastic-net for multi-omic data.
The covariance matrix used here is the block matrix $\Sigma_1$ and $P_N=10$.
``MR'' is the out-of-sample misclassification rate, ``MR CV'' the within-sample misclassification rate. The non-penalised covariates are not included when computing precision and recall.}
\label{fig:nondiagcov-few-nonpen}
\end{figure}

\begin{figure}
\centering
\includegraphics[width=.95\textwidth]{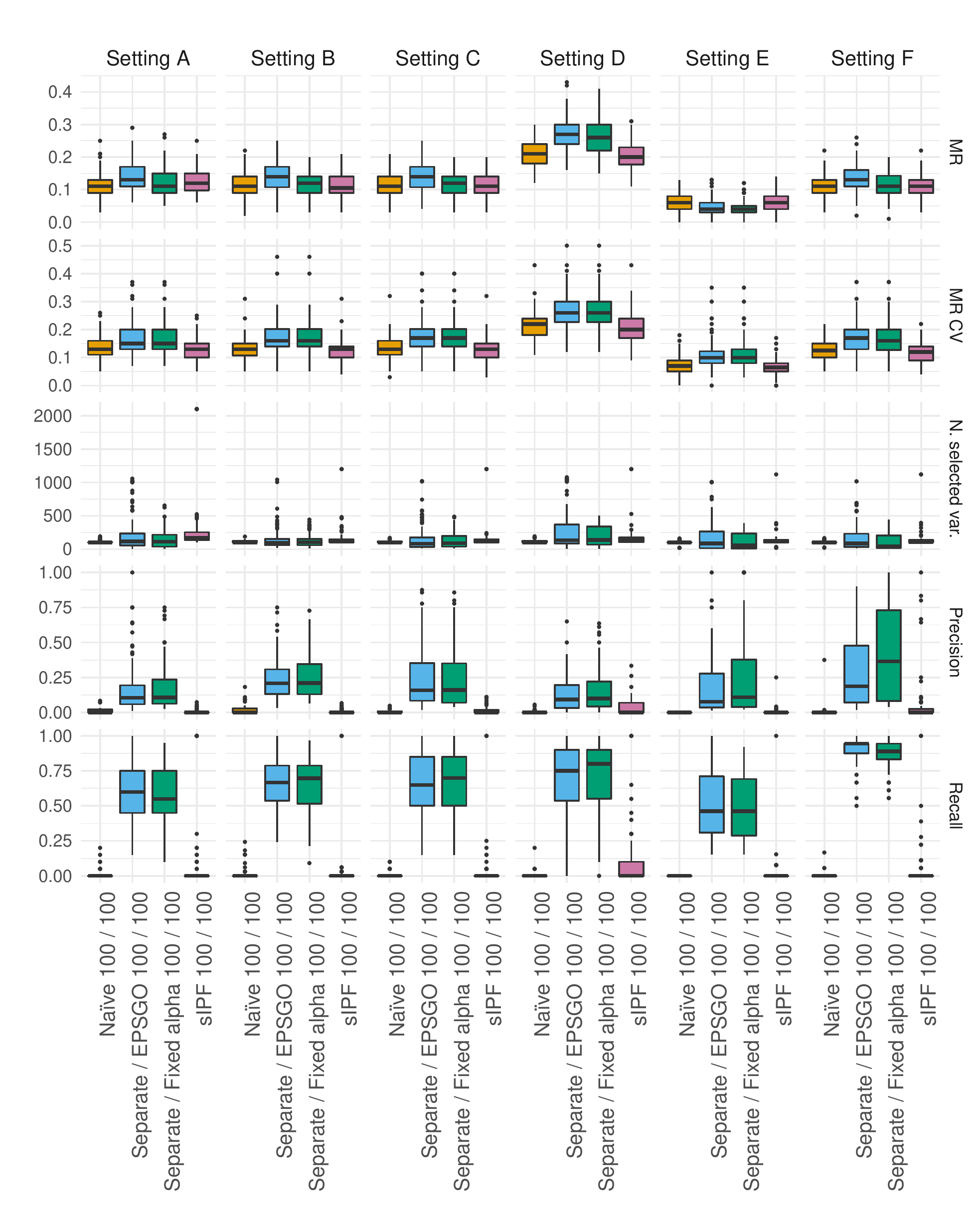}
\caption[Simulation: diagonal covariance, 100 non-penalised covariates.]{
Simulation study comparing  different variants of elastic-net for multi-omic data.
The covariance matrix used here is the diagonal matrix $\Sigma_0$ and $P_N=100$.
``MR'' is the out-of-sample misclassification rate, ``MR CV'' the within-sample misclassification rate. The non-penalised covariates are not included when computing precision and recall.}
\label{fig:diagonalcov-many-nonpen}
\end{figure}

\begin{figure}
\centering
\includegraphics[width=.95\textwidth]{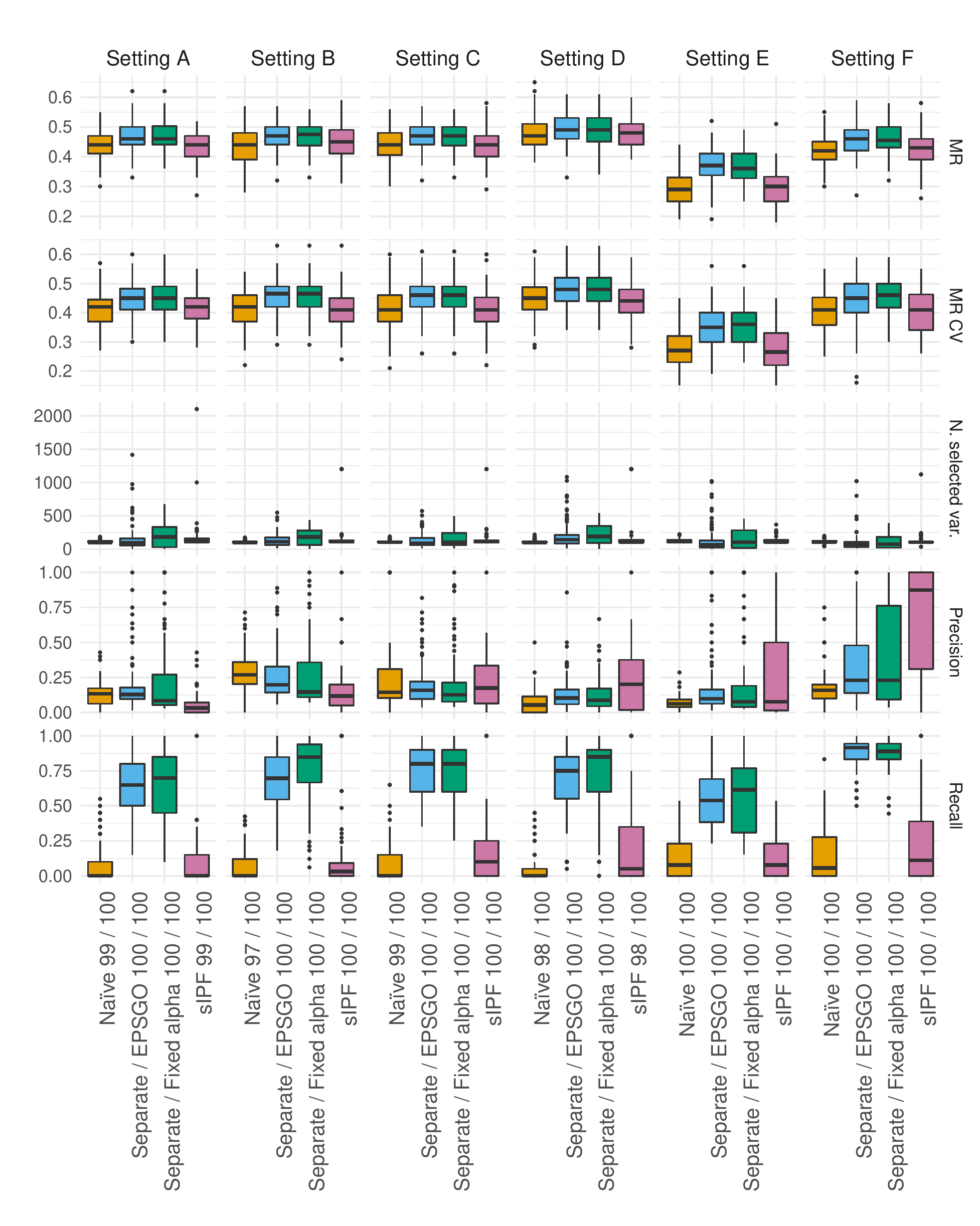}
\caption[Simulation: block diagonal covariance, 100 non-penalised covariates.]{Simulation study comparing  different variants of elastic-net for multi-omic data.
The covariance matrix used here is the block matrix $\Sigma_1$ and $P_N=100$.
``MR'' is the out-of-sample misclassification rate, ``MR CV'' the within-sample misclassification rate. The non-penalised covariates are not included when computing precision and recall.}
\label{fig:nondiagcov-many-nonpen}
\end{figure}

We repeat the experiments presented in the main paper, replacing the value of $P_N$ by 10 (Figures \ref{fig:diagonalcov-few-nonpen} and \ref{fig:nondiagcov-few-nonpen}) and 100 (Figures \ref{fig:diagonalcov-many-nonpen} and \ref{fig:nondiagcov-many-nonpen}). 

The patterns are similar to those observed in the main paper, with the number of selected variables and values of the recall becoming more and more different between the joint models (na\"{i}ve-EN and sIPF-EN) and the two-step approaches as $P_N$ increases. The univariate method cannot be run in the setting where $P_N = 100$, since the number of predictors is higher than the number of statistical units in the screening step, and a linear regression model cannot be fit.

\clearpage
\subsection{Choice of $\boldsymbol{\alpha}$}
\label{sec:simulations-choice-of-alpha}

\begin{figure}
\centering
\includegraphics[width=.95\textwidth]{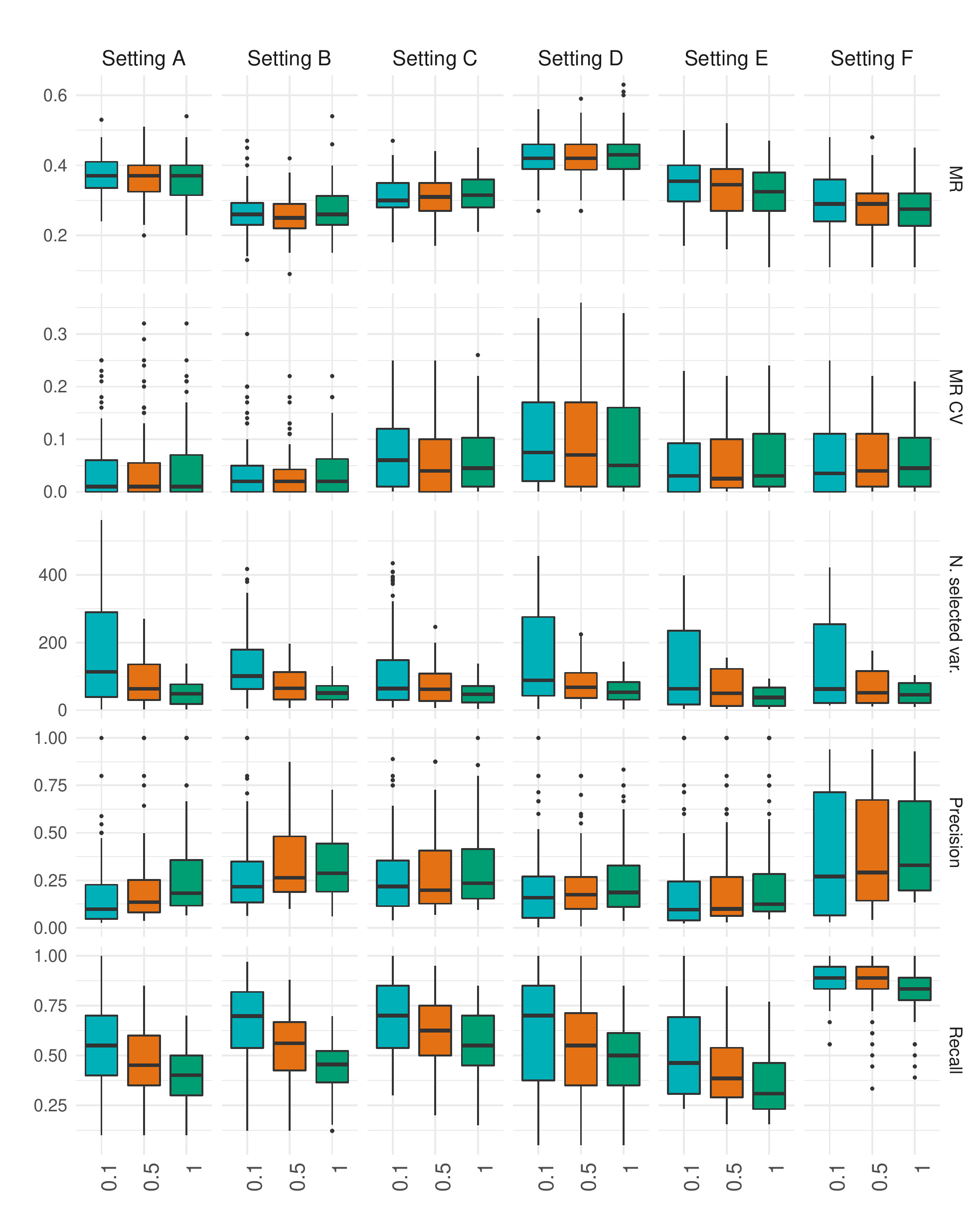}
\caption[Simulation: choice of $\alpha$. Diagonal covariance, penalised layers only.]{Simulation study comparing  different values of $\alpha$.
The covariance matrix used here is the diagonal matrix $\Sigma_0$ and $P_N=0$.
``MR'' is the out-of-sample misclassification rate, ``MR CV'' the within-sample misclassification rate. }
\label{fig:simulation-diagonalcov-allpen-choice-of-alpha}
\end{figure}

\begin{figure}
\centering
\includegraphics[width=.95\textwidth]{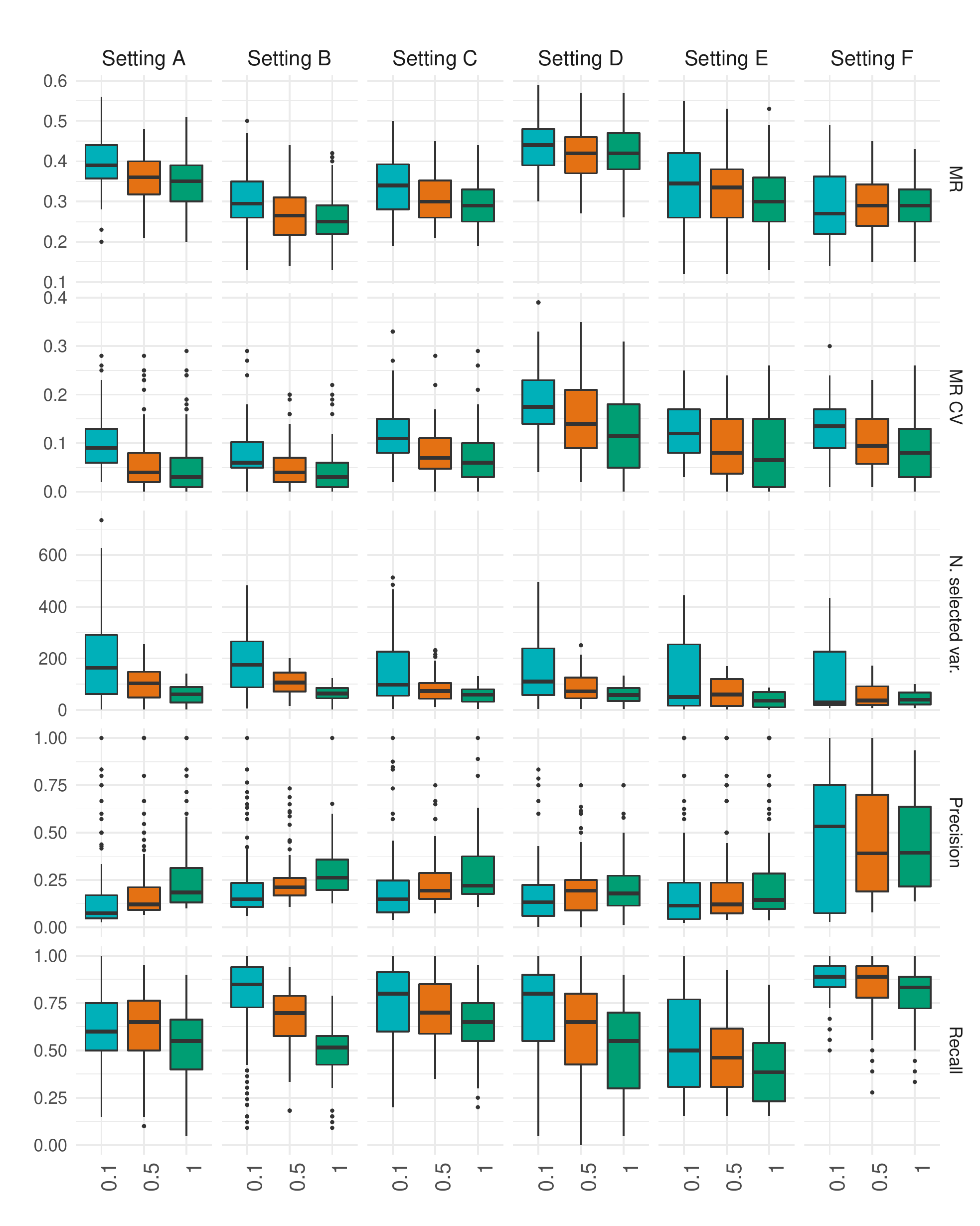}
\caption[Simulation: choice of $\alpha$. Block diagonal covariance, penalised layers only.]{Simulation study comparing  different values of $\alpha$.
The covariance matrix used here is the block matrix $\Sigma_1$ and $P_N=0$.
``MR'' is the out-of-sample misclassification rate, ``MR CV'' the within-sample misclassification rate.}
\label{fig:simulation-nondiagcov-allpen-choice-of-alpha}
\end{figure}

\begin{figure}
\centering
\includegraphics[width=.95\textwidth]{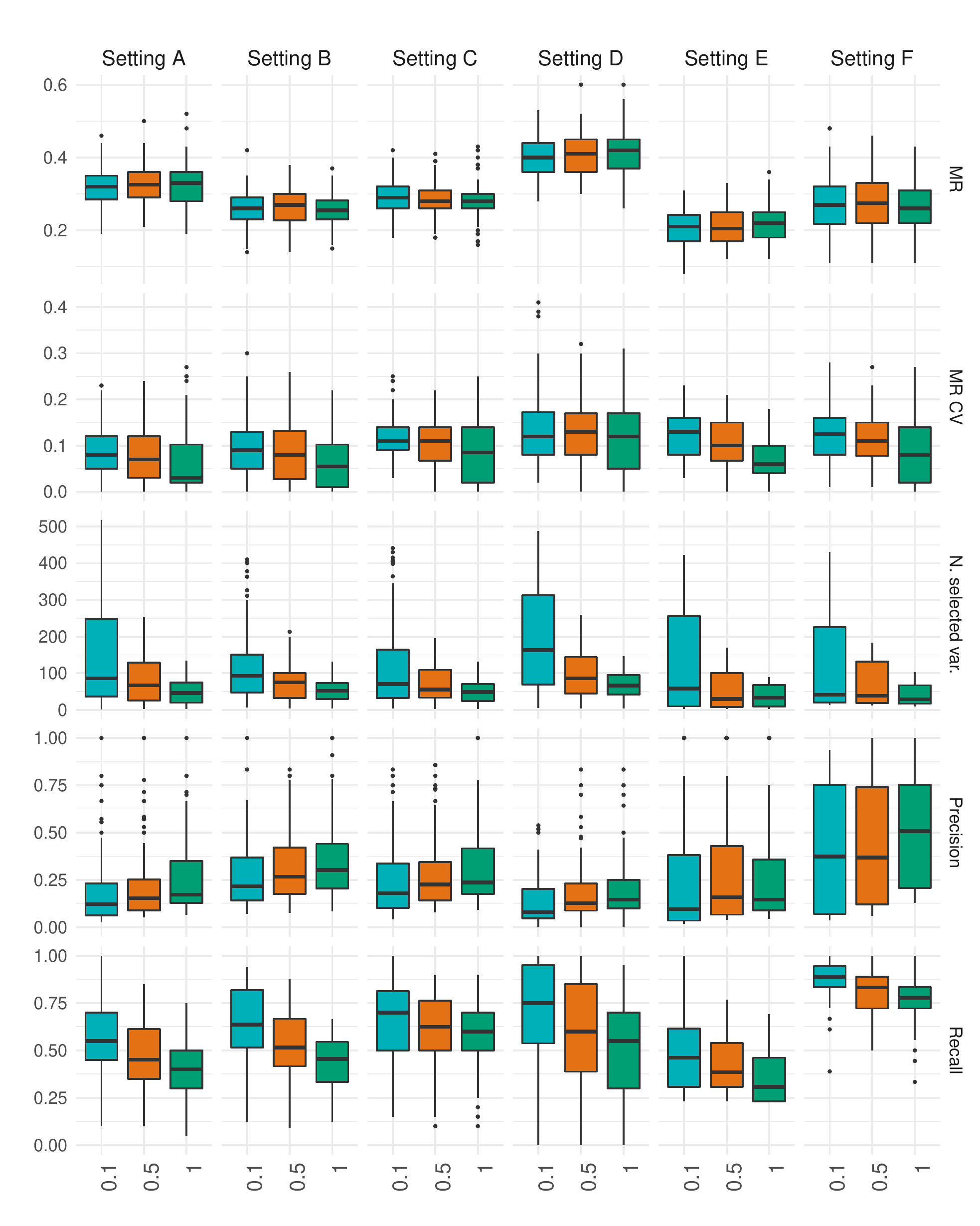}
\caption[Simulation: choice of $\alpha$. Diagonal covariance, penalised layers only.]{Simulation study comparing  different values of $\alpha$.
The covariance matrix used here is the diagonal matrix $\Sigma_0$ and $P_N=2$.
``MR'' is the out-of-sample misclassification rate, ``MR CV'' the within-sample misclassification rate. }
\label{fig:simulation-diagonalcov-two-choice-of-alpha}
\end{figure}

\begin{figure}
\centering
\includegraphics[width=.95\textwidth]{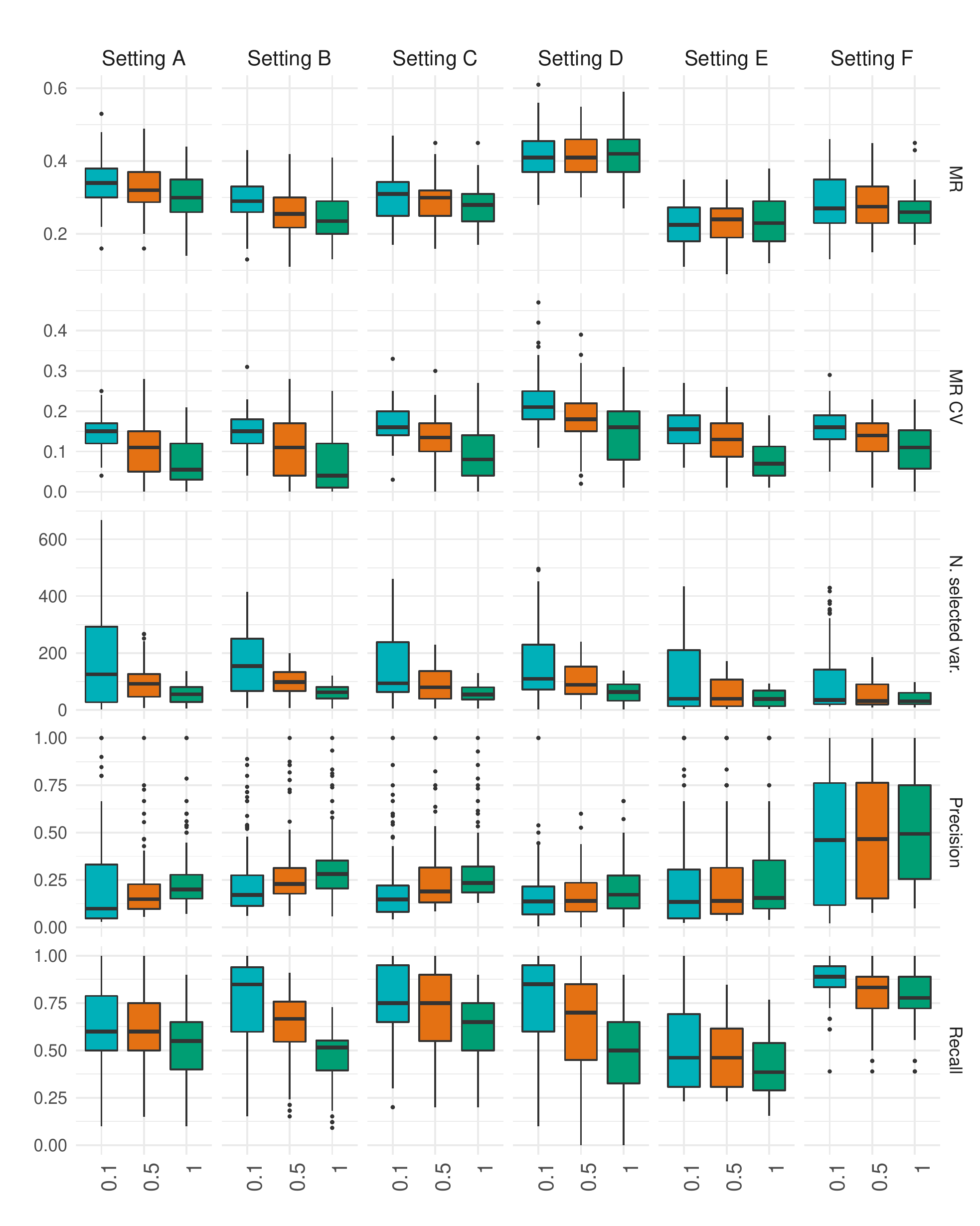}
\caption[Simulation: choice of $\alpha$. Block diagonal covariance, penalised layers only.]{Simulation study comparing  different values of $\alpha$.
The covariance matrix used here is the block matrix $\Sigma_1$ and $P_N=2$.
``MR'' is the out-of-sample misclassification rate, ``MR CV'' the within-sample misclassification rate.}
\label{fig:simulation-nondiagcov-two-choice-of-alpha}
\end{figure}

\begin{figure}
\centering
\includegraphics[width=.95\textwidth]{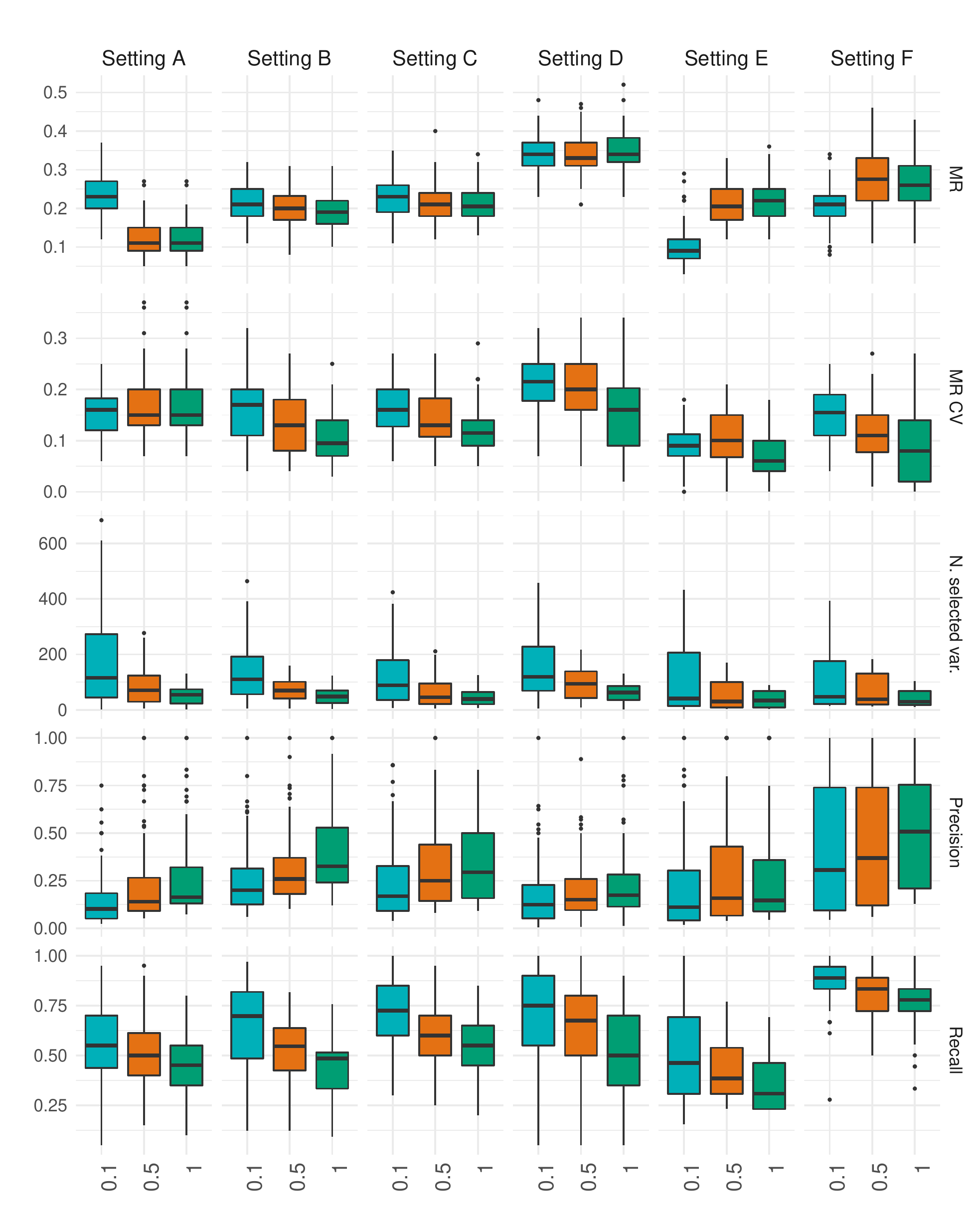}
\caption[Simulation: choice of $\alpha$. Diagonal covariance, penalised layers only.]{Simulation study comparing  different values of $\alpha$.
The covariance matrix used here is the diagonal matrix $\Sigma_0$ and $P_N=10$.
``MR'' is the out-of-sample misclassification rate, ``MR CV'' the within-sample misclassification rate. }
\label{fig:simulation-diagonalcov-few-choice-of-alpha}
\end{figure}

\begin{figure}
\centering
\includegraphics[width=.95\textwidth]{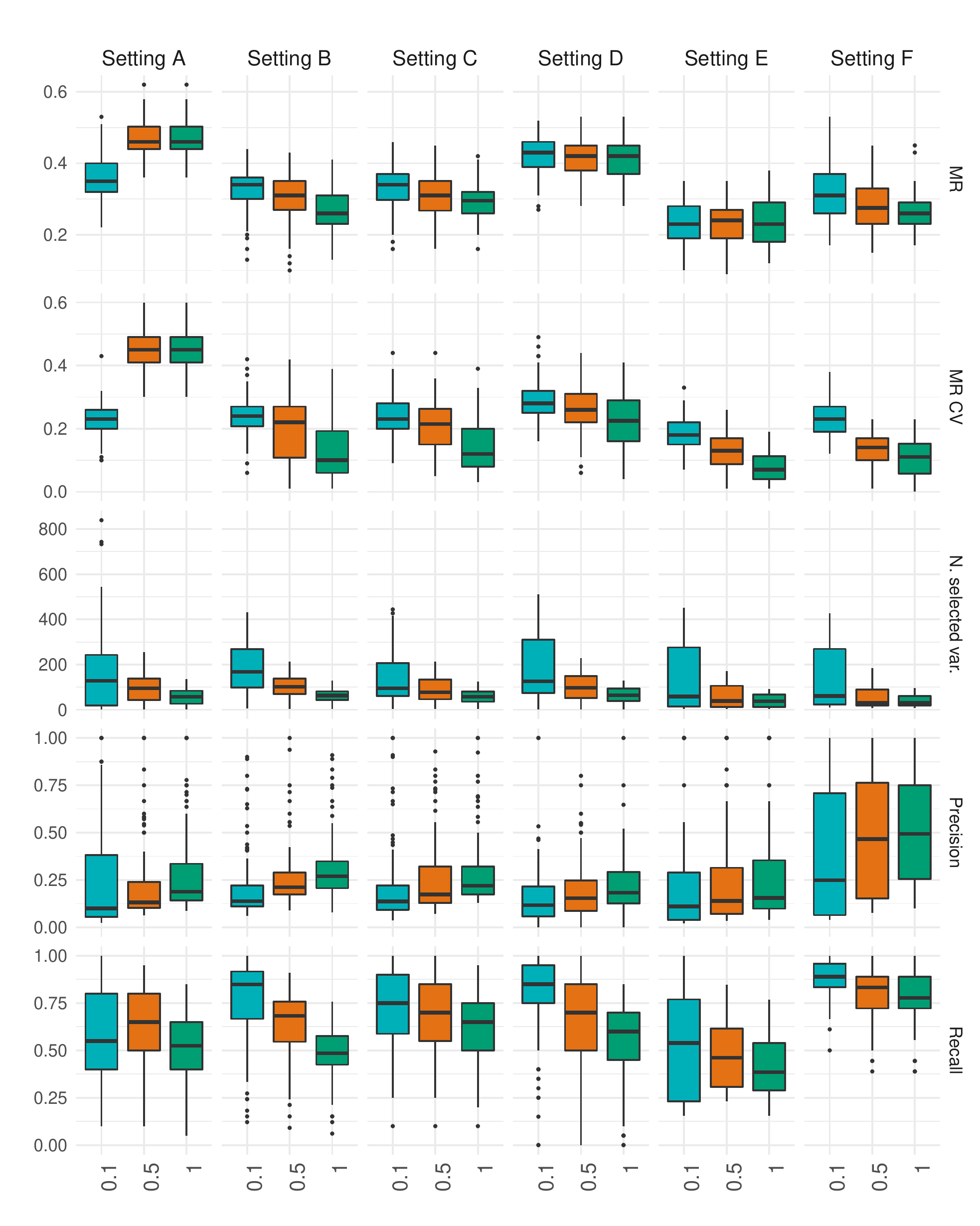}
\caption[Simulation: choice of $\alpha$. Block diagonal covariance, penalised layers only.]{Simulation study comparing  different values of $\alpha$.
The covariance matrix used here is the block matrix $\Sigma_1$ and $P_N=10$.
``MR'' is the out-of-sample misclassification rate, ``MR CV'' the within-sample misclassification rate.}
\label{fig:simulation-nondiagcov-few-choice-of-alpha}
\end{figure}

\begin{figure}
\centering
\includegraphics[width=.95\textwidth]{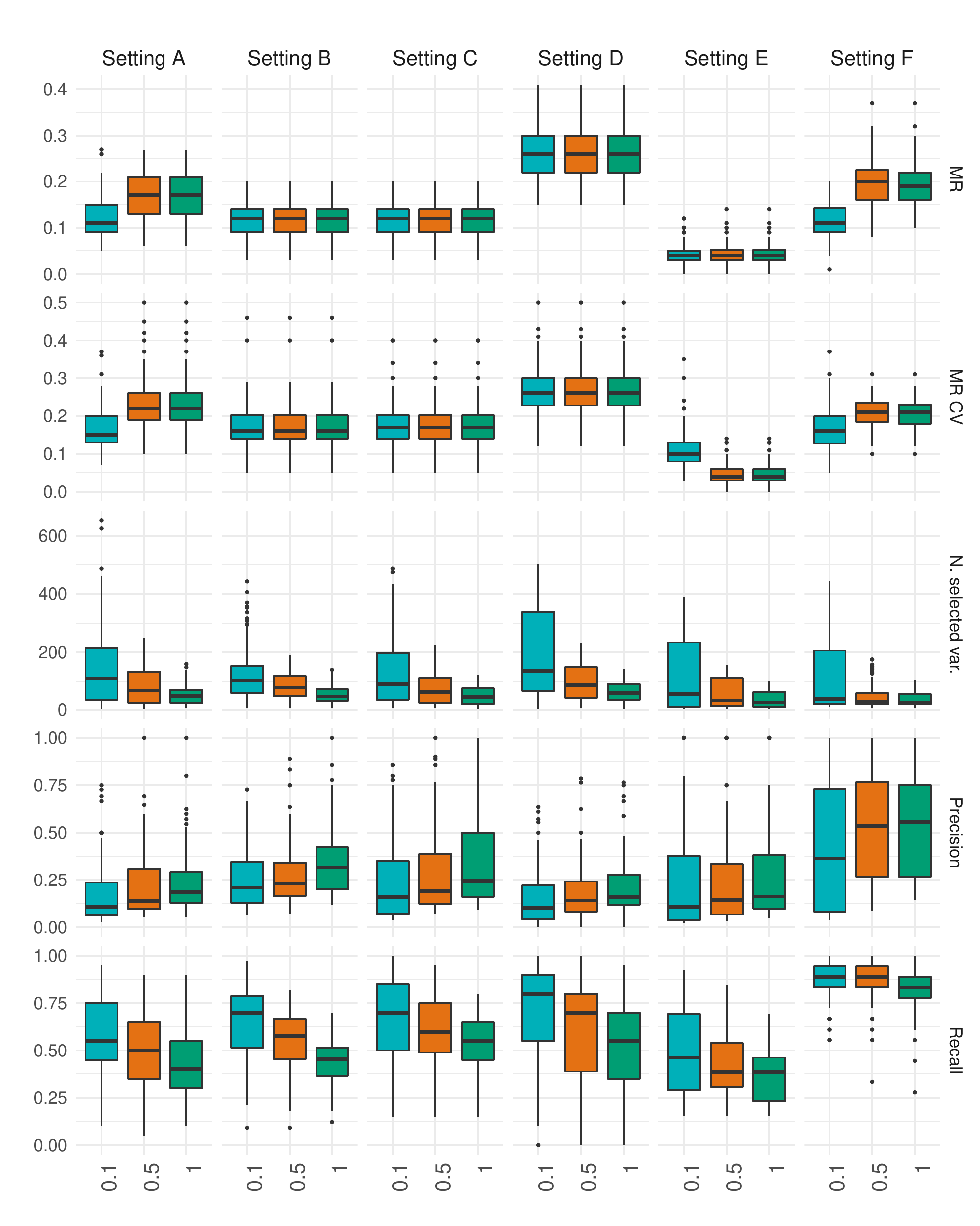}
\caption[Simulation: choice of $\alpha$. Diagonal covariance, 100 non-penalised covariates.]{Simulation study comparing  different values of $\alpha$.
The covariance matrix used here is the diagonal matrix $\Sigma_0$ and $P_N=100$.
``MR'' is the out-of-sample misclassification rate, ``MR CV'' the within-sample misclassification rate. The non-penalised covariates are not included when computing precision and recall.}
\label{fig:simulation-diagonalcov-many-nonpen-choice-of-alpha}
\end{figure}

\begin{figure}
\centering
\includegraphics[width=.95\textwidth]{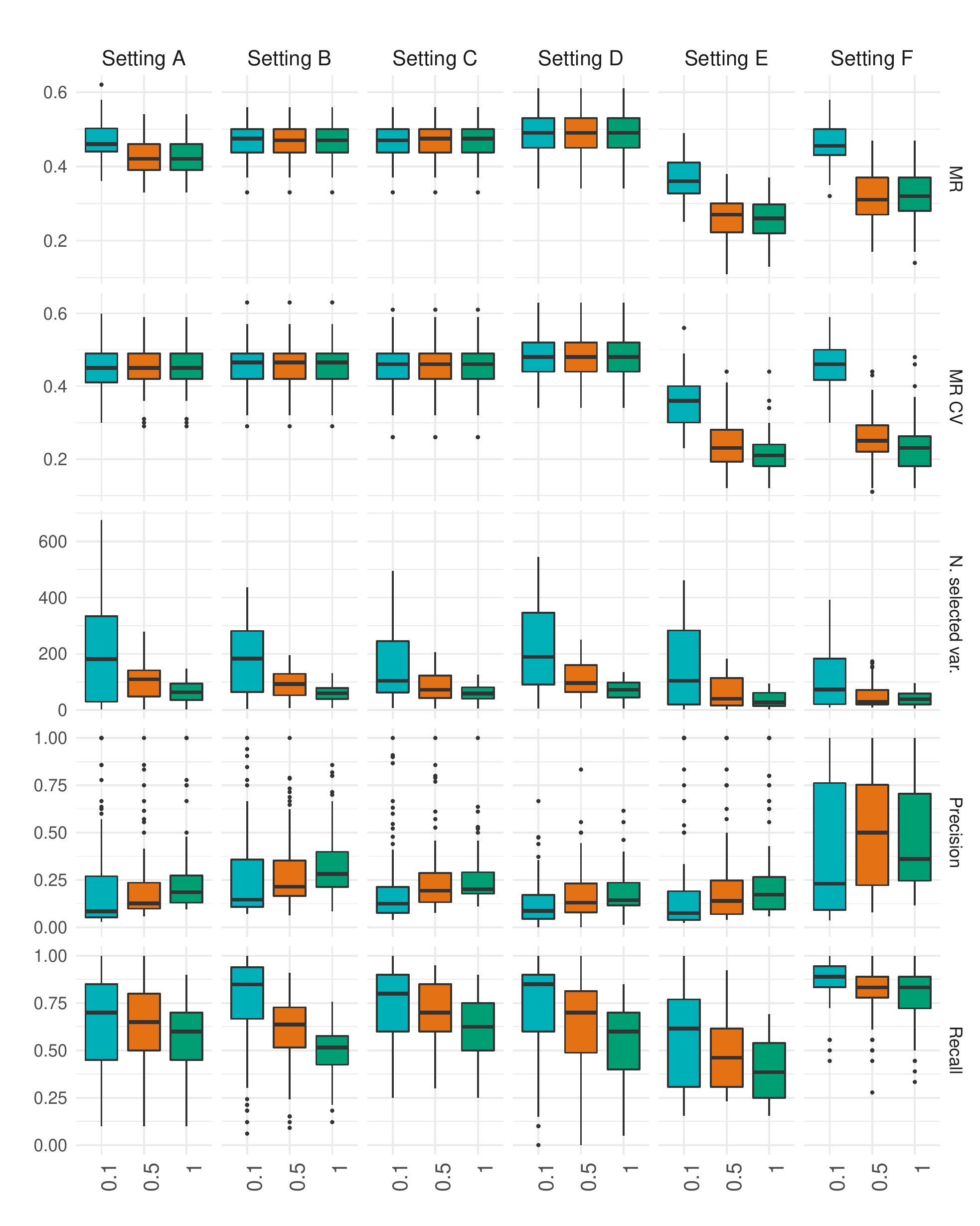}
\caption[Simulation: choice of $\alpha$. Block diagonal covariance, 100 non-penalised covariates.]{Simulation study comparing  different values of $\alpha$.
The covariance matrix used here is the block matrix $\Sigma_1$ and $P_N=100$.
``MR'' is the out-of-sample misclassification rate, ``MR CV'' the within-sample misclassification rate. The non-penalised covariates are not included when computing precision and recall.}
\label{fig:simulation-nondiagcov-many-nonpen-choice-of-alpha}
\end{figure}

We compare different values of $\alpha$ for the two-step approach proposed in the main paper, where $\alpha$ is kept fixed. We consider $\alpha=0.1, 0.5$ and $1$ and all the simulations settings compared so far: only two penalised layers (Figures \ref{fig:simulation-diagonalcov-allpen-choice-of-alpha} and \ref{fig:simulation-nondiagcov-allpen-choice-of-alpha}), two non-penalised covariates (Figures \ref{fig:simulation-diagonalcov-two-choice-of-alpha} and \ref{fig:simulation-nondiagcov-two-choice-of-alpha}), few non-penalised covariates (Figures \ref{fig:simulation-diagonalcov-few-choice-of-alpha} and \ref{fig:simulation-nondiagcov-few-choice-of-alpha}), and high number of non-penalised covariates (Figures \ref{fig:simulation-diagonalcov-many-nonpen-choice-of-alpha} and \ref{fig:simulation-nondiagcov-many-nonpen-choice-of-alpha}). 

The number of selected variables, precision and recall are as expected: the number of selected variables and the recall decrease while the precision increases as as the value of $\alpha$ is increased. The MR, however, doesn't follow a clear pattern. Table \ref{table:choice-of-alpha} shows how the out-of-sample MR varies in each simulation setting for increasing values of $\alpha$.

\begin{table}[H]
\centering
\begin{tabular}{l c c c c c c c c}
$P_N$ & 0 &0  & 2  & 2 & 10  & 10 & 100 & 100 \\
Covariance & $\Sigma_0$ & $\Sigma_1$ & $\Sigma_0$ & $\Sigma_1$ & $\Sigma_0$ & $\Sigma_1$ & $\Sigma_0$ & $\Sigma_1$ \\
\hline
Setting A & $=$ &  $\downarrow$ & $=$ & $\downarrow$ & $\downarrow$ & $\uparrow$ & $\uparrow$ & $\downarrow$ \\
Setting B & $=$ &  $\downarrow$ & $=$ & $\downarrow$ &  $\downarrow$ & $\downarrow$ & $=$ & $=$ \\
Setting C & $=$ & $\downarrow$ & $=$ & $\downarrow$ & $?$ & $\downarrow$ & $=$ & $=$ \\
Setting D &  $=$ & $\downarrow$ & $\uparrow$ & $=$ & $=$ & $=$ & $=$ & $=$\\
Setting E & $\downarrow$ & $\downarrow$ & $=$ & $=$ & $\uparrow$ & $=$ & $=$ & $\downarrow$ \\
Setting F & $\downarrow$ & $?$ & $=$ & $=$ & $\uparrow$ & $\downarrow$ & $\uparrow$ & $\downarrow$ \\
\end{tabular}
\caption{Variation in the average MR of the separate EN approach with fixed $\alpha$ when the parameter $\alpha$ is increased.}
\label{table:choice-of-alpha}
\end{table}

\clearpage

\section{Additional data analysis}
\label{sec:additional-data-analysis} 

In this section, additional information on the data analysis presented in the main paper is reported for the model discriminating between obese patients and healthy donors.
The correlations between the anthropometric and biochemical parameters and the 'omic features for the ChIP-seq, RNA-seq and methylation datasets are reported in Section \ref{sec:signature-validation}. In Section \ref{sec:comparison-uni-multivariate} are compared the variables selected by the univariate and multivariate approaches. Finally, in Section \ref{sec:univariate} we compare the multivariate approach used in the main paper to the Mann-Whitney test introduced above.

\subsection{Signature validation}
\label{sec:signature-validation}

\begin{figure}
	\centering
	\includegraphics[width=.45\textwidth]{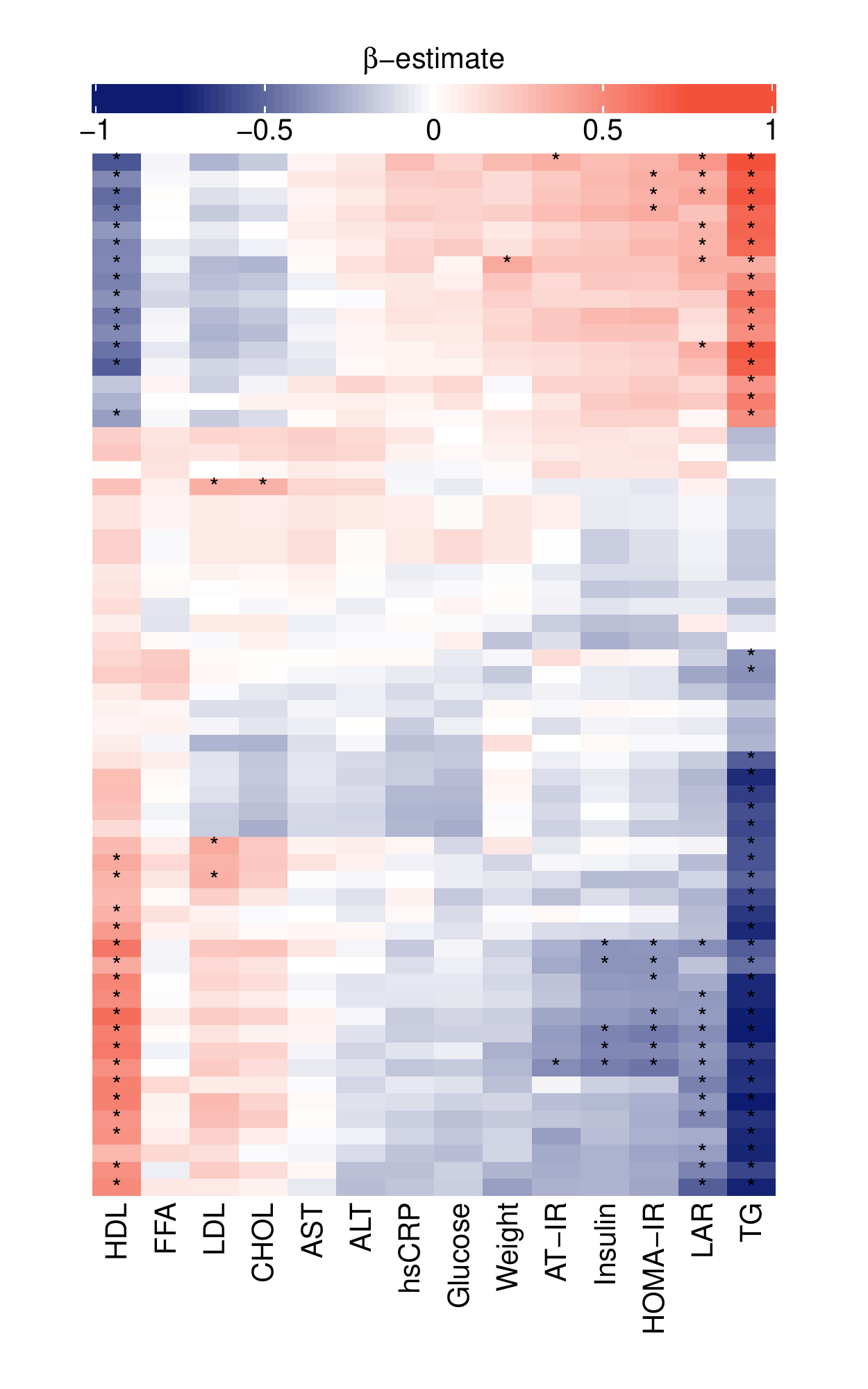}
	\caption{Association of the selected lipids with the anthropometric and biochemical parameters. Cells marked with a star represent associations that are statistically significant with a confidence level of 0.01 after correcting for multiple testing using the Benjamini-Hochberg procedure.}
\label{fig:lipids-validation}
\end{figure}

In order to validate our putative multivariate signatures of CMS, we check whether there is an association between our selected variables and the anthropometric and biochemical parameters. To do so, we fit the following regression model, that adjusts for age and sex, for each pair of anthropometric/biochemical parameter and 'omic measurement:
\begin{equation}
	\text{parameter}_i = \beta_0 + \beta_{\text{age}} \times \text{age}_i + \beta_{\text{female}} \times  \mathbb{I}(\text{female}_i) + \beta_{\text{'omic}} \times \text{'omic}_i + \epsilon,
\end{equation}
where $i = 1, \dots, N$ and $\mathbb{I}$ is the indicator function.
Figure \ref{fig:lipids-validation} shows the coefficients of the selected metabolites and lipids in each of these regressions. Those marked with a star correspond to those for which the null hypothesis is rejected for the test $H_0: \beta_{\text{'omic}} = 0 $ versus $H_1: \beta_{\text{'omic}} \not=0$ at significance level 0.01.
We control the false discovery rate (FDR) by adjusting the $p$-values using the Benjamini-Hochberg procedure. The ChIP-seq and methylation variables have no significant associations, while the RNA-seq only a few (see Supplementary Material).

\begin{figure}[h]
\centering
\begin{subfigure}[t]{.45\textwidth}
\includegraphics[width=\textwidth]{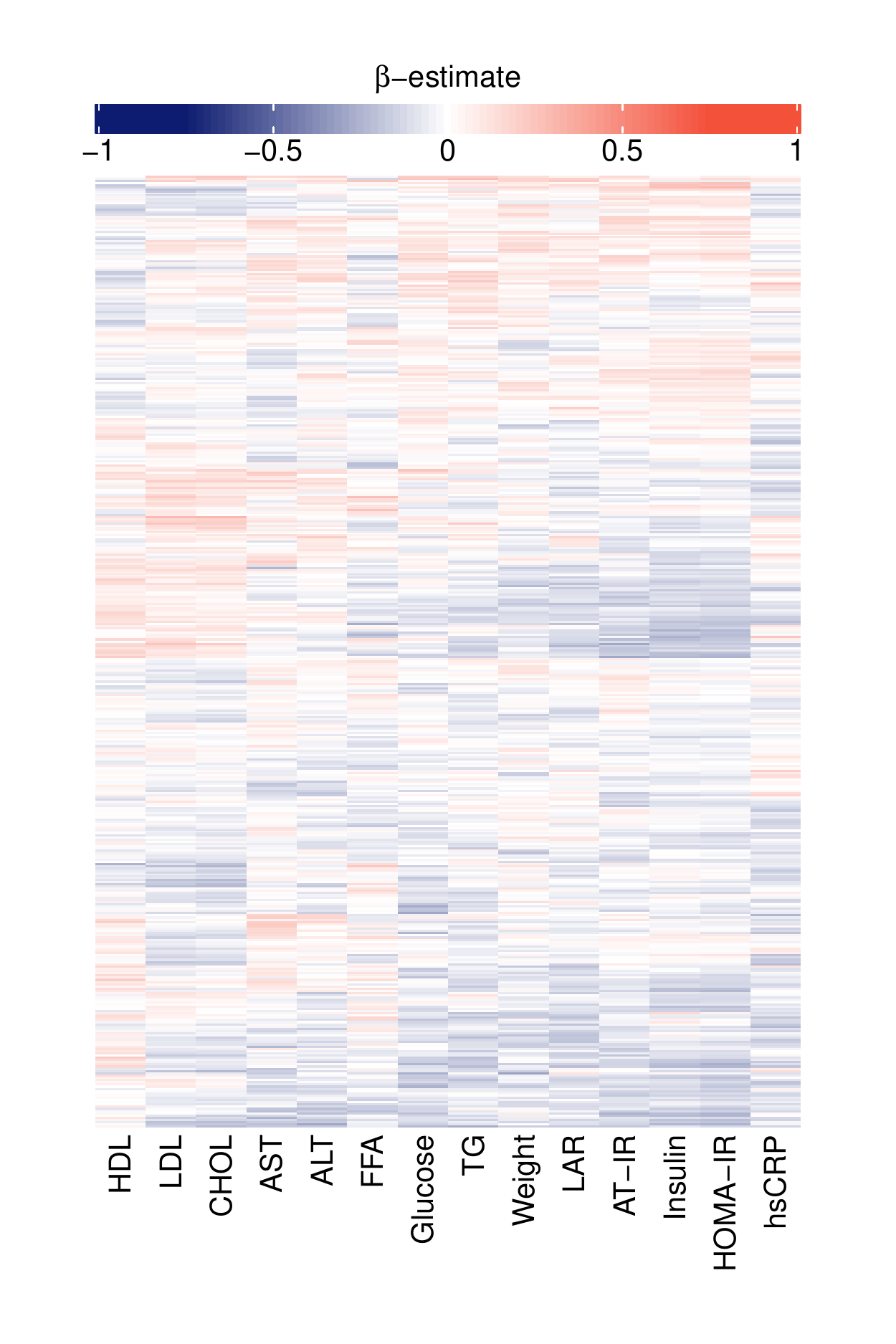}
\caption{Monocytes.}
\end{subfigure}
\begin{subfigure}[t]{.45\textwidth}
\includegraphics[width=\textwidth]{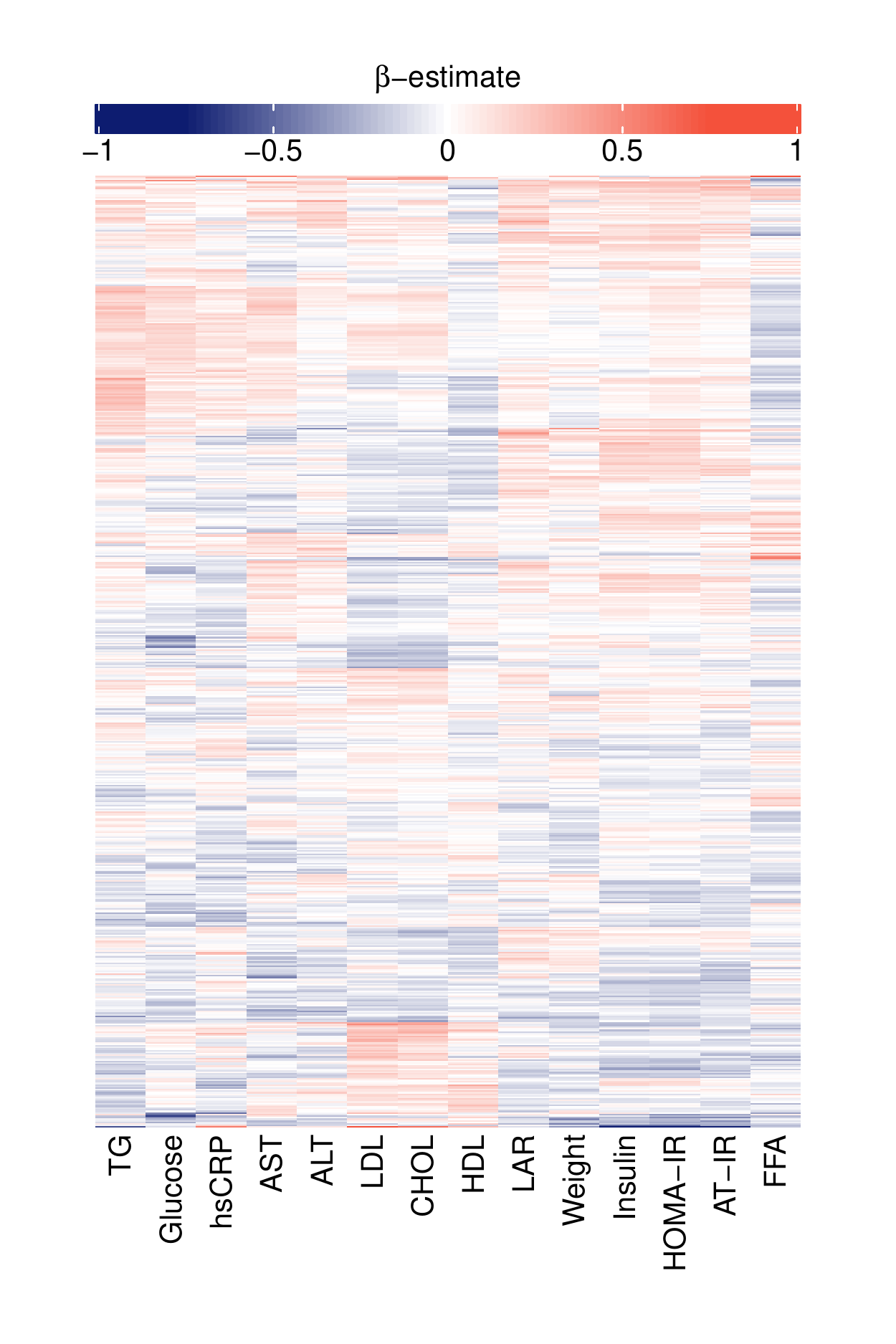}
\caption{Neutrophils.}
\end{subfigure}
\caption{ChIP-seq data.}
\label{fig:chip-seq-validation}
\end{figure}

\begin{figure}[h]
\centering
\begin{subfigure}[t]{.45\textwidth}
\includegraphics[width=\textwidth]{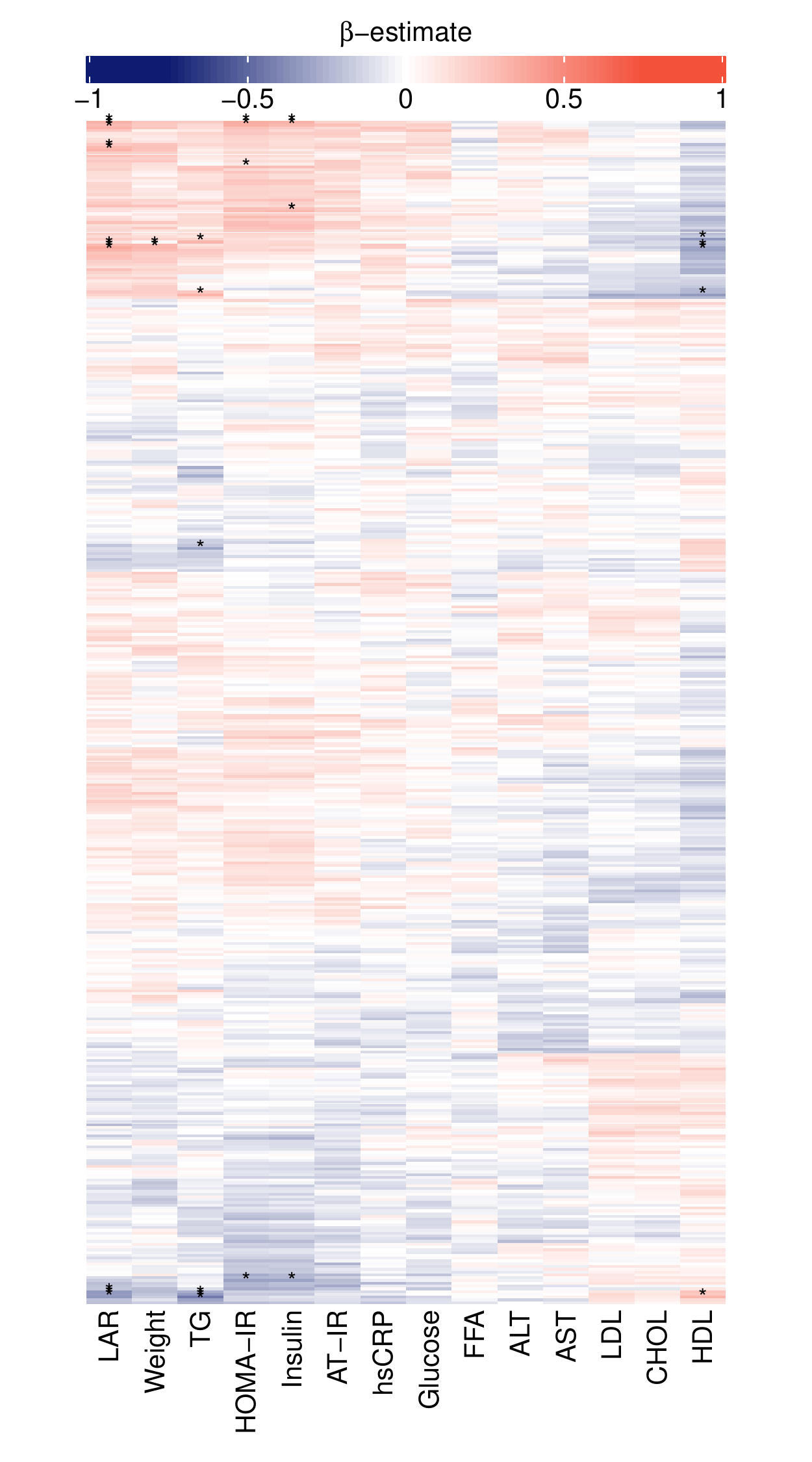}
\caption{Monocytes.}
\end{subfigure}
\begin{subfigure}[t]{.45\textwidth}
\includegraphics[width=\textwidth]{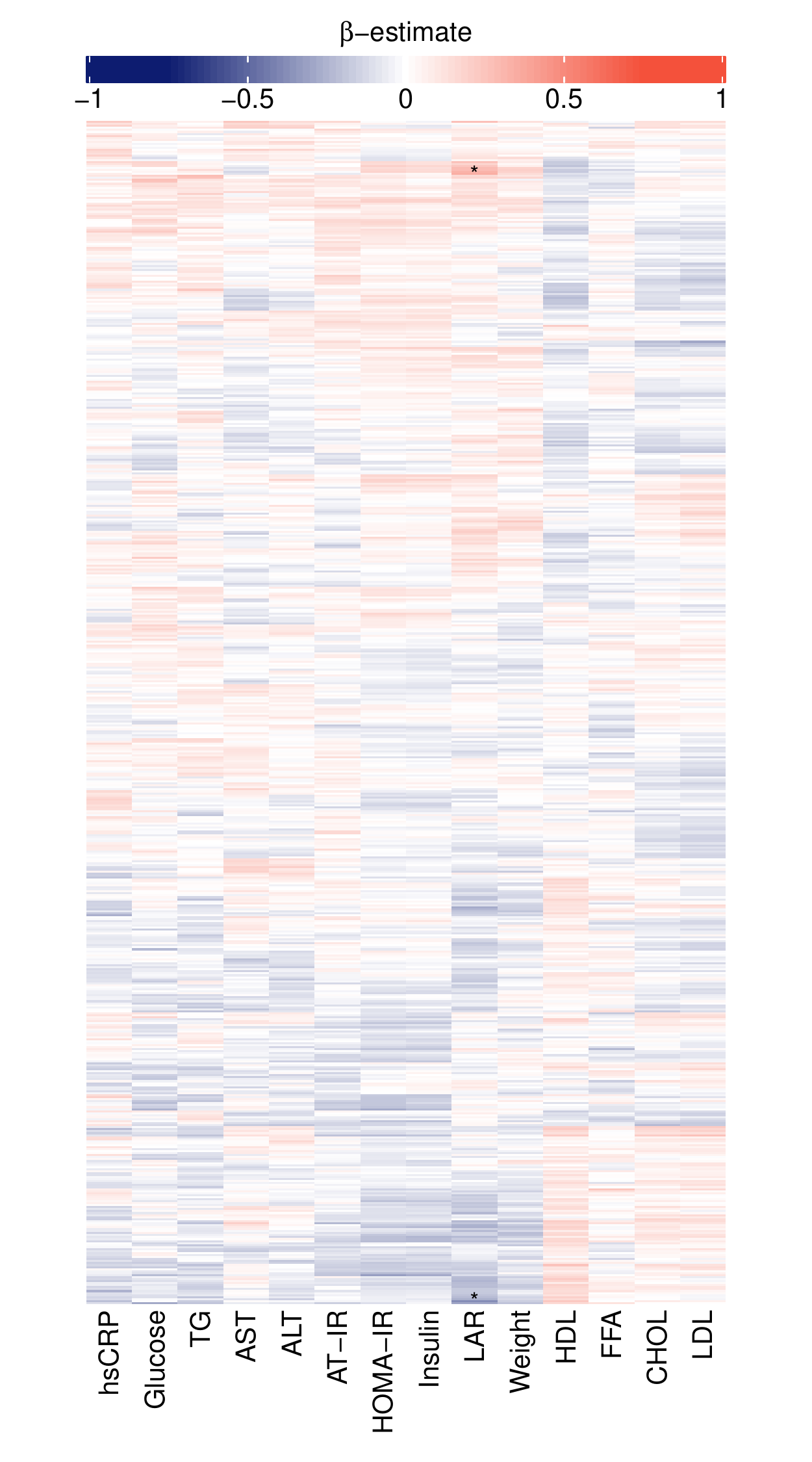}
\caption{Neutrophils.}
\end{subfigure}
\caption{RNA-seq data.}
\label{fig:rna-seq-validation}
\end{figure}

\begin{figure}[h]
	\centering
	\begin{subfigure}[t]{.45\textwidth}
		\includegraphics[width=\textwidth]{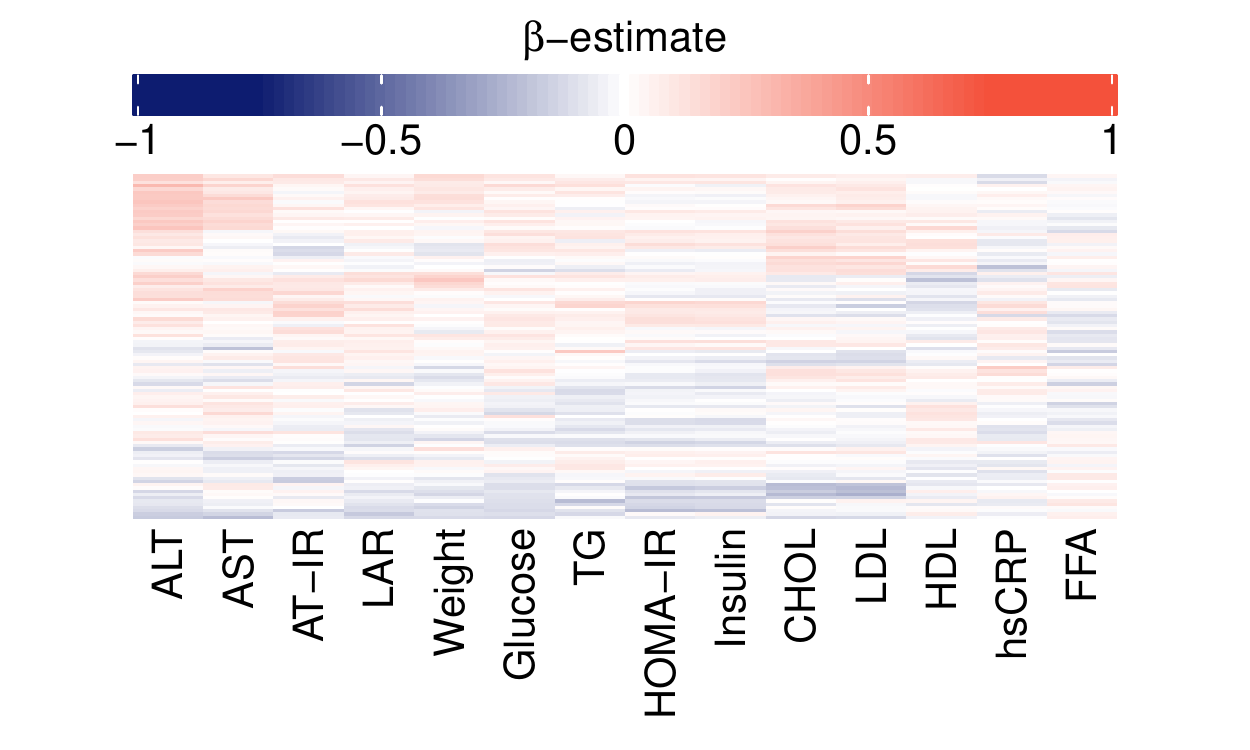}
		\caption{Monocytes.}
	\end{subfigure}
	\begin{subfigure}[t]{.45\textwidth}
		\includegraphics[width=\textwidth]{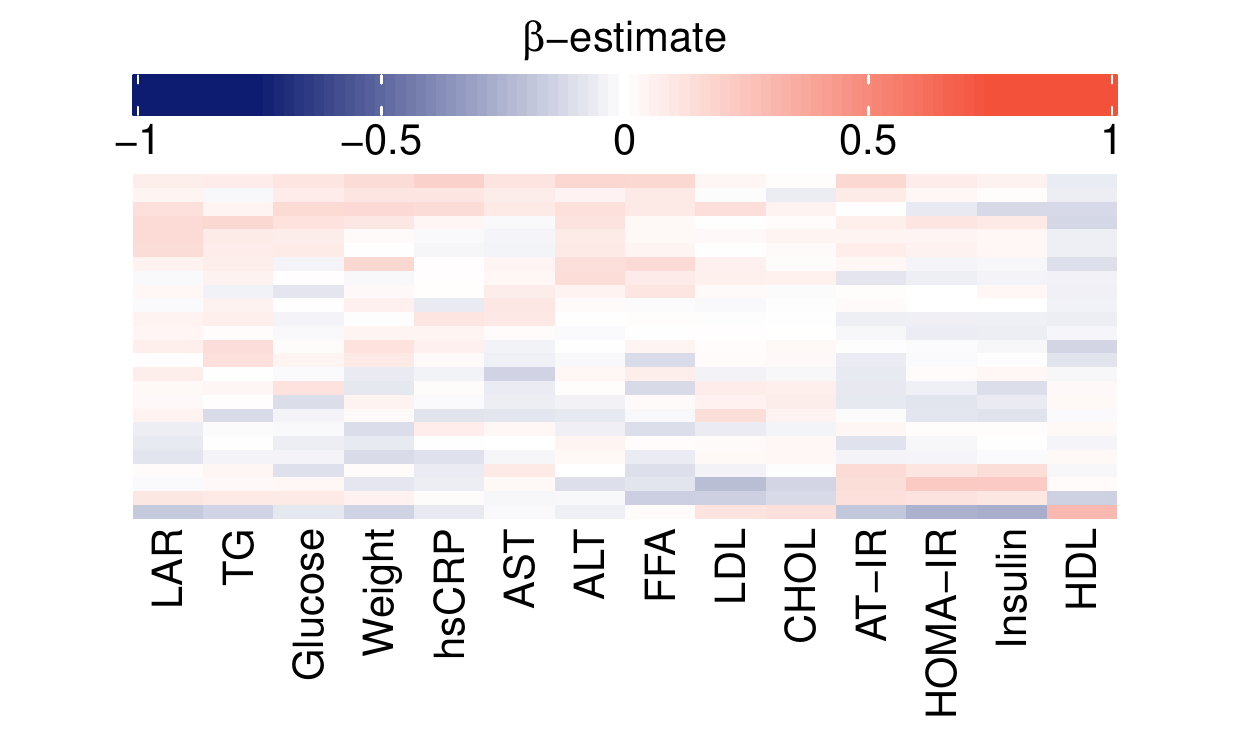}
		\caption{Neutrophils.}
	\end{subfigure}
\caption{Methylation data.}
\label{fig:met-validation}
\end{figure}

The coefficients of regression model of Equation (25) for the ChIP-seq, RNA-seq, and methylation data are shown in Figures \ref{fig:chip-seq-validation}, \ref{fig:rna-seq-validation}, and \ref{fig:met-validation}. In these datasets, the number of tests that are significant at level 0.01 after correcting for multiple testing using the Benjamini-Hochberg procedure is low.

\begin{figure}
	\centering
	\includegraphics[width=.45\textwidth]{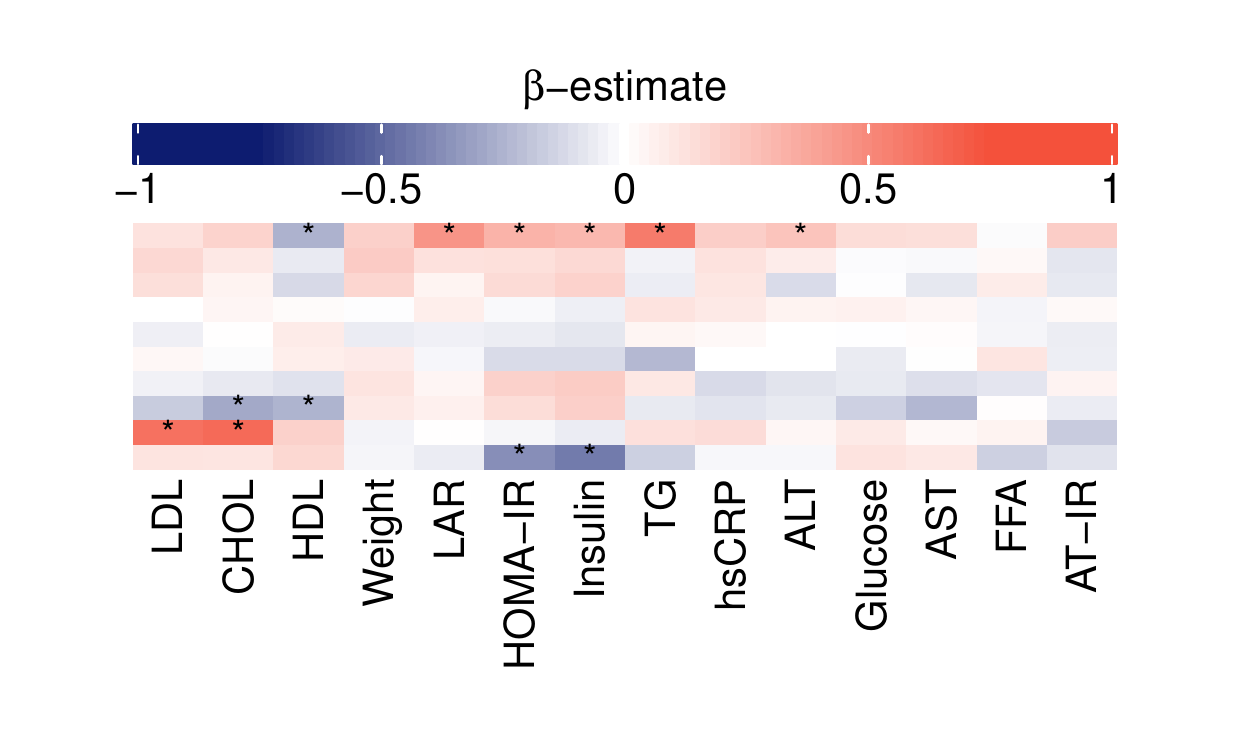}
	\caption{Association of the selected metabolites with the anthropometric and biochemical parameters. Cells marked with a star represent associations that are statistically significant with a confidence level of 0.01 after correcting for multiple testing using the Benjamini-Hochberg procedure.}
\end{figure}

\clearpage

\subsection{Comparison of the multivariate and univariate variable selection approaches}
\label{sec:comparison-uni-multivariate}

In Section 4.2 of the main paper are reported two Venn diagrams representing the intersection between the number of variables selected by the elastic-net (considering both the median and maximal sets of variables over multiple runs of 10-fold cross validation) and those selected via univariate testing. 

In Table \ref{tab:univariate-selection-intersections-bariatric} are indicated the number of variables selected by the multivariate and univariate analyses, and the cardinality of the intersections between those. Table \ref{tab:univariate-selection-intersections-lipodystrophy} contains the number of variables selected by the multivariate and univariate analyses and their intersections. 

\begin{sidewaystable}
\centering
\begin{tabular}{l c c c c c c c}
\hline
 & Max & Mode & Univ. & Max$\cap$Mode & Max$\cap$Univ. & Mode$\cap$Univ. & All\\
\hline
ChIP-seq / Monocytes & 428 & 350 & 0 & 350 & 0 & 0 & 0 \\
ChIP-seq / Neutrophils  & 611 & 611 & 0 & 611 & 0 & 0 & 0 \\
RNA-seq / Monocytes & 425 & 425 & 0 & 425 & 0 & 0 & 0 \\
RNA-seq / Neutrophils & 592 & 588 & 0 & 588 & 0 & 0 & 0 \\
Methylation / Monocytes & 106 & 45 & 0 & 45 & 0 & 0 & 0 \\
Methylation / Neutrophils & 25 & 6 & 0 & 6 & 0 & 0 & 0 \\
Metabolites & 60 & 10 & 0 & 10 & 0 & 0 & 0 \\
Lipids & 62 & 61 & 14 & 61 & 0 & 14 & 14 \\
\hline\\
\end{tabular}
\caption[Variables selected with different methods, obese patients.]{Intersections between the sets variables selected with the multivariate approach with the maximal and modal set of variables and those selected via univariate testing, model trained on the lipodystrophy patients and control donors.}
\label{tab:univariate-selection-intersections-bariatric}

\bigskip\bigskip 

\begin{tabular}{l c c c c c c c}
\hline
 & Max & Mode & Univ. & Max$\cap$Mode & Max$\cap$Univ. & Mode$\cap$Univ. & All\\
\hline
ChIP-seq / Monocytes & 582 & 577 & 0 & 577 & 0 & 0 & 0 \\
ChIP-seq / Neutrophils  & 565 & 440 & 0 & 440 & 0 & 0 & 0 \\
RNA-seq / Monocytes & 455 & 455 & 0 & 455 & 0 & 0 & 0 \\
RNA-seq / Neutrophils & 630 & 623 & 0 & 623 & 0 & 0 & 0 \\
Methylation / Monocytes & 128 & 47 & 0 & 47 & 0 & 0 & 0 \\
Methylation / Neutrophils & 71 & 0 & 0 & 0 & 0 & 0 & 0 \\
Metabolites & 232 & 131 & 0 & 131 & 0 & 0 & 0 \\
Lipids & 73 & 68 & 58 & 68 & 40 & 40 & 40 \\
\hline\\
\end{tabular}
\caption[Variables selected with different methods, lipodystrophy patients.]{Intersections between the sets variables selected with the multivariate approach with the maximal and modal set of variables and those selected via univariate testing, model trained on the lipodystrophy patients and control donors.}
\label{tab:univariate-selection-intersections-lipodystrophy}
\end{sidewaystable}

\subsection{Univariate differential analysis}
\label{sec:univariate}

We now compare the multivariate signature identification method described above to a simpler approach that uses univariate tests. For each of the 'omic dataset considered independently, we use the Mann-Whitney test \citep{mann1947test} to test for differences in distribution for each 'omic variable. In order to control the FDR, we adjust the $p$-values according the Benjamini-Hochberg procedure.
As we shall see, the univariate approach is more conservative as it provides control over the false discovery rate. On the other hand, while the multivariate approach selects variables that are jointly useful for prediction and is therefore more permissive, this is not possible with univariate methods.

We repeat the analysis for both comparisons considered above: bariatric patients versus controls and lipodystrophy patients versus controls. The Mann-Whitney test is performed using the \verb|wilcox.test| function of the \textsf{R} package \verb|stats| \citep{rcoreteam2020R}; the adjusted $p$-values are obtained using the \verb|qvalue| function of the Bioconductor package \verb|qvalue| \citep{storey2019qvalue}.

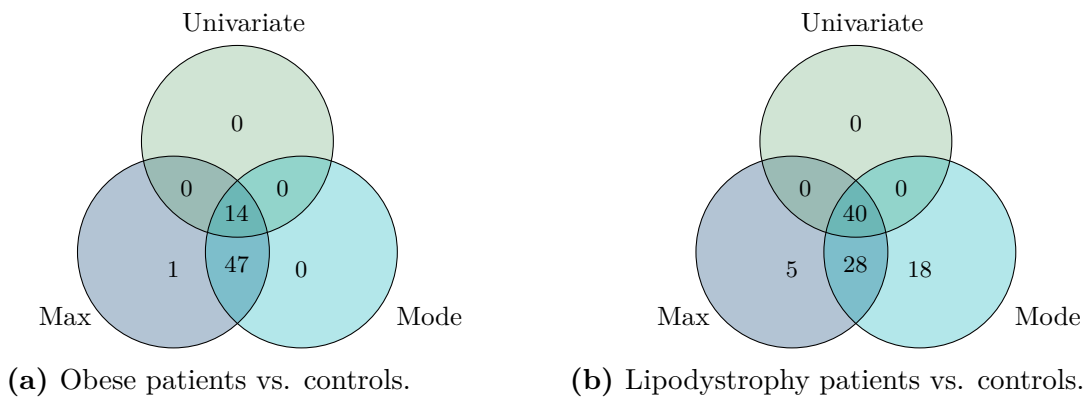
\begin{figure}[H]
\input{venn_diagrams.tex}
	
	\caption[Comparison of variables selected with different techniques.]{Venn diagram showing the intersections between the variables selected in the lipids layer with the multivariate approach with the maximal and modal set of variables and those selected via univariate testing.}
\label{fig:univariate-selection-intersections}
\end{figure}

Due to the fact that this test is very conservative and that it cannot take into account synergies between sets of variables that divide the observations into two groups, the number of selected variables is very low with respect to the multivariate setting. For example, only 14 tests are significant in the lipidomics layer, when comparing the obese patients to the control donors, and no variables show significant differences  in any of the other layers.
In Figure \ref{fig:univariate-selection-intersections} are shown the number of variables that have been selected in the lipidomics layer with the multivariate approach (both choosing the largest set of selected variables and the one that is selected most frequently) which are compared to those selected by univariate testing. The corresponding  information for the other seven layers can be found in the Supplementary Material.

\clearpage
\bibliographystyle{apalike}
\addcontentsline{toc}{section}{Bibliography}
\bibliography{supplement}

%% file: venn_diagrams.tex
\centering
\def\firstcircle{(0,0) circle (1.5cm)}
\def\secondcircle{(60:2cm) circle (1.5cm)}
\def\thirdcircle{(0:2cm) circle (1.5cm)}

	\begin{subfigure}[b]{.49\textwidth}
		\begin{tikzpicture}[scale=0.85]
    			\begin{scope}
        		\fill[cambridgeblue2dark, fill opacity=0.3] \firstcircle;
        		\fill[cambridgegreencore, fill opacity=0.3] \secondcircle;
        		\fill[cambridgebluecore, fill opacity=0.3] \thirdcircle;
        		\draw \firstcircle node[below] {1}; 
        		\draw \secondcircle node [above] {0}; 
        		\draw \thirdcircle node [below] {0}; 
        		\node[text width=1cm] at (0.7,1) {0};
        		\node[text width=1cm] at (2.2,1) {0};
        		\node[text width=1cm] at (1.4, -0.2) {47};
        		\node[text width=1cm] at (1.4, 0.6) {14};
        		\node[text width=4cm, align = center] at (1.1,3.6) {\small Univariate};
        		\node[text width=4cm, align = center] at (-1.7,-1) {\small Max};
        		\node[text width=4cm, align = center] at (4,-1) {\small Mode};
   			\end{scope}
		\end{tikzpicture}
		\caption{Obese patients vs. controls.}
	\end{subfigure}
	\begin{subfigure}[b]{.49\textwidth}
		\begin{tikzpicture}[scale=0.85]
    			\begin{scope}
        		\fill[cambridgeblue2dark, fill opacity=0.3] \firstcircle;
        		\fill[cambridgegreencore, fill opacity=0.3] \secondcircle;
        		\fill[cambridgebluecore, fill opacity=0.3] \thirdcircle;
        		\draw \firstcircle node[below] {5}; 
        		\draw \secondcircle node [above] {0}; 
        		\draw \thirdcircle node [below] {18}; 
        		\node[text width=1cm] at (0.7,1) {0};
        		\node[text width=1cm] at (2.2,1) {0};
        		\node[text width=1cm] at (1.4, -0.2) {28};
        		\node[text width=1cm] at (1.4, 0.6) {40};
        		\node[text width=4cm, align = center] at (1.1,3.6) {\small Univariate};
        		\node[text width=4cm, align = center] at (-1.7,-1) {\small Max};
        		\node[text width=4cm, align = center] at (4,-1) {\small Mode};
    			\end{scope}
			\end{tikzpicture}
			\caption{Lipodystrophy patients vs. controls.}
	\end{subfigure}